\documentclass[aps,prb,reprint,superscriptaddress]{revtex4-1}
\usepackage{amssymb,amsmath,mathptmx}
\usepackage{graphicx}
\usepackage{dcolumn}
\usepackage{bm}
\usepackage{hyperref}
\usepackage{xcolor}
\usepackage[utf8x]{inputenc}
\usepackage{array}
\newcolumntype{C}{>{\centering\arraybackslash}p{1em}}

\newcommand{\be}{\begin{equation}}
\newcommand{\ee}{\end{equation}}
\newcommand{\bea}{\begin{eqnarray}}
\newcommand{\eea}{\end{eqnarray}}
\begin{document}


\title{Inversion in a four-terminal superconducting device on the
  quartet line:\\I. Two-dimensional metal and the quartet beam
  splitter } \author{R\'egis M\'elin}

\affiliation{Univ. Grenoble-Alpes, CNRS, Grenoble INP\thanks{Institute
    of Engineering Univ. Grenoble Alpes}, Institut NEEL, 38000
  Grenoble, France}

\begin{abstract}
  In connection with the recent Harvard group experiment on
  graphene-based four-terminal Josephson junctions containing a
  grounded loop, we consider biasing at opposite voltages on the
  quartet line and establish lowest-order perturbation theory in the
  tunnel amplitudes between a two-dimensional (2D) metal and four
  superconducting leads in the dirty limit. We present in addition
  general nonperturbative and nonadiabatic results.  The critical
  current on the quartet line $I_c(\Phi/\Phi_0)$ depends on the
  reduced flux $\Phi/\Phi_0$ {\it via} interference between the
  three-terminal quartets (3TQ) and the nonstandard four-terminal
  split quartets (4TSQ). The 4TSQ result from synchronizing two
  Josephson junctions by exchange of two quasiparticles ``surfing'' on
  the 2D quantum wake, and this mechanism is already operational at
  equilibrium. Perturbation theory in the tunnel amplitudes shows that
  the 3TQ are $\pi$-shifted but the 4TSQ are $0$-shifted if the
  contacts have linear dimension which is large compared to the
  elastic mean free path. We establish the gate voltage dependence of
  the quartet critical current oscillations $I_c(\Phi/\Phi_0)$.  It is
  argued that ``{\it Observation of $I_c(0)\ne I_c(1/2)$}'' implies
  ``{\it Evidence for the four-terminal 4TSQ}'' for finite bias
  voltage on the quartet line and arbitrary interface
  transparencies. This statement relies on physically-motivated
  approximations leading to the Ambegaokar-Baratoff-type formula for
  the quartet critical current-flux relation. It is concluded that the
  recent experiment mentioned above finds evidence for the
  four-terminal 4TSQ.
\end{abstract}
\maketitle

\section{Introduction}
A superconductor such as Aluminum is characterized by a macroscopic
phase variable $\varphi$ and a gap $\Delta$ separating the collective
BCS ground state from the first quasiparticles. A BCS superconductor
supports dissipationless supercurrent flow in response to phase
gradients.

BCS theory assigns given numerical values to the phase $\varphi$ of a
single superconductor, even if $\varphi$ is a nongauge-invariant
quantity which cannot be observed under any experimental condition.
BCS theory also yields absence of the Meissner effect, {\it i.e.}
BCS superconductors do not repel magnetic field. These paradoxes were
resolved \cite{Anderson-RPA,Anderson-mass} by the so-called Higgs
mechanism {\it i.e.}  a theory of superconductivity which takes
Coulomb interactions into account, and describes the dynamics of the
collective modes in the so-called ``mexican-hat'' potential.

Following the seminal works \cite{Anderson-RPA,Anderson-mass} on gauge
invariance mentioned above, Josephson calculated \cite{Josephson} the
supercurrent through a tunnel junction connecting two superconductors
$S_1$ and $S_2$ with phases $\varphi_1$ and $\varphi_2$. The phase
$\varphi$ of a single superconductor is not gauge invariant, thus it
is not observable. The difference $\varphi_1-\varphi_2$ between the
phases of $S_1$ and $S_2$ is gauge-invariant. The latter is observable
as the following dissipationless current through a
superconductor-insulator-superconductor $S_1IS_2$ Josephson junction:
\begin{equation}
      I=I_c^{(2T)}\sin(\varphi_1-\varphi_2)
\label{eq:Josephson-current-phase}
,
\end{equation}
which has maximal value set by the two-terminal critical current
$I_c^{(2T)}$.

Eq.~(\ref{eq:Josephson-current-phase}) describes the tunneling of
single Cooper pairs between the superconductors $S_1$ and
$S_2$. Composite objects made of two or more Cooper pairs tunnel in
the same quantum event at larger interface transparency. The
possibility of two-Cooper pair tunneling yields the
$\sin\left(2(\varphi_1-\varphi_2)\right)$ term in the following
equation:
\begin{eqnarray}
  \label{eq:Josephson-ordre2-higher-order-1}
  I&=&\left[I_c^{(1),1}+I_c^{(2),1}+...\right]\sin(\varphi_1-\varphi_2)\\
  \label{eq:Josephson-ordre2-higher-order-2}
  &+& \left[I_c^{(2),2}+...\right]
  \sin\left(2(\varphi_1-\varphi_2)\right)\\
  \label{eq:Josephson-ordre2-higher-order-4}
  &+& ...
\end{eqnarray}
Due to their internal structure, both Cooper pairs are coupled to each
other by the Fermi exclusion principle since they are located within
the same coherence volume $\sim \xi^3$ in the same time window
$\tau_\Delta=\hbar/\Delta$, where the zero-energy coherence length is
$\xi=\xi_{ball}(0)$ in the ballistic limit:
\begin{equation}
  \label{eq:xi-ball-0}
    \xi_{ball}(0)=\frac{\hbar v_F}{\Delta}
,
\end{equation}
with $v_F$ the Fermi velocity.

The expansion given by
Eqs.~(\ref{eq:Josephson-ordre2-higher-order-1})-(\ref{eq:Josephson-ordre2-higher-order-4})
shows fast convergence under usual experimental conditions:
Eqs.~(\ref{eq:Josephson-ordre2-higher-order-1})-(\ref{eq:Josephson-ordre2-higher-order-4})
are usually dominated by $I_c^{(1),1}$, such that $|I_c^{(1),1}|\gg
|I_c^{(2),1}|$ and $|I_c^{(1),1}|\gg |I_c^{(2),2}|$.

The following paper demonstrates that {\it multiterminal Josephson
  junctions} offer a playground for investigating the physics of
two-Cooper pair tunneling in connection with interpretation of a
recent experiment realized in the Harvard group
\cite{Harvard-group-experiment}.

Namely, the terms similar to $I_c^{(1),1}+I_c^{(2),1}+...$ in
Eq.~(\ref{eq:Josephson-ordre2-higher-order-1}) {becomes ac in the
  multiterminal Josephson effect (typically in the range of a GHz or
  $10$~GHz), thus not contributing to the dc-current response.} This
offers experimental { signal controlled solely by higher-order
  dc-contributions similar to $I_c^{(2),2}+...$ in
  Eq.~(\ref{eq:Josephson-ordre2-higher-order-2}), without the terms
  similar to $I_c^{(1),1}+I_c^{(2),1}+...$ in the dc-current.}

Concerning multiterminal Josephson junctions, it was shown in
Refs.~\onlinecite{Freyn,Melin1} that nonstandard effects appear in
``supercurrent splitting'', if three superconducting leads are
connected at distance shorter than $\sim\xi$. The {three-terminal
  quartets (3TQ) and higher-order resonances such as the
  three-terminal sextets and octets} were predicted to be revealed
upon voltage biasing $(S_a,S_b,S_c)$ at $(V_a,V_b,V_c)${, see
  Refs.~\onlinecite{Freyn,Melin1}.}  This four-fermion quartet
resonance can be viewed as being ``glued'' by the interfaces, in
absence of pre-existing quartets in the bulk of BCS
superconductors.  Namely, energy conservation puts constraint on the
bias voltages $V_a$ and $V_b$ which have to be ``on the quartet line''
$V_a+V_b=0$ in the $(V_a,V_b)$ voltage plane ($S_c$ being grounded at
$V_c=0$). The predicted
\cite{Freyn,Jonckheere,Rech,Melin1,Melin2,Melin3,Melin-finite-frequency-noise,Doucot}
Josephson anomaly on the quartet line $V_a+V_b=0$ originates from
quantum-mechanically synchronizing the three superconductors
$(S_a,S_b,S_c)$, {\it via} the following gauge-invariant static
combination of their respective macroscopic phase variables:
\begin{equation}
  \label{eq:phiq}
  \varphi_{q,\,3T}=\varphi_a+\varphi_b-2\varphi_c
  .
\end{equation}
The Josephson relations imply that the phase combination $\varphi_a(t)
+ \varphi_b(t) - 2 \varphi_c(t)$ is time $t$-independent, with
$\varphi_a(t) = 2eVt/\hbar + \varphi_a$, $\varphi_b(t) = -2eVt/\hbar +
\varphi_b$, and $\varphi_c(t)=\varphi_c$. The previous difference
$\varphi_1-\varphi_2$ between the phases $\varphi_1$ and $\varphi_2$
of $S_1$ and $S_2$ enters the two-terminal dc-Josephson current-phase
relation given by Eq.~(\ref{eq:Josephson-current-phase}). Conversely,
in a three-terminal Josephson junction, the nonstandard combination
given by Eq.~(\ref{eq:phiq}) implies that the 3TQ current $I_q$ is
given by
\begin{equation}
  \label{eq:Iq-tunnel}
  I_q=I_{c,\,q} \sin\varphi_{q,\,3T}
\end{equation} in the limit of tunnel contacts. Eq.~(\ref{eq:Iq-tunnel})
depends on the phases of the three superconductors through the 3TQ
phase $\varphi_{q,\,3T}$ in Eq.~(\ref{eq:phiq}), not only on the
two-body $\varphi_1-\varphi_2$ entering
Eq.~(\ref{eq:Josephson-current-phase}) for the two-terminal
dc-Josephson effect.

\begin{figure}[htb]
  \includegraphics[width=.8\columnwidth]{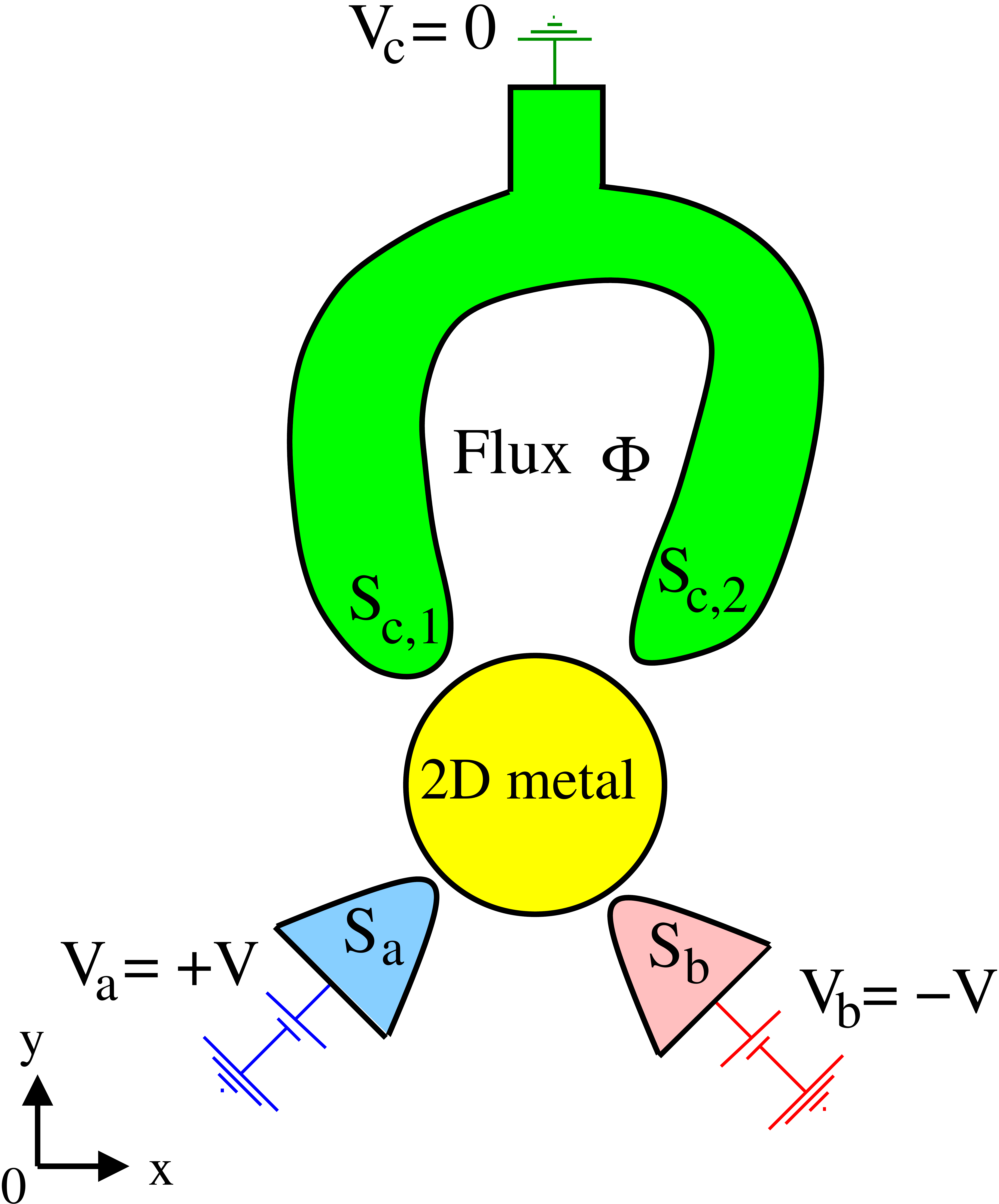}
  \caption{{\it The four-terminal superconducting device:} The
    superconducting leads $S_a$, $S_b$ and $S_c$ are voltage-biased at
    $(V_a,V_b,V_c)$, with $V_a=-V_b\equiv V$ on the quartet line and
    $S_c$ is grounded at $V_c=0$. The loop in $S_c$ terminates at the
    contact points $S_{{c_1}}$ and $S_{{c_2}}$ on the 2D metal used to
    describe the sheet of graphene gated away from the Dirac points in
    the Harvard group experiment \cite{Harvard-group-experiment}. The
    loop is threaded by the magnetic flux~$\Phi$.
    \label{fig:device}}
\end{figure}

The prediction of the 3TQ was confirmed experimentally by the Grenoble
group \cite{Lefloch} (with a metallic structure) {and by the} Weizmann
Institute group \cite{Heiblum} (with a semiconducting nanowire double
quantum dot). The recent Harvard group experiment
\cite{Harvard-group-experiment} provides evidence for unanticipated
features of the quartets in the graphene-based four-terminal device
schematically shown on figure~\ref{fig:device}, in connection with the
additional parameter provided by the flux $\Phi$ in the loop.

\begin{figure*}[htb]
    \includegraphics[width=\textwidth]{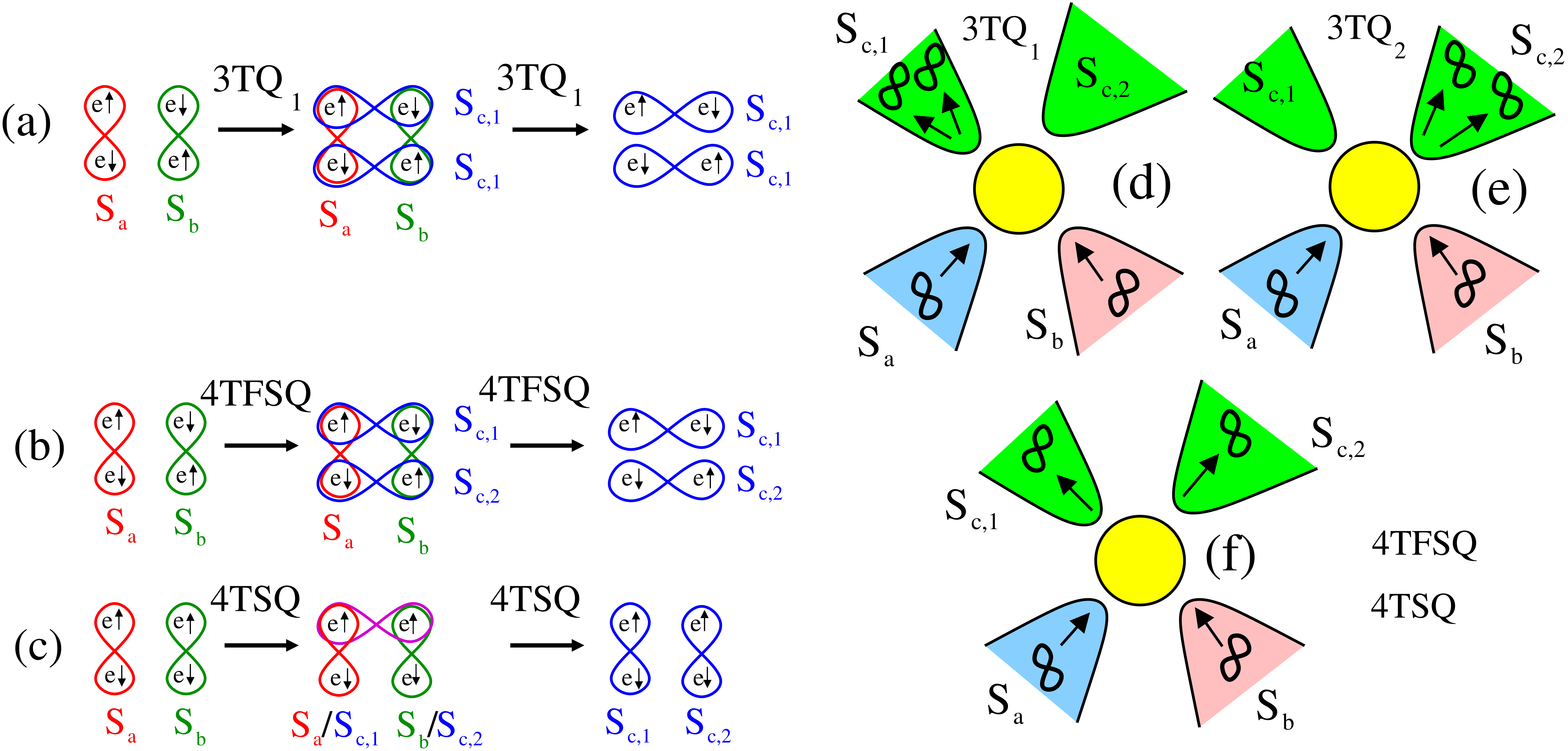}
  \caption{{\it Artist view of the microscopic processes:} The
    three-terminal quartets (3TQ$_1$, panels a and d), the 3TQ$_2$
    (panel e), the four-terminal statistical fluctuations of the split
    quartet current (4TFSQ, panel b and f) and the four-terminal split
    quartets (4TSQ, panel c and f). The two pairs taken from $S_a$ and
    $S_b$ biased at $\pm V$ exchange partners according to the
    ``intermediate state'' represented schematically on panels a, b
    and c for the three-terminal 3TQ$_1$ and the four-terminal 4TFSQ
    and 4TSQ respectively. Two Cooper pairs are transmitted together
    into $S_{{c_1}}$ (3TQ$_1$ on panels a and d), or into $S_{c,2}$
    (3TQ$_2$ on panel e). Alternatively, a single Cooper pair is
    transmitted into $S_{{c_1}}$ and another one into $S_{{c_2}}$ by
    the four-terminal 4TFSQ on panels b and f, and by the 4TSQ on
    panels c and f. A four-particle resonance is produced for the 3TQ
    and the 4TFSQ on panels a and b, {\it i.e.} the two Cooper pairs
    from $S_a$ and $S_b$ recombine after exchanging partners. The 4TSQ
    on panel c involve interchanging a quasiparticle ``surfing'' on
    the quantum wake between the $S_{c,1}$ and $S_{c,2}$ contacts,
    from two Cooper pairs originating from $S_a$ and $S_b$. The
    microscopic mechanism is different for the 3TQ and the 4TFSQ
    (panels a and b), or for the 4TSQ (panel c).
    \label{fig:artist-view}
  }
\end{figure*}

The four-terminal Josephson junction in figure~\ref{fig:device} is an
opportunity for investigating interference in the quartet current, in
the spirit of a superconducting quantum interference device
(SQUID)\cite{SQUID}. Several experiments on multiterminal devices
containing loops have been proposed recently
\cite{Rech,Pillet,Pillet2} in absence of voltage biasing, {\it i.e.}
at equilibrium, where all parts of the circuit are grounded.

The device on figure~\ref{fig:device} was proposed recently
\cite{Nazarov1,Nazarov2} to probe Weyl points and nontrivial
topology. The voltage biasing conditions are different in
Refs.~\onlinecite{Nazarov1,Nazarov2} for topology and
Refs.~\onlinecite{Freyn,Melin1} for the quartets: the voltages are
incommensurate in Refs.~\onlinecite{Nazarov1,Nazarov2}, so as to sweep
the $(\varphi_a,\varphi_b)$ Brillouin zone of the superconducting
phases. Experiments related to the theoretical proposal on topology
\cite{Nazarov1,Nazarov2} were attempted recently
\cite{manip-topology1,manip-topology2,manip-topology3}.

{ Coming back to the Harvard group experiment
  \cite{Harvard-group-experiment}, the emergence of quartet anomaly in
  four-terminal configurations naturally raises the question of making
  the theory of the quartets with four terminals, instead of three
  terminals in the previous theoretical
  \cite{Freyn,Jonckheere,Rech,Melin1,Melin2,Melin3,Melin-finite-frequency-noise,Doucot}
  and experimental \cite{Lefloch,Heiblum} investigations.}  {In this
  sequence of papers~I, II and III, the
  $(S_a,S_b,S_{{c_1}},S_{{c_2}})$ four-terminal device is biased at
  $(V_a,V_b,V_c,V_c)$, where $V_c=0$ is the reference voltage of the
  grounded $S_c$ containing a loop threaded by the magnetic flux
  $\Phi$ and terminated by $S_{{c_1}}$ and $S_{{c_2}}$ (see
  figure~\ref{fig:device}). Our strategy in this series of papers }is
to develop a theory which is intended to interpret the following
unexpected features reported by the Harvard group
\cite{Harvard-group-experiment}:

(i) A quartet Josephson anomaly appears on the $V_a+V_b=0$ quartet
line, once one of the elements of the conductance matrix is plotted in
color as a function of the $(V_a,V_b)$ voltages
\cite{Harvard-group-experiment}. This is compatible with the
theoretical prediction of the quartets for three superconducting
terminals\cite{Freyn,Melin1}, and with the previous Grenoble
  \cite{Lefloch} and Weizmann Institute \cite{Heiblum} group
  experiments.

(ii) In addition, the four-terminal Harvard group experiment
\cite{Harvard-group-experiment} demonstrates oscillations of the
quartet current as a function of the reduced flux $\Phi/\Phi_0$ in the
loop.

(iii) An ``inversion'' appears \cite{Harvard-group-experiment} in a
low bias voltage window, if the experimental data for the amplitude of
the quartet anomaly are plotted as a function of
$\Phi/\Phi_0$. Namely, the quartet anomaly is stronger at
$\Phi/\Phi_0=1/2$ than at $\Phi/\Phi_0=0$ even if superconductivity
should naively be stronger at $\Phi/\Phi_0=0$ than at
$\Phi/\Phi_0=1/2$. The following paper~I addresses a theoretical
description of ``{\it Inversion in $I_c(\Phi/\Phi_0)$ between
  $\Phi/\Phi_0=0$ and $\Phi/\Phi_0=1/2$}'' {on the basis of
  perturbation theory in the tunnel amplitudes within the simplest
  $V=0^+$ adiabatic limit. In addition, the model is generalized
  beyond the perturbative and adiabatic regimes.}

{(iv) Gating away from the Dirac point in the Harvard group experiment
  \cite{Harvard-group-experiment} favors $\pi$-periodicity of
  $I_c(\Phi/\Phi_0)$ with respect to $2\pi$-periodicity. The following
  paper~I turns out to be compatible with this observation.}

(v) A small voltage scale $V_*$ emerges\cite{Harvard-group-experiment}
in the bias voltage $V$-dependence of the quartet signal. Paper~II
addresses how inversion is produced by increasing the bias voltage~$V$
on the quartet line, in the simple situation of a 0D quantum dot.
Paper~III addresses whether a ``Floquet mechanism'' similar to that of
paper~II can extrapolate to the 2D metal of paper~I, in connection
with answering the question of why the voltage $V_*$ for the inversion
is much smaller than the gap in the Harvard group experiment
\cite{Harvard-group-experiment}.

{In short, the progression between the three papers is about different
  levels of the modeling: Paper~I starts from the four-terminal split
  quartets (4TSQ) treated in perturbation in the tunnel amplitudes and
  in the $V=0^+$ adiabatic limit with a 2D metal. Paper~I also
  addresses how the nonstandard four-terminal quartets can be
  generalized to arbitrary device parameters. Paper~II addresses the
  full Floquet theory at finite bias voltage for a zero-dimensional
  (0D) quantum dot, {\it i.e.} how the Floquet spectra and populations
    can produce inversion in $I_c(\Phi/\Phi_0)$ between
    $\Phi/\Phi_0=0$ and $\Phi/\Phi_0=1/2$. Paper~III combines
    papers~I and~II and {addresses finite bias voltage for a} 2D metal
    of paper~I within physically-motivated approximations.}

{The results of the following paper~I which are not presented as
  theoretical support in the experimental Harvard group
  paper\cite{Harvard-group-experiment} are the following:}

{(i) Rigorous microscopic calculation for the sign and the amplitude
  of the critical currents through a 2D metal within perturbation
  theory in the tunnel amplitudes and in the adiabatic limit, taking
  disorder in the superconducting leads in the dirty limit into
  account.}

{(ii) Physically-motivated approximations for addressing the
  nonstandard 4TSQ at arbitrary interface transparencies and finite
  bias voltage.}

In the following paper~I, we propose a simple model for the Harvard
group experiment \cite{Harvard-group-experiment} (see
sections~\ref{sec:themodel}, ~\ref{sec:themethods},
~\ref{sec:perturbation-V0+}, \ref{sec:interference-Q-4TSQ}
and~\ref{sec:why-2D}), and next, the model is analyzed in connection
with this experiment (see
sections~\ref{sec:<>},~\ref{sec:gate-voltage0}
and~\ref{sec:beyond-perturbation-theory}).

The detailed structure of paper~I is this
following. Section~\ref{sec:3papers} summarizes the three papers of
the series. The model and the methods are presented in
sections~\ref{sec:themodel} and~\ref{sec:themethods} respectively.
The three-terminal 3TQ and the four-terminal 4TSQ current-phase
relations are next calculated from perturbation theory in the tunnel
amplitudes combined to the adiabatic limit, see
section~\ref{sec:perturbation-V0+}.
Section~\ref{sec:interference-Q-4TSQ} deals with the interference
between the three-terminal 3TQ and the four-terminal 4TSQ. The
importance of two space dimensions is pointed out in
section~\ref{sec:why-2D}, in connection with the 2D quantum wake.
Section~\ref{sec:<>} shows that ``{\it Relative shift of $\pi$ between
  the three-terminal 3TQ and the four-terminal 4TSQ}'' implies ``{\it
  Inversion in the critical current $I_c(\Phi/\Phi_0)$ between the
  reduced flux values $\Phi/\Phi_0=0$ and $\Phi/\Phi_0=1/2$}''. The
consequence of the model for the gate voltage dependence of the
magnetic field oscillations is discussed in
section~\ref{sec:gate-voltage0} in connection with the Harvard group
experimental paper
\cite{Harvard-group-experiment}. Section~\ref{sec:beyond-perturbation-theory}
discusses arbitrary interface transparencies and finite bias voltage
within the proposed approximations.  A summary and final remarks are
provided in section~\ref{sec:conclusions}.

\section{The three papers of the series}
\label{sec:3papers}

In this section, we present the three papers of the
series. Specifically, the following items (A), (B) and (C) detail
which features of the Harvard group experiment
\cite{Harvard-group-experiment} will be addressed and explained in
which paper I, II or III.
  
(A) The following ``paper I'' starts with the simplest predictive
approach, {\it i.e.} perturbation theory in the interface
transparencies in the adiabatic limit where
$(S_a,S_b,S_{{c_1}},S_{{c_2}})$ are biased at $(V,-V,0,0)$ on the
quartet line, with $V=0^+$. In the context of Cooper pair splitting in
a three-terminal normal metal-superconductor-normal metal ($NSN$)
device, similar perturbative approach \cite{Hekking,Melin-Feinberg}
turned out to usefully uncover the important elementary processes of
{``elastic cotunneling'' \cite{Hekking,Melin-Feinberg} and ``crossed
  Andreev reflection'' \cite{Hekking,Feinberg,Melin-Feinberg}.}
Concerning the four-terminal Josephson junction on
figure~\ref{fig:device}, the following perturbative calculations
reveal the 3TQ \cite{Freyn,Melin1} interfering with the nonstandard
4TSQ.

More precisely, perturbation theory and the adiabatic limit lead to
the three processes which are shown in figure~\ref{fig:artist-view}:

(a) {\it The three-terminal 3TQ$_1$, 3TQ$_2$} in which two pairs
(from $S_a$ and from $S_b$ biased at $\pm V$ respectively) exchange
partners and recombine as two outgoing pairs transmitted at the same
contact {with $S_{{c_1}}$ for the 3TQ$_1$ (or at the contact with
  $S_{{c_2}}$ for the 3TQ$_2$), see figures~\ref{fig:artist-view}a, d
  and e.}

(b) {\it The four-terminal statistical fluctuations of the split
  quartets} (4TFSQ) take one pair from $S_a$, another one from $S_b$
biased at $\pm V$ respectively. Both of them split and recombine as
one pair transmitted into $S_{{c_1}}$ and another one into
$S_{{c_2}}${, see figures~\ref{fig:artist-view}b and f}. The 4TFSQ
solely contribute to small sample-to-sample statistical fluctuations of the
supercurrent.

(c) {\it The four-terminal split quartets} (4TSQ) exchange a
quasiparticle between two pairs taken from $S_a$ and $S_b$ biased at
$\pm V$ respectively. The 4TSQ realize a ``four-terminal quartet beam
splitter'', namely, they take two pairs from $S_a$ and $S_b$, make
their wave-function overlap and transmit a pair into $S_{{c_1}}$ and
another one into $S_{{c_2}}$ in the outgoing state, see
figures~\ref{fig:artist-view}c and f. Contrary to the 4TFSQ of the
previous item~(b), the four-terminal 4TSQ turn out to be robust
against averaging their critical current in the presence of
multichannel contacts.

{It is demonstrated that the three-terminal 3TQ$_1$, 3TQ$_2$} (see the
above item a ) and the four-terminal 4TSQ (the above item c) are $\pi$-
and $0$-shifted respectively, due to the minus sign in the
wave-function of a Cooper pair for the former, and {to the additional
  exchange of two quasiparticles} {\it via} the quantum wake for the
latter.  The critical current is larger at $\Phi/\Phi_0=1/2$ than at
$\Phi/\Phi_0=0$, {\it i.e.} the model of this paper~I produces the
inversion between $\Phi/\Phi_0=0$ and $\Phi/\Phi_0=1/2$ which is also
obtained in the Harvard group experiment
\cite{Harvard-group-experiment}.

{In addition, an approximation on disorder is implemented to address general
  values of the parameters, {\it i.e.} finite bias voltage on the
  quartet line and arbitrary interface transparencies.}

Now, we provide items (B) and (C) summarizing the goals of papers~II
and~III of this series, in connection with explaining the Harvard
group experiment\cite{Harvard-group-experiment}:

{(B) We} propose in the next paper~II a ``Floquet level and population
mechanism'' by which an inversion between $\Phi/\Phi_0=0$ and
$\Phi/\Phi_0=1/2$ is produced by tuning the bias voltage $V$ on the
quartet line. Most of the description in this paper~II is based on a
{simplified} 0D quantum dot configuration supporting a single level at
zero energy. The paper~II {relies on a combination of analytical
  theory and extensive numerical calculations.} {An interesting} link
is established {in paper~II}, which relates the inversion in the
critical current between $\Phi/\Phi_0=0$ and $\Phi/\Phi_0=1/2$ to
repulsion between the Floquet levels as a function of the bias voltage
$V$ on the quartet line.  Robustness of the inversion is established
with respect to crossing-over from weak to strong Landau-Zener
tunneling by changing the couplings between the dot and the
superconducting leads, and with respect to introducing several energy
levels in a multilevel quantum dot. It turns out that the
complementary ``Floquet mechanism'' of paper~II for the inversion
tuned by the voltage $V$ is different in nature from what is proposed
here in the paper~I.

{(C) The last paper~III of the series ``merges'' the papers~I and~II
  into an approximation scheme for the effect of bias voltage within
  the model proposed here in paper~I.} A link is established to the
proximity effect, however taking the specificities of the three- and
four-terminal 3TQ and 4TSQ through a 2D metal into account. In order
to illustrate this point, let us consider a two-terminal normal
metal-superconductor ($NS$) Andreev interferometer containing a loop
in its $N$ part. Electrons with charge $-e$ from $N$ are
Andreev-reflected as holes with charge $e$ and a Cooper with charge
$-2e$ is transmitted into $S$. Doubling the charge for the quartet
mechanism, a pair of electron-like quasiparticles with charge $-2e$
can be reflected as a pair of hole-like quasiparticles with charge
$2e$ while two Coopers with charge $-4e$ are transmitted into
$S_c$. We investigate in paper~III whether this can produce inversion
in the critical current on the quartet line between $\Phi/\Phi_0=0$
and $\Phi/\Phi_0=1/2$ (for instance in connection with figure 3c in
Ref.~\onlinecite{Nazarov-electroflux}). In addition, we obtain
emergence of a small energy scale which is compatible with the
observation \cite{Harvard-group-experiment} of a small voltage scale
$V_*$ in the variations of the critical current with bias voltage $V$.

The above items (A), (B) and (C) summarize the main motivations for
investigating the three complementary mechanisms of papers~I, II and
III. Now, we come specifically to the following paper~I.

\section{The model}

\label{sec:themodel}

{ This section presents the model of the following paper~I.
  The Hamiltonians are provided in subsection~\ref{sec:H-I}. The voltage
  biasing conditions are given in
  subsection~\ref{sec:voltage-biasing-condition}. The critical current
  on the quartet line is defined in
  subsection~\ref{sec:relevant-quantity}, in connection with making
  the link between our calculations and the Harvard group experiment
  \cite{Harvard-group-experiment}.}
\subsection{The Hamiltonians}
\label{sec:H-I}

 {The assumptions of the model are presented in this subsection. The
   essential features of the Harvard group experiment
   \cite{Harvard-group-experiment} are listed} in
 subsection~\ref{sec:ingredients}. The Hamiltonians are presented
 next, first the BCS Hamiltonian of the superconducting leads (see
 subsection~\ref{sec:H-BCS}), next the Hamiltonian of the 2D metal used
 to model the sheet of graphene (see subsection~\ref{sec:H-2D}), and
 finally the term of the Hamiltonian describing the contacts between
 the 2D metal and the superconducting leads (see
 subsection~\ref{sec:tunneling-Hamiltonian}).

\begin{figure*}[htb]
  \includegraphics[width=\textwidth]{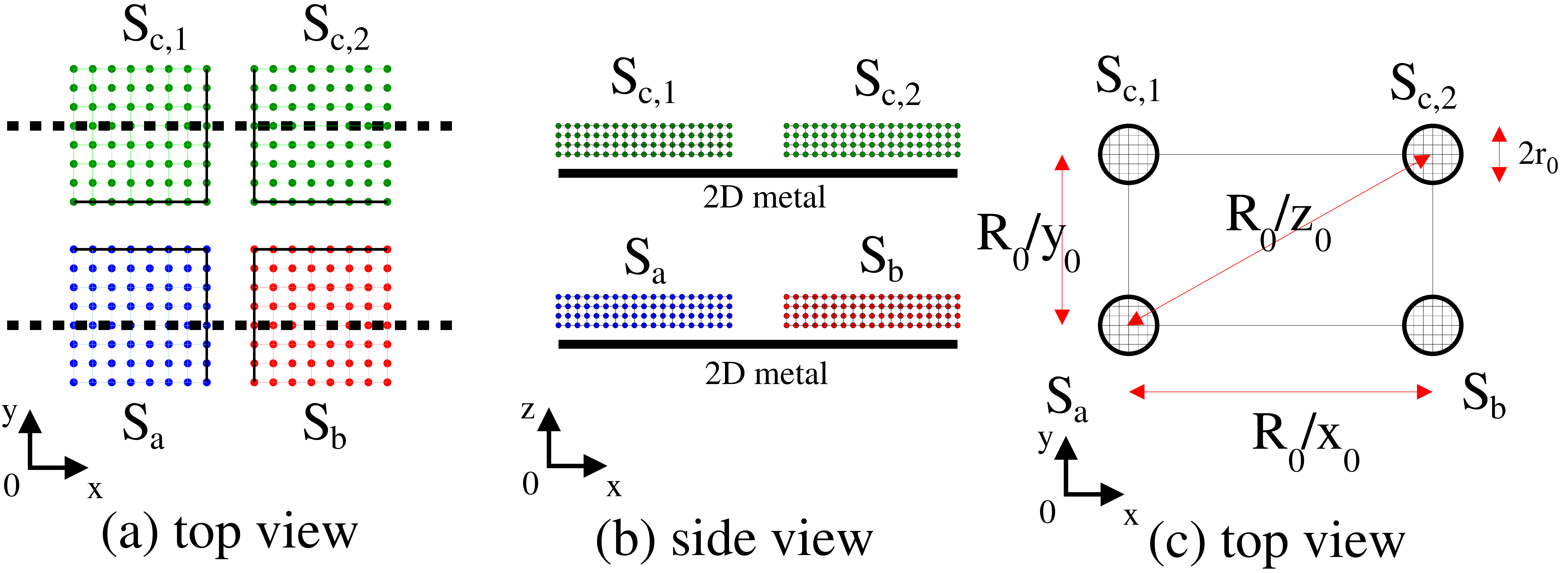}
  \caption{{{\it The four-terminal superconducting device:} Panel a
      shows a schematic top view of the Harvard group device
      \cite{Harvard-group-experiment}, with four superconducting
      terminals $S_a$, $S_b$, $S_{c,1}$ and $S_{c,2}$ evaporated on
      top of the sheet of graphene which is in the $(x,y)$ plane of
      the figure . The grounded loop connecting $S_{c,1}$ to $S_{c,2}$
      is not shown on panel a. Panel b shows cuts in the $(y,z)$ plane
      along the dashed lines on panel a. Panel c shows in the $(x,y)$
      plane a schematic top view of the geometry considered in this
      paper~I, with superconducting contacts having radius $r_0$,
      which can be smaller or larger than the zero-energy dirty-limit
      BCS coherence length.  The separation between the
      superconducting contacts is $R_0/x_0$ along $x$-axis, $R_0/y_0$
      along $y$ axis and $R_0/z_0$ along the diagonals. The
      superconducting leads $S_a$ and $S_b$ are biased at $\pm V$
      while $S_{c,1}$ and $S_{c,2}$ belong to the same superconducting
      loop defined in $S_c$ which is grounded at $V_c=0$ and contains
      a loop pierced by flux $\Phi$, see figure~\ref{fig:device}.}
    \label{fig:device2}}
\end{figure*}

\subsubsection{The essential ingredients}
\label{sec:ingredients}

We start with presenting which ingredients of the Harvard group
experiment \cite{Harvard-group-experiment} are important to our
theoretical description. The model relies on the following facts:

(i) {The superconductors are connected on a 2D metal
  which consists of graphene gated away from the Dirac point, see
  figures~\ref{fig:device} and~\ref{fig:device2}.}

(ii) The experiment involves four terminals instead of three in the
previous theoretical
\cite{Freyn,Jonckheere,Rech,Melin1,Melin2,Melin3,Melin-finite-frequency-noise,Doucot}
and experimental \cite{Lefloch,Heiblum} {papers, see
  figures~\ref{fig:device} and~\ref{fig:device2}.}

{The discussion starts with two limiting cases for the
  device parameters:}

(a) The limit of low-transparency interfaces between the 2D metal and
the superconducting leads.

(b) The $V=0^+$ adiabatic limit with voltage-biasing on the quartet
line.

{The theory is next generalized to arbitrary interface transparencies
  and finite bias voltage within a physically motivated approximation
  regarding disorder.}

{The assumptions about the geometry are illustrated in
  figure~\ref{fig:device2}. Panels a and b show the geometry of the
  Harvard group experiment \cite{Harvard-group-experiment}, with four
  superconducting contacts $S_a$, $S_b$, $S_{c,1}$ and $S_{c,2}$
  evaporated on top of the sheet of graphene. Panel a shows top view
  of the experimental configuration in the plane of the
  $(x,y)$-coordinates. Panel b shows side views in the $(y,z)$-plane,
  {\it i.e.} cuts along the dashed lines on panel
  a. Figure~\ref{fig:device2}c represents the $(x,y)$-plane top view
  of the model considered in this paper~I, in which four
  superconducting leads $S_a$, $S_b$, $S_{c,1}$ and $S_{c,2}$ form
  contacts of radius $r_0$ on the 2D metal, where $r_0$ can be smaller
  or larger than the zero-energy dirty-limit BCS coherence length. The
  separation between the contacts on figure~\ref{fig:device2}c
  corresponds to the parameters $R_0/x_0$ and $R_0/y_0$ along the $x$-
  and $y$-axis directions respectively, and to $R_0/z_0$ along the
  diagonals.}

\subsubsection{BCS Hamiltonian of the superconducting leads}
\label{sec:H-BCS}

Now, we present the standard BCS Hamiltonian of each superconducting
lead taken individually. In zero flux $\Phi/\Phi_0=0$, all
superconducting leads are described by
\begin{eqnarray}
  \label{eq:H-BCS1}
        {\cal H}_{BCS}&=&-W \sum_{\langle i,j \rangle}
        \sum_{\sigma=\uparrow,\downarrow} \left(c_{i,\sigma}^+
        c_{j,\sigma}+ c_{j,\sigma}^+ c_{i,\sigma}\right) \\&-&
        \Delta \sum_i \left(e^{i\varphi} c_{i,\uparrow}^+
        c_{i,\downarrow}^+ + e^{-i\varphi} c_{i,\downarrow}
        c_{i,\uparrow}\right) ,
        \label{eq:H-BCS2}
\end{eqnarray}
where the summation $\sum_{\langle i,j \rangle}$ runs over the pairs
of nearest neighbors {on a 3D tight-binding cubic lattice while
  $\sum_i$ runs over the tight-binding sites.} The notation
$\sigma=\uparrow, \downarrow$ stands for the spin. The first term in
Eq.~(\ref{eq:H-BCS1}) is the kinetic energy. The second term given by
Eq.~(\ref{eq:H-BCS2}) is the BCS mean field pairing with
superconducting gap $\Delta$. The macroscopic superconducting phase
variable is generically denoted by $\varphi$ in Eq.~(\ref{eq:H-BCS2}),
and it takes the values $\varphi_a$, $\varphi_b$,
{$\varphi_{c,1}$ or $\varphi_{c,2}$} according to which of the
superconducting lead $S_a$, {$S_b$, $S_{c,1}$ or $S_{c,2}$} is
considered.

A magnetic field in the loop is taken into account in the following
gauge:
\begin{eqnarray}
  \label{eq:gauge1}
  \varphi_{c,\,1}&=&\varphi_c-\frac{\Phi}{2}\\ \varphi_{c,\,2}
  &=&\varphi_c+\frac{\Phi}{2}
  \label{eq:gauge2}
  ,
\end{eqnarray}
with a phase gradient along the loop $S_c$ terminated by $S_{{c_1}}$ and
$S_{{c_2}}$, supposed to have perimeter large compared to the
superconducting coherence length.

\subsubsection{Hamiltonian of the 2D metal}
\label{sec:H-2D}

Now, the 2D metal Hamiltonian is presented, see the yellow region on
figure~\ref{fig:device}:
\begin{equation}
  \label{eq:HLR}
  {\cal H}_{2D\,metal}=-W \sum_{\langle i,j \rangle}
  \sum_{\sigma=\uparrow,\downarrow} \left(c_{i,\sigma}^+ c_{j,\sigma}+
  c_{j,\sigma}^+ c_{i,\sigma}\right)
  ,
\end{equation}
where the summation $\sum_{\langle i,j \rangle}$ runs over pairs
of neighbors on a {2D tight-binding lattice.}

In the following calculations, the 2D metal is considered to be infinite
in the $x$- and $y$-axis directions, which is compatible with the
large sheet of graphene used in Harvard group experiment
\cite{Harvard-group-experiment}, having typical dimension $\sim
10\,\mu$m.

{We simply take the continuum limit for a 2D Fermi gas with circular
  Fermi surface, parameterized by the single Fermi wave-vector $k_F$
  {and the band-width $W$.} The assumption of circular Fermi surface
  can be realized approximately {from the generic tight-binding
    Hamiltonians} given by Eq.~(\ref{eq:HLR}) at low or high filling,
  and it provides a useful phenomenological basis for describing a
  sheet of graphene gated away from the Dirac point, with a minimal
  number of parameters and only two essential ingredients: spin-$1/2$
  fermions and 2D.}

{In spite of its simplicity, it turns out that this circular Fermi
  surface 2D Fermi gas Hamiltonian will be well suited for addressing
  how the gate voltage on the sheet of graphene in the Harvard group
  experiment \cite{Harvard-group-experiment} couples to the signal on
  the quartet line. Approaching the Dirac points with gate voltage
  could be interesting for future experiments, which would require
  taking into account the additional ingredient of the full dispersion
  relation of graphene, including the Dirac cones.  }

\subsubsection{Tunneling between the superconductors and the 2D metal}
\label{sec:tunneling-Hamiltonian}
{ Now, we present the tunnel Hamiltonian between the 2D metal and each
  of the superconducting lead {$S_N$} among
  $\{S_a,S_b,S_{c_1},S_{c_2}\}$.  This coupling Hamiltonian consists
  of hopping between both sides of the junction: {
  \begin{equation}
    \label{eq:J}
          {\cal H}_{T,\,N}=-J_N \sum_{\langle i_N,j_N \rangle}
          \sum_{\sigma=\uparrow, \downarrow} \left( c_{j_N,\sigma}^+
          c_{i_N,\sigma} + c_{i_N,\sigma}^+ c_{j_N,\sigma} \right) ,
\end{equation}}
where the summation {$\sum_{\langle i_N,j_N \rangle}$} runs over the
pairs of sites on both sides of the interfaces.}

{The notations used throughout the paper for labeling the interfaces
  between the 2D metal and the four $S_a$, $S_b$, $S_{c,1}$ and
  $S_{c,2}$ superconducting leads are the following: We denote by
  $a_p$, $b_p$, $c_{1,p}$ and $c_{2,p}$ the tight-binding sites on the
  superconducting side of the contacts, and by $\alpha_p$, $\beta_p$,
  $\gamma_{c_1,p}$ and $\gamma_{c_2,p}$ their counterpart on the 2D
  metal.}

\subsection{Voltage biasing conditions}
\label{sec:voltage-biasing-condition}

The voltage biasing conditions are made explicit in this subsection.
The four-terminal $(S_a,S_b,S_{{c_1}},S_{{c_2}})$ device in
figure~\ref{fig:device} is voltage-biased on the quartet line at
$(V_a,V_b,V_c,V_c),$ with $V_a=-V_b\equiv V$ and $V_c=0$. We implement
the $V=0^+$ adiabatic limit combined to perturbation theory in
$\{J_N\}$ in the following sections~\ref{sec:perturbation-V0+} ,
\ref{sec:interference-Q-4TSQ}, \ref{sec:wake-synchronization},
\ref{sec:<>} and~\ref{sec:gate-voltage0}, see Eq.~(\ref{eq:J}) for
$J_N$. In addition, section~\ref{sec:beyond-perturbation-theory}
addresses the more general conditions of finite bias voltage~$V$ on the
quartet line and arbitrary interface transparencies, within the
physically-motivated approximation for disorder introduced in
section~\ref{sec:connection-ball}.
  
\subsection{A relevant physical quantity}
\label{sec:relevant-quantity}

In this subsection, we present the definition of the critical current
on the quartet line. This quantity is measured in the Harvard group
experiment \cite{Harvard-group-experiment}, and it is evaluated
theoretically in the following papers~I, II and III. The ``critical
current on the quartet line'' $I_c(V,\Phi/\Phi_0)$ is called in short
as ``the critical current'':
\begin{equation}
  \label{eq:Ic-V-Phi}
  I_c(V,\Phi/\Phi_0)=\mbox{Max}_{\varphi_{q,\,3T}} I_S(\varphi_{q,\,3T},V,\Phi/\Phi_0)
  ,
\end{equation}
where $I_c(V,\Phi/\Phi_0)$ is gauge-invariant and the quartet phase
$\varphi_{q,\,3T}$-sensitive $I_S(\varphi_{q,\,3T},V,\Phi/\Phi_0)$ can
be calculated in any gauge. This is why it is legitimate to use the
specific gauge given by Eqs.~(\ref{eq:gauge1})-(\ref{eq:gauge2}).

\section{The methods}
\label{sec:themethods}

{ This section introduces the methods used in this paper I.}{ The
  calculation of the currents is presented in
  subsection~\ref{sec:the-currents}.
  Subsection~\ref{sec:perturbative-expansion-principle} deals with
  their perturbative expansion in the tunnel
  amplitudes. Superconducting diffusion modes are next
  introduced in
  subsection~\ref{sec:mode-averaging}. Subsection~\ref{sec:connection-ball}
  presents the approximations on disorder which will be used in
  section~\ref{sec:beyond-perturbation-theory} to address arbitrary
  interface transparencies and finite bias voltage on the quartet
  line.}

\subsection{Calculation of the current}
\label{sec:the-currents}

This subsection explains the method to evaluate the currents from
the Keldysh Green's functions. Subsection~\ref{sec:bare} presents the
bare Green's functions in absence of the tunnel coupling between the
different leads. The Dyson equations are next presented in
subsection~\ref{sec:Dyson-equilibrium}.
Subsection~\ref{sec:finite-bias-voltage} deals with how the current is
expressed with Keldysh Green's function. The transport formula is next
specialized to the equilibrium and adiabatic limits in
subsection~\ref{sec:specialization}.

\subsubsection{Bare Green's functions}
\label{sec:bare}
{In this subsection, we present the Green's functions in
  absence of the tunnel coupling between the different parts of the
  circuit, {\it i.e.} the bare Green's functions with $J_N=0$ in
  Eq.~(\ref{eq:J}).}

{The two-component Bogoliubov-de Gennes wave-functions
  for spin-up electrons and spin-down holes yield $2\times 2$ matrix
  advanced (or retarded) Green's function describing propagation
  between the tight-binding sites ${\bf x}_1$ and ${\bf x}_2$ at times
  $t_1$ and $t_2$:
  \begin{eqnarray}
    \label{eq:gA-Nambu}
    &&\hat{g}^A_{{\bf x}_1,{\bf x}_2}(t_1,t_2)=-i\theta(t_1-t_2)\\
    \nonumber
   && \left(
    \begin{array}{cc}
    \langle \left\{c_{{\bf x}_1,\uparrow}(t_1) , c^+_{{\bf x}_2,\uparrow}(t_2) \right\}\rangle
      &
      \langle \left\{c_{{\bf x}_1,\uparrow}(t_1) , c_{{\bf x}_2,\downarrow}(t_2) \right\}\rangle\\
        \langle \left\{c^+_{{\bf x}_1,\downarrow}(t_1) , c^+_{{\bf x}_2,\uparrow}(t_2) \right\}\rangle &
          \langle \left\{c^+_{{\bf x}_1,\downarrow}(t_1) , c_{{\bf x}_2,\downarrow}(t_2) \right\}\rangle
    \end{array}
    \right)
    ,
  \end{eqnarray}
where $\left\{A,B\right\}=AB+BA$ is an anticommutator between the
fermionic creation or annihilation operators $A$ and
$B$. Eq.~(\ref{eq:gA-Nambu}) is useful in connection {with the Dyson
  equations, and it can be used to address the time-periodic dynamics
  underlying the emergence of a dc-current of
  quartets\cite{Freyn,Melin1}, as well as arbitrary device parameters
  ({\it i.e.} arbitrary interface transparencies and finite bias
  voltage on the quartet line).  }

{Considering the 2D metal [see the Hamiltonian given by
    Eq.~(\ref{eq:HLR}), taken at low or high filling] and Fourier
  transforming from the time variables $t_1$ and $t_2$ to frequency
  $\omega$, Eqs.~(\ref{eq:Green2D-1})-(\ref{eq:Green2D-2}) in
  Appendix~\ref{app:2D-metal-Greens-function} imply the following
  limiting long-distance behavior of the Green's function for
  $k_FR\gg1$:}
{
  \begin{eqnarray}
  \label{eq:gA-asymptotics}
  g^{A,(1,1)}_{2D\,metal}(R,\omega)
  &=&g^{A,(2,2)}_{2D\,metal}(R,\omega)\\\label{eq:gA-asymptotics-2}
  &\simeq& \frac{i}{W\sqrt{k_F R}}\cos\left(k_F
  R-\frac{\pi}{4}\right)\\ \label{eq:gR-asymptotics-2}
  g^{R,(1,1)}_{2D\,metal}(R,\omega) &=&
  g^{R,(2,2)}_{2D\,metal}(R,\omega)\\ &\simeq& -\frac{i}{W\sqrt{k_F
      R}}\cos\left(k_FR-\frac{\pi}{4}\right)
  \label{eq:gR-asymptotics}
  ,
\end{eqnarray}
where the $(1,1)$ and $(2,2)$ labels in the superscript denote the
Nambu ``spin-up electron'' and ``spin-down hole'' components
respectively. The notation $R=|{\bf x}_1-{\bf x}_2|$ stands for the
separation between ${\bf x}_1$ and ${\bf x}_2$ in real
space. Eqs.~(\ref{eq:gA-asymptotics})-(\ref{eq:gR-asymptotics}) assume
that the separations $R_0/x_0$, $R_0/y_0$ and $R_0/z_0$ between the
contacts are small compared to the zero-energy ballistic-limit
coherence length given by Eq.~(\ref{eq:xi-ball-0}), see
figure~\ref{fig:device2} for the notations $R_0/x_0$, $R_0/y_0$ and
$R_0/z_0$. Taking the short-junction limit $\Delta R / v_F \ll 1$
amounts to substituting the electron and hole wave-vectors $k=k_e$ and
$k=k_h$ in Eqs.~(\ref{eq:gA-asymptotics})-(\ref{eq:gA-asymptotics-2})
and Eqs.~(\ref{eq:gR-asymptotics-2})-(\ref{eq:gR-asymptotics})
respectively with the Fermi wave-vector $k_F$, without accounting for
their different energy-$\omega$ dependence $k_{e,h}=k_F\pm\omega/v_F$.}

{A sanity check of
  Eqs.~(\ref{eq:gA-asymptotics})-(\ref{eq:gR-asymptotics}) is provided
  in section~I of the Supplemental Material \cite{supplemental}, for a
  double junction between a 2D normal metal and 3D normal
  leads. In particular, it is mentioned at the end of section~I in
  the Supplemental Material \cite{supplemental} that
  Eqs.~(\ref{eq:gA-asymptotics})-(\ref{eq:gR-asymptotics}) imply the
  magnetic proximity effect at a 2D metal-3D ferromagnet interface,
  namely, a magnetization in induced in the 2D metal.}

{Considering now the superconducting leads [see the
    Hamiltonian given by Eqs.~(\ref{eq:H-BCS1})-(\ref{eq:H-BCS2})],
  the ballistic nonlocal Green's function of the 3D superconductor $S_N$
  with gap $\Delta$ and phase $\varphi_N$ is the following:
  \begin{eqnarray}
    \nonumber
  &&\hat{g}^A_{{\bf x}_1,{\bf x}_2}(\omega)=
  \frac{1}{W} \frac{1}{k_F R}
  \exp\left\{\left(-\frac{\left|{\bf x}_1 - {\bf x}_2\right|}{\xi_{ball}
    (\omega-i\eta)}\right)\right\}\\
\label{eq:gA-supra-general-ballistique}
&&  \left[\frac{\sin(k_F R)}{\sqrt{\Delta^2-(\omega-i\eta)^2}}
  \left(\begin{array}{cc} -(\omega-i \eta) & \Delta e^{i\varphi_N}\\
    \Delta e^{-i\varphi_N}&-(\omega-i\eta)\end{array}\right)\right.\\
  &&\left. \left. + \cos(k_F R) \left(\begin{array}{cc} -1 & 0 \\ 0 & 1
  \end{array}\right)\right]\right\}
  \nonumber
  .
  \end{eqnarray}
  The Dynes parameter $\eta\ll\Delta$ is viewed as a requirement for
  making the difference between the ``advanced'' and ``retarded''
  Green's functions, or as a phenomenological parameter to capture
  the experimental line-width broadening and relaxation in
  superconductors\cite{Melin3,Kaplan,Dynes,Pekola1,Pekola2}}.
{In addition, the ballistic-limit BCS coherence length
  appearing in Eq.~(\ref{eq:gA-supra-general-ballistique}) is given by
  \begin{equation}
    \label{eq:xi-ball}
  \xi_{ball}(\omega-i\eta)=\frac{\hbar
    v_F}{\sqrt{\Delta^2-(\omega-i\eta)^2}}
  ,
\end{equation}}
which goes to Eq.~(\ref{eq:xi-ball-0}) if $\omega-i\eta\rightarrow 0$.

{At equilibrium {\it i.e.} if $V=0$, the hopping amplitude $\hat{J}_N$
  between the 2D metal and the superconducting lead $S_N$ is given by
  the diagonal $2\times 2$ Nambu matrix
  \begin{equation}
    \label{eq:Nambu-Jp}
    \hat{J}_N=\left(
    \begin{array}{cc}
      J_N & 0 \\ 0 & -J_N
    \end{array}\right)
    .
  \end{equation}
}

\subsubsection{Dyson equations at equilibrium}
\label{sec:Dyson-equilibrium}
{Now, we consider $J_N\ne 0$ in Eq.~(\ref{eq:Nambu-Jp}) and {start
    with equilibrium conditions, {\it i.e.}  all leads are grounded at
    $V=0$ and the superconductors are phase-biased.} All parts of the
  circuit then have identical chemical potential taken as the energy
  reference.}

{The fully dressed advanced and retarded Green's functions $
  \hat{G}^A$ and $\hat{G}^R$ describe the 2D metal connected by finite
  hopping amplitudes $\{\hat{J}_N\}$ to the superconducting
  leads. Their values are obtained from the Dyson equations, which
  take the following form in a compact notation:
  \begin{equation}
    \label{eq:otimes}
    \hat{G}^{A,R} = \hat{g}^{A,R} + \hat{g}^{A,R}
    \otimes \hat{J} \otimes \hat{G}^{A,R}
    ,
  \end{equation}
  where the symbol $\otimes$ is a convolution over time variables
  [such the time variables $t_1$ and $t_2$ in Eq.~(\ref{eq:gA-Nambu})]
  which becomes a simple product after Fourier transforming to the
  frequency/energy $\omega$. A summation over all possible
  tight-binding sites between the 2D metal and the superconducting
  leads is carried out according to
  \begin{eqnarray}
    \label{eq:otimes-expanded}
    \hat{G}^{A,R}_{\alpha_r,\beta_s}(\omega) &=&
    \hat{g}^{A,R}_{\alpha_r,\beta_s}(\omega) + \sum_{\gamma_p}
    \hat{g}^{A,R}_{\alpha_r,\gamma_p}(\omega) \hat{J}_{\gamma_p,c_p}
    \hat{G}^{A,R}_{c_p,\beta_s}(\omega)\\ \label{eq:Dyson-2d-order}
    &=&\hat{g}^{A,R}_{\alpha_r,\beta_s}(\omega)\\
    \nonumber
    &+& \sum_{\gamma_p,\gamma_q}
    \hat{g}^{A,R}_{\alpha_r,\gamma_p}(\omega) \hat{J}_{\gamma_p,c_p}
    \hat{g}^{A,R}_{c_p,c_q}(\omega) \hat{J}_{c_q,\gamma_q}
    \hat{G}^{A,R}_{\gamma_q,\beta_s}(\omega)
    ,
  \end{eqnarray}
  where a closed set of linear equations is obtained at second order
  for $\{\hat{G}^{A,R}_{\alpha_r,\beta_s}\}$ in
  Eq.~(\ref{eq:Dyson-2d-order}).}

\subsubsection{Finite bias voltage on the quartet line}
\label{sec:finite-bias-voltage}
{Finite bias voltage $V\ne 0$ on the quartet line implies a single
  Josephson frequency $2eV/\hbar$ for the considered four-terminal
  Josephson junction biased at opposite voltages, see
  subsection~\ref{sec:finite-bias-voltage}. The periodic time dynamics
  is encoded in the Nambu tunnel amplitudes between the 2D metal and
  the superconducting leads $\{S_N\}$: Eq.~(\ref{eq:Nambu-Jp}) is
  replaced by
  \begin{equation}
    \hat{J}_N(t)=
    \left(\begin{array}{cc}
      J_N \exp(ieV_N t/\hbar) & 0 \\
      0 & - J_N \exp(-ieV_N t/\hbar)
    \end{array} \right)
    ,
  \end{equation}
  where $V_N$ is the voltage $V_N=0,\,\pm V$ at which superconducting
  lead $S_N$ is biased.  At finite voltage $V$, and after Fourier
  transforming from time $t$ to frequency $\omega$, the ``advanced''
  and ``retarded'' Green's functions in
  Eq.~(\ref{eq:otimes})-(\ref{eq:Dyson-2d-order}) become infinite
  matrices having labels in the extended space of the harmonics of the
  Josephson frequency, in addition to being matrices in Nambu.}

{The fully dressed Keldysh Green's function takes the
  form\cite{Caroli,Cuevas}
  \begin{equation}
    \label{eq:expression-de-Gpm}
    \hat{G}^{+,-}= \left(\hat{I}+\hat{G}^R\otimes \hat{J}\right)\otimes
    \hat{g}^{+,-} \otimes\left(\hat{I} + \hat{J}\otimes \hat{G}^A \right)
    .
\end{equation}}

{ The ``bare'' Keldysh Green's function is given by
  \begin{equation}
    \hat{g}^{+,-}(\omega)=n_F(\omega) \left[ \hat{g}^A(\omega) -
      \hat{g}^R(\omega)\right]
    ,
  \end{equation}
  where $n_F(\omega)$ is the Fermi-Dirac distribution function, which
  reduces to the step function $n_F(\omega)=\theta(-\omega)$ in the
  limit of zero temperature.
}

{The current $I_{\alpha \rightarrow a}$ flowing from
  the 2D metal to the superconducting lead $S_a$ at the $\alpha-a$
  contact is given by \cite{Caroli,Cuevas}
  \begin{eqnarray}
    \label{eq:I1}
    - I_{\alpha \rightarrow a}&=&\frac{e}{\hbar} \sum_p \int d\omega
    \left\{ \left[\hat{J}_{a_p,\alpha_p}
      \hat{G}^{+,-}_{\alpha_p,a_p}\right]_{(1,1)/(0,0)}(\omega)\right. \\&&-
    \left[\hat{J}_{a_p,\alpha_p}
      \hat{G}^{+,-}_{\alpha_p,a_p}\right]_{(2,2)/(0,0)}(\omega)\\&&
    -\left[\hat{J}_{\alpha_p,a_p}
      \hat{G}^{+,-}_{a_p,\alpha_p}\right]_{(1,1)/(0,0)}(\omega)\\&& +
    \left.  \left[\hat{J}_{\alpha_p,a_p}
      \hat{G}^{+,-}_{a_p,\alpha_p}\right]_{(2,2)/(0,0)}(\omega)\right\}
\label{eq:I4}
    ,
  \end{eqnarray}
  where ``$(1,1)$'' or ``$(2,2)$'' in the first pair of labels
  correspond to the Nambu components, as in the above equations. The
  notation $(n,m)=(0,0)$ in the second pair of labels denotes the
  static dc-component in the extended space of the harmonics of the
  Josephson frequency $(neV/\hbar,meV/\hbar)$. The variable $p$ in
  Eqs.~(\ref{eq:I1})-(\ref{eq:I4}) runs over the tight-binding sites
  at the interface between the 2D metal and the superconductors, see
  figure~\ref{fig:device2} for the geometry of the contacts.}
Eqs.~(\ref{eq:expression-de-Gpm})-(\ref{eq:I4}) are the starting point
of the demonstration of the generalized Ambegaokar-Baratoff formula at
finite bias voltage~$V$ on the quartet line, see the forthcoming
subsection~\ref{sec:AB-V-1}.

\subsubsection{Specializing to equilibrium and the adiabatic limit}
\label{sec:specialization}
{ Now, we come back to the equilibrium limit $V=0$. The
  Keldysh Green's function given by Eq.~(\ref{eq:expression-de-Gpm})
  simplifies as
\begin{equation}
  \label{eq:Gpm-equilibrium}
  G^{+,-}_{eq}(\omega)=n_F(\omega) \left[
    \hat{G}^A(\omega)
    -\hat{G}^R(\omega)\right]
  .
\end{equation}
Inserting Eq.~(\ref{eq:Gpm-equilibrium}) into
Eqs.~(\ref{eq:I1})-(\ref{eq:I4}) for the current as a function of
$\hat{G}^{+,-}$ leads to the equilibrium current though the
multichannel ``$\alpha,a$'' contact:
\begin{eqnarray}
  \nonumber &&-I_{\alpha\rightarrow a,eq}=\frac{e}{\hbar}\sum_p \int
  d\omega n_F(\omega) \left\{\right.\\ \label{eq:I1-eq}
  &&\left.\left[\hat{J}_{a_p,\alpha_p}
    \left(\hat{G}^{A}_{\alpha_p,a_p}-\hat{G}^{R}_{\alpha_p,a_p}\right)\right]_{(1,1)}(\omega)\right. \\&&-
  \left[\hat{J}_{a_p,\alpha_p}
    \left(\hat{G}^{A}_{\alpha_p,a_p}-\hat{G}^{R}_{\alpha_p,a_p}\right)\right]_{(2,2)}(\omega)\\&&
  -\left[\hat{J}_{\alpha_p,a_p}\left(
    \hat{G}^{A}_{a_p,\alpha_p}-\hat{G}^{R}_{a_p,\alpha_p}\right)\right]_{(1,1)}(\omega)\\&&
  + \left.  \left[\hat{J}_{\alpha_p,a_p}\left(
    \hat{G}^{A}_{a_p,\alpha_p}-\hat{G}^{R}_{a_p,\alpha_p}\right)\right]_{(2,2)}(\omega)\right\}
  \label{eq:I4-eq}
  ,
\end{eqnarray}
where $\alpha_p$ and $a_p$ label the tight-binding sites on the 2D
metal and superconducting sides
respectively. Eqs.~(\ref{eq:I1-eq})-(\ref{eq:I4-eq}) are the starting
point of the perturbative expansion of the current in powers of
$J_0/W$, see the forthcoming section~\ref{sec:perturbation-V0+}.

The matrices $\hat{J}$ [defined by Eq.~(\ref{eq:Nambu-Jp})] and
$\hat{G}$ [defined by Eq.~(\ref{eq:otimes})] appearing in
Eq.~(\ref{eq:I1-eq})-(\ref{eq:I4-eq}) are $2\times 2$ in Nambu, and
the ``$(1,1)$'' or ``$(2,2)$ Nambu components of their product is
evaluated according to the labels in the subscript.}

{The equilibrium current $I_{\alpha\rightarrow a,eq}$
  given by Eqs.~(\ref{eq:I1-eq})-(\ref{eq:I4-eq}) depends on all
  superconducting phase variables $\varphi_a$, $\varphi_b$,
  $\varphi_{c_1}$ and $\varphi_{c_2}$. Gauge invariance implies that
\begin{equation}
  \label{eq:I-eq}
  I_{\alpha\rightarrow a,eq}=I_{\alpha\rightarrow a}\left(
  \varphi_a+\alpha,\varphi_b+\alpha,\varphi_{c_1}+\alpha,\varphi_{c_2}+\alpha
  \right)
\end{equation}
is independent on $\alpha$ because a global superconducting phase is
not measurable.}

{ At finite bias voltage~$V$ on the quartet line, the phase variables
  are given by $\varphi_a = \varphi^{(0)}_a+\psi$, $\varphi_b =
  \varphi^{(0)}_a-\psi$, $\varphi_{c_1} = \varphi_{c_1}^{(0)}$ and
  $\varphi_{c_2} =\varphi_{c_2}^{(0)}$, where $\psi=2eVt$ is linear in
  the time variable $t$. Assuming in addition adiabatic voltage
  biasing at $V=0^+$ leads to slow time-dependence of the variable
  $\psi$. Then, the adiabatic-limit current is obtained by averaging
  Eq.~(\ref{eq:I-eq}) over $\psi$:
\begin{eqnarray}
  \label{eq:I-adiab}
 && I_{\alpha\rightarrow a,adiab}=\int
  \frac{d\psi}{2\pi}\\
  &&I_{\alpha\rightarrow a}
    \left(\varphi_a^{(0)}
    +\psi+\alpha,
    \varphi_b^{(0)}-\psi+\alpha,
    \varphi_{c_1}^{(0)}+\alpha, \varphi_{c_2}^{(0)}+\alpha\right)
    .
    \nonumber
\end{eqnarray}
Energy conservation puts the constraint that, on the quartet line,
$I_{\alpha\rightarrow a,adiab}$ in Eq.~(\ref{eq:I-adiab}) depends only
on the gauge-invariant quartet phase variable
$\varphi_{q,\,3T}=\varphi_a^{(0)}+\varphi_b^{(0)}-2\varphi_c^{(0)}
\equiv \varphi_a+\varphi_b-2\varphi_c$. Gauge invariance implies that
the current $I_{\alpha\rightarrow a,adiab}$ is independent on
$\alpha$, similarly to the previous Eq.~(\ref{eq:I-eq}) corresponding
to equilibrium with $V=0$.}

\subsection{Perturbative expansion of the adiabatic current}
\label{sec:perturbative-expansion-principle}

{This subsection presents how the Dyson Eq.~(\ref{eq:otimes}) is used
  in the forthcoming section~\ref{sec:perturbation-V0+} to produce a
  systematic expansion of the current in powers of the tunnel
  amplitudes $\{J_N\}$ between the 2D metal and the superconductors
  $\{S_N\}$. Iterating Eq.~(\ref{eq:otimes}) produces the series
  \begin{eqnarray}
    \label{eq:pert1}
    G&=&g\\ &+&g\otimes J \otimes g\\ &+&g\otimes J \otimes g \otimes
    J \otimes g\\ &+& g\otimes J \otimes g \otimes J \otimes g \otimes
    J \otimes g\\ &+& g\otimes J \otimes g \otimes J \otimes g \otimes
    J \otimes g \otimes J \otimes g\\ &+& g\otimes J \otimes g \otimes
    J \otimes g \otimes J \otimes g \otimes J \otimes g \otimes J
    \otimes g\\ &+& g\otimes J \otimes g \otimes J \otimes g \otimes J
    \otimes g \otimes J \otimes g \otimes J \otimes g \otimes J
    \otimes g \\ &+& ...  \label{eq:pert2}
    ,
  \end{eqnarray}
  which is inserted into Eqs.~(\ref{eq:I1-eq})-(\ref{eq:I4-eq}) for
  the equilibrium current.

  At each order $J_a^{2 m_a} J_b^{2 m_b} J_{{c_1}}^{2 m_{{c_1}}}
  J_{{c_2}}^{2 m_{{c_2}}}$ in the tunnel amplitudes $\{J_N\}$, the
  expansion given by Eqs.~(\ref{eq:pert1})-(\ref{eq:pert2}) produces a
  finite number of ``closed loop diagrams'' contributing to the
  dc-current, where $m_a$, $m_b$, $m_{{c_1}}$ and $m_{{c_2}}$ are four
  positive integers.}

{As seen from Eqs.~(\ref{eq:I1-eq})-(\ref{eq:I4-eq})
  and from Eqs.~(\ref{eq:pert1})-(\ref{eq:pert2}), this diagrammatic
  expansion has a simple structure, due to the fact that all terms in
  the Hamiltonian are quadratic, see
  Eqs.~(\ref{eq:H-BCS1})-(\ref{eq:H-BCS2}), Eq.~(\ref{eq:HLR}) and
  Eq.~(\ref{eq:J}). The diagrams consist of alternations between:}

{(i) The tunnel amplitudes in and out the 2D metal,
  see Eq.~(\ref{eq:Nambu-Jp}).}

{(ii) Propagation through the 2D metal [see
    Eqs.~(\ref{eq:gA-asymptotics})-(\ref{eq:gR-asymptotics})] or
  through one of the superconducting leads [see
    Eq.~(\ref{eq:gA-supra-general-ballistique})].}

{Some of the relevant diagrams are shown schematically on the
  forthcoming figures~\ref{fig:structure-diagrams},
  \ref{fig:diagrams-quartets-current-2}, \ref{fig:nambu-schematics},
  \ref{fig:diagrams-split-quartets}
  and~\ref{fig:diagrams-downmixing}.}

{The equilibrium current is obtained as a series of diagrams which are
  labeled by the four positive integers
  $(m_a,m_b,m_{{c_1}},m_{{c_2}})$ mentioned above. Assuming identical
  tunnel amplitudes $J_0\equiv J_a=J_b=J_{{c_1}}=J_{{c_2}}$ for all
  contacts produces the prefactor $(J_0)^m$, with
  $m=m_a+m_b+m_{{c_1}}+m_{{c_2}}$. For instance, the three-terminal
  3TSQ$_1$, 3TQ$_2$ appear at the order $m=8$, see the forthcoming
  subsection~\ref{sec:butterfly-quartet-diagrams}. The four-terminal
  4FTSQ and 4TSQ appear at the order $m=12$, see the forthcoming
  subsections~\ref{sec:4TFSQ-process} and~\ref{sec:the4TSQ}.}

{Each Green's function propagating through any superconducting lead $S_N$ is
  within the electron-electron, hole-hole, electron-hole or
  hole-electron Nambu channel. Each electron-hole or hole-electron
  conversion produces $\exp\left(\pm i \varphi_N\right)$, where
  $\varphi_N$ is the macroscopic phase variable of the superconductor
  $S_N$ (which is among $\{S_a,S_b,S_{c_1},S_{c_2}\}$).} {To each
  diagram is thus associated the overall factor
  \begin{equation}
    \label{eq:nanbnc}
\exp\left[i\left(  n_a \varphi_a + n_b \varphi_b + n_{c_1} \varphi_{c_1}
  + n_{c_2} \varphi_{c_2} \right)\right]
  ,
\end{equation}
where $(n_a,n_b,n_{c_1},n_{c_2})$ are four (positive or negative)
integers counting the number and the sign of the electron-hole or
hole-electron conversions in the leads $\{S_a,S_b,S_{c_1},S_{c_2}\}$
respectively, within a given quantum process.}

{Voltage biasing at $V_a=-V_b\equiv V$ on the quartet line (see
  subsection~\ref{sec:voltage-biasing-condition}) implies a constraint
  on $(n_a,n_b,n_{c,1},n_{c,2})$ coming from conservation of energy
  between:}

{ (i) The energy $n_a eV_a$ of the $n_a$ pairs taken
  from $S_a$, and the energy $n_a eV_b$ of the $n_b$ pairs taken from
  $S_b$, and}
  
{ (ii) The energy $(n_{c_1}+n_{c_2}) e V_c=0$ of the
  $n_{c_1}+n_{c_2}$ pairs transmitted into $S_{c_1}$ and $S_{c_2}$
  which are both grounded at $V_c=0$.}

{Energy conservation on the quartet line implies $n_a e V_a + n_b e
  V_b= 0$ and thus $n_a=n_b$. }

{In addition, gauge invariance puts the constraint
  $n_a+n_b+n_{c,1}+n_{c,2}=0$, which is compatible with
  Eqs.~(\ref{eq:I-eq}) and~(\ref{eq:I-adiab}) being independent on
  $\alpha$.}

\subsection{The superconducting diffusion modes}
\label{sec:mode-averaging}

{Now, we discuss the importance of disorder in the superconductors
  supposed to be in the dirty limit, {\it i.e.} the elastic mean free
  path $l_e$ is much shorter than the ballistic-limit coherence length
  $\xi_{ball}(0)$ given by Eq.~(\ref{eq:xi-ball-0}). This realistic
  assumption puts a severe constraints on the diagrammatic
  perturbation theory: The nonlocal Green's functions are gathered in
  a pair-wise manner in a real-space representation, even those
  crossing the ballistic 2D metal. In addition, small disorder in the
  2D metal in the form of nonmagnetic impurities helps gathering the
  Green's function in a pair-wise manner. It is likely that the 4TSQ
  are robust against introducing a small concentration of nonmagnetic
  impurities in the 2D metal, assuming localization length which is
  larger than the separation between the contacts. Clarifying this
  issue in future work requires understanding the fate of the quantum
  wake in the presence of disorder, see section~\ref{sec:why-2D} for
  the quantum wake in the absence of disorder.}

{Considering a superconductor in the dirty limit, the
  disorder-averaged single-particle Nambu Green's function oscillates
  with the Fermi wave-vector $k_F$ [see
    Eq.~(\ref{eq:gA-supra-general-ballistique})] and its envelope
  decays exponentially over the elastic mean free path
  \cite{Abrikosov}. This puts constraint of locality on the
  ``unpaired'' single-particle Green's function at each
  superconducting lead $S_N$.}

{Second, the superconducting diffusion modes are defined as pairs of
  single-particle Green's functions which scatter together on the same
  realization of the disorder. The range of the superconducting
  diffusion modes reaches the dirty-limit coherence length at subgap
  energies, which is much larger than the elastic mean free path for a
  superconductor such as Aluminum in the dirty limit.}

{The calculation of the superconducting diffusion modes in the dirty
  limit generalizes Ref.~\onlinecite{Smith-Ambegaokar}, see
  Appendix~\ref{app:dirty-limit}.}{ The superconducting diffusion
  modes have four Nambu labels $(\tau_1,\tau_2,\tau_3,\tau_4)$
  attached to them, see Appendix~\ref{app:dirty-limit}.  The resulting
  $2^4=16$ terms are provided by
  Eqs.~(\ref{eq:mode-EC})-(\ref{eq:g11g12-2}). They take the following
  form in the ladder approximation:
  \begin{eqnarray}
    \label{eq:diff-mode1}
&& \int \frac{d{\bf k}}{(2\pi)^3} \overline{g_{\tau_1,\tau_2}({\bf
        k},\omega) g_{\tau_3,\tau_4} ({\bf k}+{\bf
        q},\omega)}\\ &=&\frac{1}{16\pi W}
    \frac{1}{2\sqrt{|\Delta|^2-(\omega-i\eta)^2}+{\cal D} q^2}
    F_{(\tau_1,\tau_2)}^{(\tau_3,\tau_4)}
    \left(\frac{\omega-i\eta}{|\Delta|},\varphi_N\right) ,\nonumber
    .
  \end{eqnarray}
  where ${\bf k}$ and ${\bf q}$ are the wave-vectors, ${\cal D}=v_F^2
  \tau/3$ is the diffusion constant with $\tau$ the elastic scattering
  time, and $\varphi_N$ is the superconducting phase variable of the
  superconducting lead $S_N$. The function
  $F_{(\tau_1,\tau_2)}^{(\tau_3,\tau_4)}$ appearing in
  Eq.~(\ref{eq:diff-mode1}) is deduced from
  Eqs.~(\ref{eq:mode-EC})-(\ref{eq:g11g12-2}) in
  Appendix~\ref{app:dirty-limit}, for instance:
  \begin{eqnarray}
    \label{eq:F-3TQ}
  F_{(1,2)}^{(1,2)}\left(\frac{\omega-i\eta}{|\Delta|},\varphi_N\right)
  &=& \frac{|\Delta|^2}{|\Delta|^2-(\omega-i\eta)^2}
  \exp\left(2i\varphi_N\right)\\ F_{(1,1)}^{(1,2)}\left(\frac{\omega-i\eta}{|\Delta|},\varphi_N\right)
  &=& \frac{(\omega-i\eta)|\Delta|}{|\Delta|^2-(\omega-i\eta)^2}
  \exp\left(i\varphi_N\right)
  .
  \label{eq:F-4TSQ}
  \end{eqnarray} 
  The $\overline{g_{(1,2)}g_{(1,2)}}$ superconducting diffusion mode
  in Eq.~(\ref{eq:F-3TQ}) is relevant to the three-terminal 3TQ$_1$,
  3TQ$_2$. Conversely, $\overline{g_{(1,1)}g_{(1,2)}}$ given by
  Eq.~(\ref{eq:F-4TSQ}) is relevant to the four-terminal 4TSQ, as well
  as to the normal metal-superconductor-superconductor ($NSS$) double
  junction considered in Ref.~\onlinecite{NSS}.  Eq.~(\ref{eq:F-3TQ})
  and Eq.~(\ref{eq:F-4TSQ}) are deduced from the corresponding
  Eqs.~(\ref{eq:g12g12-1}) and~(\ref{eq:g11g12-1}) in
  Appendix~\ref{app:dirty-limit}.}

{Fourier transforming Eq.~(\ref{eq:diff-mode1}) from the wave-vector
  ${\bf q}$ to the real-space coordinate $R$ leads to
  \begin{eqnarray}
    \label{eq:modeA}
    &&
    \overline{g_{\tau_1,\tau_2}g_{\tau_3,\tau_4}}(R,\omega)
    = \frac{c}{W^2}\left(\frac{\xi_{dirty}(\omega-i\eta)}{l_e}\right)\\
    &\times& F_{(\tau_1,\tau_2)}^{(\tau_3,\tau_4)}
    \left(\frac{\omega-i\eta}{|\Delta|},\varphi_N\right)
    \exp\left(-\frac{R}{\xi_{dirty}(\omega-i\eta)}\right)
    ,\nonumber
  \end{eqnarray}
  where $c$ is a constant of order unity and
  \begin{equation}
    \label{eq:xi-dirty-def}
  \xi_{dirty}(\omega)\sim \sqrt{l_e \xi_{ball}(\omega)}
\end{equation}
  denotes the superconducting coherence length in the dirty limit.}

{Next, we integrate Eq.~(\ref{eq:modeA}) over the separation $R=|{\bf
    x}_1-{\bf x}_2|$ between the tight-binding sites ${\bf x}_1$ and
  ${\bf x}_2$ at the interface. We distinguish between the following
  two situations:}

{(i) If $r_0\agt \xi_{dirty}(0)$, then
  \begin{equation}
    \label{eq:scaling-with-xi}
\langle \langle g_{\tau_1,\tau_2} g_{\tau_3,\tau_4} \rangle \rangle \simeq
\frac{c'}{2W^2}\frac{\xi_{dirty}(\omega-i\eta)}{l_e}
F_{(\tau_1,\tau_2)}^{(\tau_3,\tau_4)}
\left(\frac{\omega-i\eta}{|\Delta|},\varphi_N\right) ,
  \end{equation}
  where the contact radius $r_0$ is shown on
  figure~\ref{fig:device2}c, $\xi_{dirty}(\omega)$ is given by
  Eq.~(\ref{eq:xi-dirty-def}), $c'$ is a constant of order unity and
  $\langle\langle ... \rangle\rangle$ stands for summation of
  $\overline{g_{\tau_1,\tau_2} g_{\tau_3,\tau_4}}$ over the separation
  $R$ in Eq.~(\ref{eq:modeA}).}

{(ii) Conversely, the assumption $r_0\alt \xi_{dirty}(0)$ leads to
  \begin{equation}
        \label{eq:scaling-with-r0}
    \langle \langle g_{\tau_1,\tau_2} g_{\tau_3,\tau_4} \rangle
    \rangle \simeq \frac{d'}{2 W^2}\frac{r_0}{l_e}
    F_{(\tau_1,\tau_2)}^{(\tau_3,\tau_4)}
    \left(\frac{\omega-i\eta}{|\Delta|},\varphi_N\right) ,
  \end{equation}
  where $d'$ is another constant of order unity.}

{The scaling in
  Eq.~(\ref{eq:scaling-with-xi})-(\ref{eq:scaling-with-r0}) is linear
  in the dirty-limit coherence length or in the radius $r_0$ of the
  contact. This is consistent with the observation that the
  intersection between the 2D Brownian surfaces (resulting from
  scattering on disorder in the superconducting lead $S_N$) and the 2D
  interfaces generically forms a 1D object.}

\begin{figure}[htb]
    \includegraphics[width=\columnwidth]{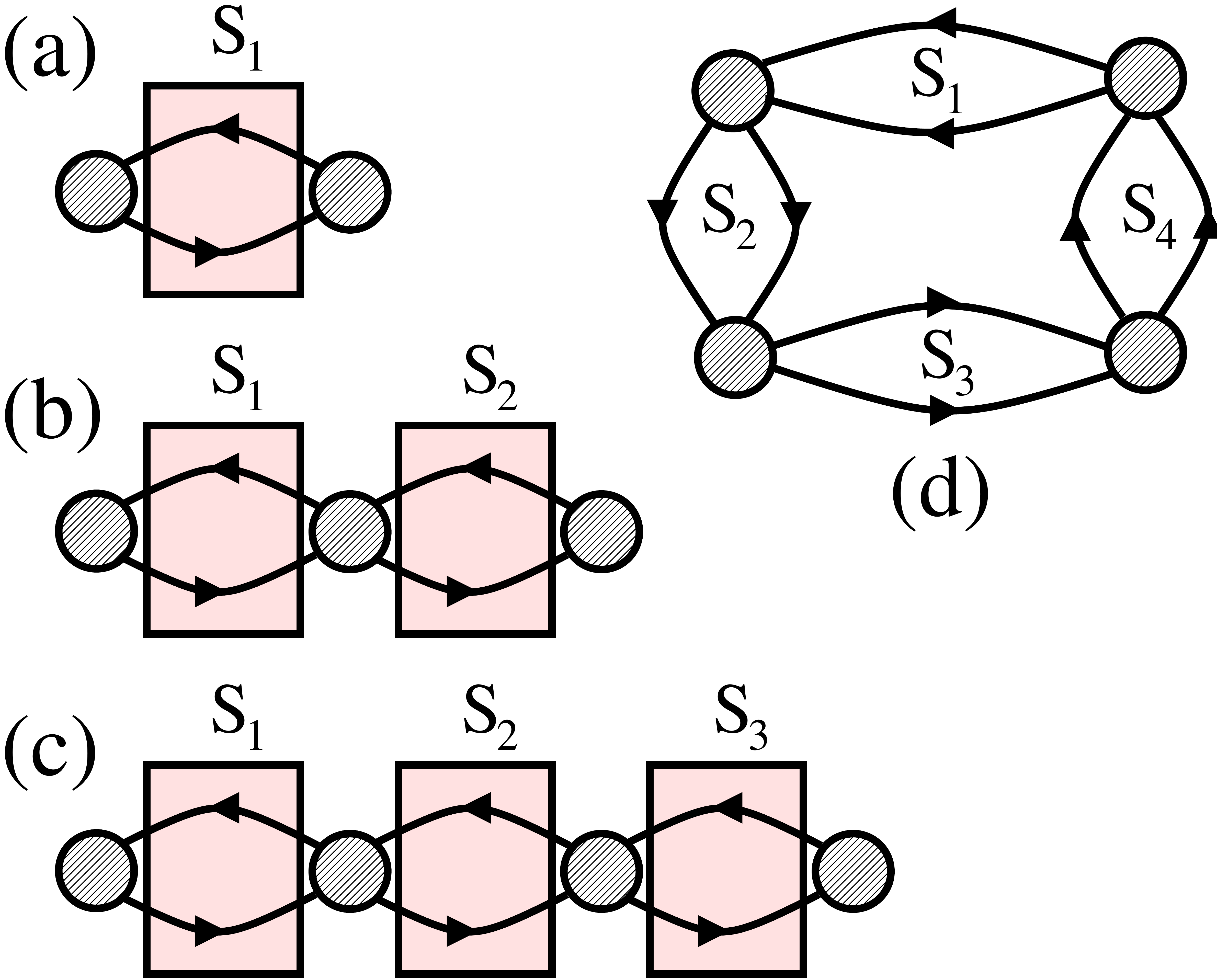}
    \caption{{\it Structure of the diagrammatic series:} The
      ``superconducting diffusion modes'' are made of pairs of
      nonlocal Green's in the superconducting leads denoted by $S_1$,
      ..., $S_4$ on the figure, which are among
      $\{S_a,\,S_b,\,S_{c,1},\,S_{c,2}\}$. The nonlocal
      superconducting modes connect the ``nodes'' corresponding to the
      dashed circled area. The nodes contain dressing by higher-order
      tunneling processes taking place locally between the 2D metal
      and the superconductors, and nonlocal transmission through the
      2D metal. Panels a, b and c show the ``diffuson-like diagrams''
      with superconducting diffusion modes formed of Green's functions
      oriented in opposite directions.  Panel c shows a
      ``weak-localization-like diagram'' with the same orientation for
      the pairs of nonlocal superconducting Green's function.
    \label{fig:structure-diagrams}}
\end{figure}

\begin{figure}[htb]
  \includegraphics[width=.8\columnwidth]{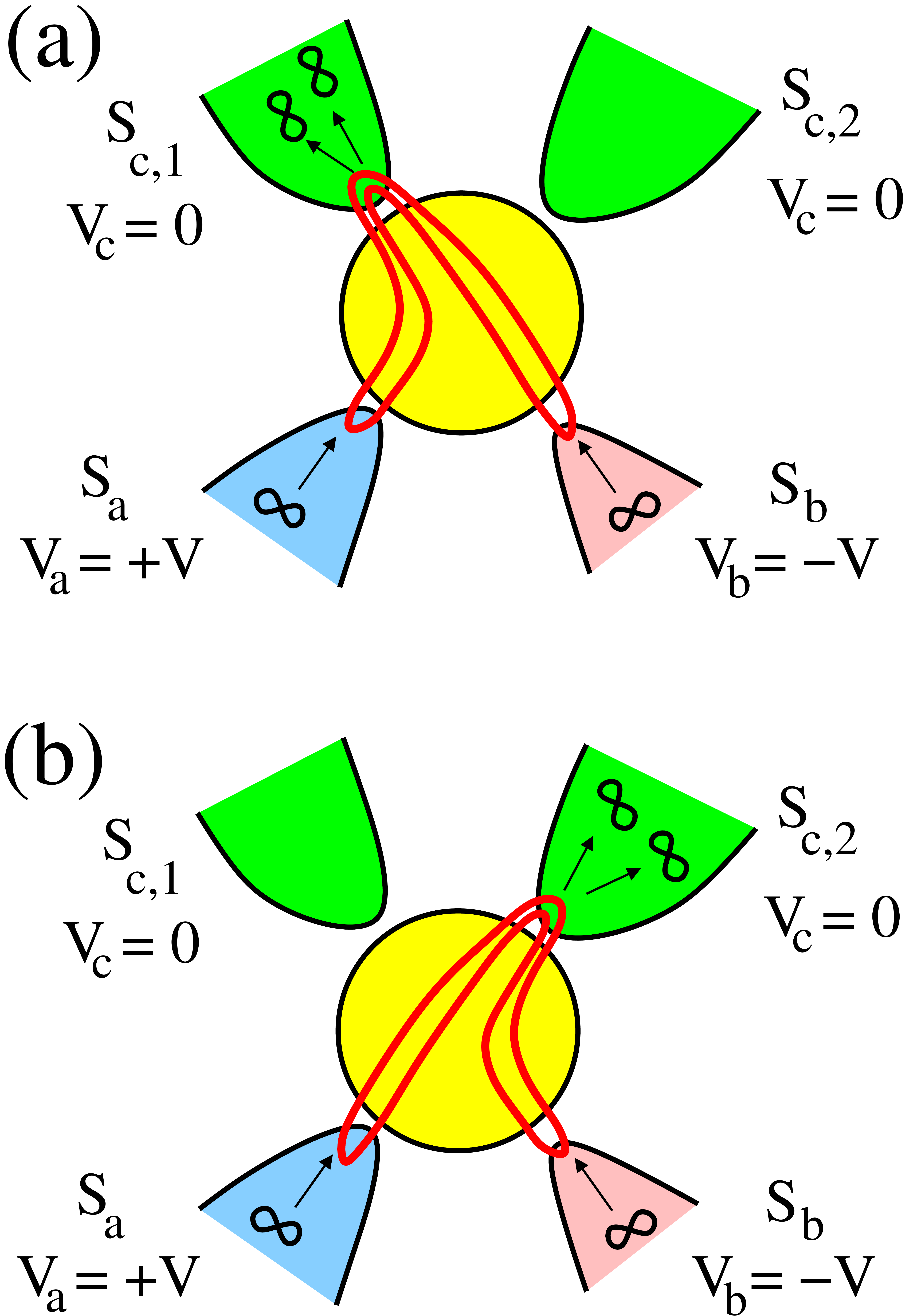}
  \caption{{\it The lowest-order three-terminal 3TQ$_1$, 3TQ$_2$
      diagrams in a real space representation} (on panels a and b
    respectively): Two pairs are taken from $(S_a,S_b)$ biased at
    $(V,-V)$. After making a quartet from taking the square of the
    wave-function of a pair, the two outgoing Cooper pairs are
    transmitted into the grounded $S_{{c_1}}$ for the 3TQ$_1$ (panel
    a) or into $S_{{c_2}}$ for the 3TQ$_2$ (panel b). The 3TQ$_1$ and
    the 3TQ$_2$ current-phase relations are given by
    Eqs.~(\ref{eq:quartet-current-1}) and~(\ref{eq:quartet-current-2})
    respectively. \label{fig:diagrams-quartets-current-2}}
\end{figure}

\begin{figure*}[htb]
  \begin{minipage}{.625\textwidth}
    \includegraphics[width=.95\textwidth]{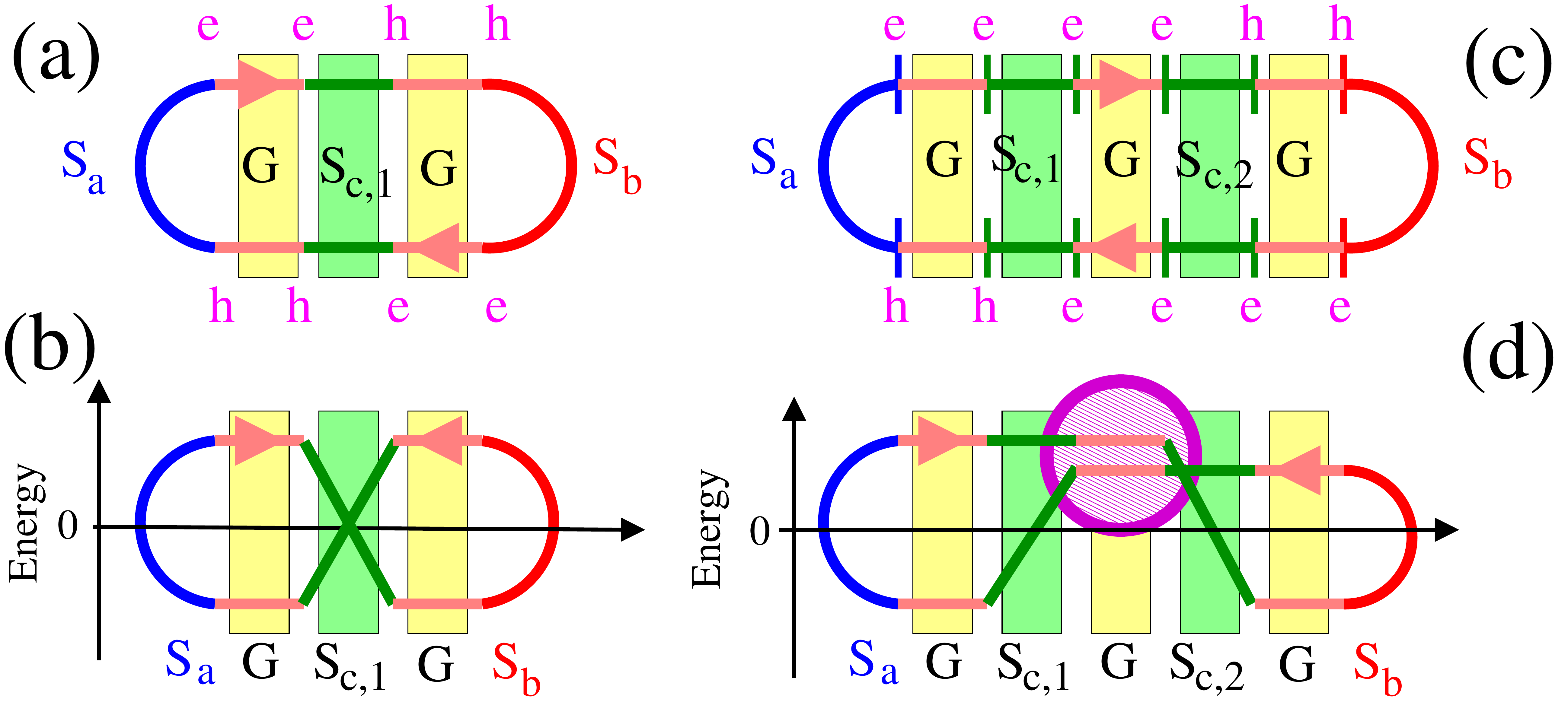}
    \end{minipage}\begin{minipage}{.37\textwidth}
  \caption{{\it Diffuson and energy pictures for the three-terminal
      3TQ$_1$ and for one of the contributions to the four-terminal
      4TSQ:} Panels a and b represent the three-terminal 3TQ$_1$ in
    the diffuson and in the energy pictures respectively. Panels c and
    d show similar representations for the four-terminal 4TSQ. The
    sequence of spin-up electron (e) and spin-down hole (h) Nambu
    labels is indicated on panels a and c. The highlighted section of
    the four-terminal 4TSQ diagram on panel d shows long range
    propagation over the mesoscopic phase coherence length $l_\varphi$
    in between $S_{{c_1}}$ and $S_{{c_2}}$.
    \label{fig:nambu-schematics}}
  \end{minipage}
\end{figure*}

\subsection{Approximation on disorder for
  finite bias voltage and arbitrary interface transparencies}
\label{sec:connection-ball}

{Now, we present a technical introduction to the calculations of the
  forthcoming section~\ref{sec:beyond-perturbation-theory} about the
  interplay between disorder in the superconducting leads, arbitrary
  interface transparencies and finite bias voltage~$V$ on the quartet
  line. We start with what we coin ``the model I'' consisting of the
  four-terminal device on figures~\ref{fig:device}
  and~\ref{fig:device2} with superconductors in the dirty limit
  connected to the 2D metal by clean interfaces, see the tunnel
  Hamiltonian given by Eq.~(\ref{eq:J}).

{Within this model~I, we consider expansion of the current as the
  closed-loop diagrams mentioned above in
  subsection~\ref{sec:perturbative-expansion-principle}. After forming
  the superconducting diffusion modes of
  subsection~\ref{sec:mode-averaging}, these diagrams consist of the
  elements shown on figure~\ref{fig:structure-diagrams}:}

{(i) The ``superconducting diffusion modes'' are pairs of nonlocal
  superconducting Green's functions propagating together in the
  superconductors over the dirty-limit coherence length given by
  Eq.~(\ref{eq:xi-dirty-def}).}

{(ii) The superconducting diffusion modes of item (i) bridge
  between the ``nodes'' shown by the dashed circles on
  figure~\ref{fig:structure-diagrams}. The nodes contain dressing by
  processes taking place locally between the 2D metal and the
  superconductors or nonlocal transmission through the 2D metal.}

We consider now ``the model~II'' as the approximation to the ``model
I''{, see the following Hamiltonian for tunneling between the 2D metal
  and the superconductors within model~II:
\begin{equation}
      \label{eq:HTNeff}
            {\cal H}_{T,\,N,\,eff}=-\sum_{\langle i,j \rangle} \sum_{\sigma=\uparrow,
              \downarrow} \left(
            J_{i\rightarrow j} c_{j,\sigma}^+ c_{i,\sigma}
            +J_{j\rightarrow i} c_{i,\sigma}^+ c_{j,\sigma}
            \right)
            ,
\end{equation}
where the summation $\sum_{\langle i,j \rangle}$ runs over the pairs
of sites on both sides of the contacts. The amplitude for hopping from
$i$ (in the 2D metal layer) to $j$ (the corresponding site in the
superconducting lead) is a complex number with a random phase:
\begin{eqnarray}
  \label{eq:tij1}
  J_{i\rightarrow j}&=&J_0 \exp\left( i \psi_{i\rightarrow
    j}\right)\\
  J_{j\rightarrow i}&=&J_0 \exp\left( i
  \psi_{j\rightarrow i}\right) ,
  \label{eq:tij2}
\end{eqnarray}
where $\psi_{i\rightarrow j} =-\psi_{j\rightarrow i}$ and
$\psi_{i\rightarrow j}$ is uniformly distributed in between $0$ and
$2\pi$. The variables $\psi_{i\rightarrow j}$ and $\psi_{k\rightarrow
  l}$ are uncorrelated if $i,j\ne
k,l$. Eqs.~(\ref{eq:tij1})-(\ref{eq:tij2}) automatically imply
$\langle\langle \left(J_{i\rightarrow j}\right)^2 \rangle\rangle=0$,
which produces a vanishingly small value for the weak
localization-like diagrams \cite{Melin-wl} which intersect the
interface with only two Green's functions. These weak
localization-like diagrams would not be washed out if disorder is
introduced in the amplitudes $|J_{i\rightarrow j}|=|J_{j\rightarrow
  i}|$ instead of the random phases $\psi_{i\rightarrow
  j}=-\psi_{j\rightarrow i}$ in Eqs.~(\ref{eq:tij1})-(\ref{eq:tij2}).}

However, the weak-localization-like diagrams which intersect the
interfaces with four Green's functions at the same tight-binding site
are not washed out by the random $\psi_{i\rightarrow j}$ in Eqs.~(\ref{eq:tij1})-(\ref{eq:tij2}). This is because $\langle\langle|J_{i\rightarrow j}|^4
\rangle\rangle\ne 0$ can be written as $\langle\langle
\left(J_{i\rightarrow j}\right)^2\left(\overline{J_{i\rightarrow
    j}}\right)^2 \rangle\rangle$, where the terms $\left(J_{i\rightarrow
  j}\right)^2$ and $\left(\overline{J_{i\rightarrow j}}\right)^2$ match
both ends of a weak-localization-like loop.

Now, we provide two additional remarks:

(i) Eqs.~(\ref{eq:mode-EC})-(\ref{eq:g11g12-2}) and
Eqs.~(\ref{eq:clean1})-(\ref{eq:clean7}) in the dirty and ballistic
limits respectively have the same dependence on energy-$\omega$, apart
from different prefactors, see Appendices~\ref{app:dirty-limit}
and~\ref{app:clean-limit} respectively.

(ii) The opposite signs of the $\langle\langle{g_{(1,1)}g_{(1,2)}}
\rangle\rangle$ modes in the dirty and ballistic limits (see
subsection~\ref{sec:explanation} of Appendix~\ref{app:clean-limit}) is
not relevant to the four-terminal 4TSQ, because the
$\langle\langle{g_{(1,1)}g_{(1,2)}}\rangle\rangle$ modes come in pairs
within each 4TSQ diagram. Their product has thus necessarily positive
sign.

Based on these remarks on the structure of perturbation theory in the
presence of superconductors in the dirty limit, we propose now ``the
model~III'' which is practically implemented in the forthcoming
calculations of section~\ref{sec:beyond-perturbation-theory} and
includes the same weak-localization-like diagrams as model~I, such as
those on figure~\ref{fig:structure-diagrams}. The model~III makes use
of the nondisordered interfaces of model~I combined to the ballistic
limit Green's functions of model~II.

Specifically, in the model~III, the interfaces are described by
Eq.~(\ref{eq:J}) and now, the $\{k_F R_{k,l}\}$ oscillations at the
scale of the Fermi wave-vector are averaged out in the expression of
the critical currents, where $R_{k,l}$ denotes the separation between
pairs of tight-binding sites at the four interfaces within each part
of the circuit, see
Eqs.~(\ref{eq:gA-asymptotics})-(\ref{eq:gR-asymptotics}) and
Eq.~(\ref{eq:gA-supra-general-ballistique}).

These arguments show that replacing model~I by model~III can be
considered as being legitimate as a physically-motivated approximation
to simulate disorder in the superconductors, {\it i.e.}  to gather the
superconducting Green's functions in a pair-wise manner.  The
model~III is used in the forthcoming
section~\ref{sec:beyond-perturbation-theory} in absence of other known
method to address the interplay between disorder averaging, arbitrary
interface transparencies and finite bias voltage on the quartet line,
taking also the 2D metal into account. Now, we proceed with presenting
our results in themselves.

\section{Current-phase relations of the three-terminal 3TQ and
  the four-terminal 4TSQ}
\label{sec:perturbation-V0+}


In this section, we present a simple model for the microscopic
processes contributing to the $\varphi_{q,\,3T}$-sensitive current on
the quartet line. {The gauge is given by
  Eqs.~(\ref{eq:gauge1})-(\ref{eq:gauge2}), and we calculate the
  currents in perturbation in the tunnel amplitudes and in the
  adiabatic limit.} { Subsection~\ref{sec:butterfly-quartet-diagrams}
  deals with the three-terminal quartets (3TQ$_1$) at the order
  $(J_0/W)^8$, see figure~\ref{fig:artist-view}d.}  {Similarly, the
  3TQ$_2$ at the order $(J_0/W)^8$ are shown in
  figure~\ref{fig:artist-view}e.}  {Subsection~\ref{sec:4TFSQ-process}
  describes the ``four-terminal statistical fluctuations of the split
  quartets'' (4TFSQ) at the order $(J_0/W)^8$, see
  figure~\ref{fig:artist-view}f.}  {Subsection~\ref{sec:the4TSQ}
  presents the four-terminal split quartets (4TSQ) at the order
  $(J_0/W)^{12}$, see figure~\ref{fig:artist-view}f.}

{We microscopically calculate the current-phase relations:

  (i) Eqs.~(\ref{eq:quartet-current-1})-(\ref{eq:quartet-current-2}) for
  the three-terminal 3TQ$_1$ and the 3TQ$_2$.

  (ii) Eq.~(\ref{eq:split-quartet-current}) for the four-terminal 4TFSQ.

  (iii) Eqs.~(\ref{eq:I-(2)})-(\ref{eq:Ic-4TSQ-result-2}) for the
  four-terminal 4TSQ with multichannel contacts.

{These perturbative expansions nontrivially show that the
  three-terminal 3TQ$_1$, 3TQ$_2$ current-phase relations are
  $\pi$-shifted and the four-terminal 4TSQ are $0$-shifted if the
  contact geometry is such that $r_0\gg l_e$, where $r_0$ is shown on
  figure~\ref{fig:device2}c.}
  
\subsection{The Three-Terminal quartets (3TQ$_1$ and 3TQ$_2$)}
\label{sec:butterfly-quartet-diagrams}

\subsubsection{Microscopic calculation of the three-terminal 3TQ$_1$, 3TQ$_2$
  critical currents}

{Now, we consider the three-terminal 3TQ$_1$, 3TQ$_2$ of
  Refs.~\onlinecite{Freyn,Melin1} (see also
  figure~\ref{fig:diagrams-quartets-current-2}), and evaluate them at
  the order $(J_0/W)^8$ in perturbation in the tunnel amplitudes for
  the 2D metal which is relevant to the Harvard group experiment
  \cite{Harvard-group-experiment}. The three-terminal 3TQ$_1$, 3TQ$_2$
  transmit four fermions into the same superconducting lead, {\it
    i.e.} into $S_{c,\,1}$ for the 3TQ$_1$ (see
  figure~\ref{fig:diagrams-quartets-current-2}a) or into $S_{c,\,2}$
  for the 3TQ$_2$ (see
  figure~\ref{fig:diagrams-quartets-current-2}b).}

  The first term
  $\left[\hat{J}_{a,\alpha}\hat{G}^A_{\alpha,a}\right]_{(1,1)}$ in the
  equilibrium current given by Eq.~(\ref{eq:I1-eq}) takes the
  following form, at the lowest order $m=8$ in an expansion in $(J_0/W)^m$
  and in the adiabatic limit:
\begin{widetext}
  \begin{eqnarray}
    \label{eq:order8-1}
J_0 \langle\langle G_{[8],S_{c,1},(\alpha,a),(1,1)}^{A,(-1,-1,2,0)}
\rangle\rangle &=& J_0^{8} \langle\langle
g_{\alpha,\gamma_{c_1}}^{A,(1,1)} g_{c_1,c_1}^{A,(1,2)}
g_{\gamma_{c_1},\beta}^{A,(2,2)} g_{b,b}^{A,(2,1)}
g_{\beta,\gamma_{c_1}}^{A,(1,1)} g_{c_1,c_1}^{A,(1,2)}
g_{\gamma_{c_1},\alpha}^{A,(2,2)} g_{a,a}^{A,(2,1)} \rangle\rangle\\
\label{eq:order8-2}
&=& J_0^{8}\langle\langle g_{a,a}^{A,(2,1)} \rangle\rangle
\langle\langle g_{b,b}^{A,(2,1)} \rangle\rangle \langle\langle
\left(g_{c_1,c_1}^{A,(1,2)}\right)^2 \rangle\rangle \langle\langle
g_{\alpha,\gamma_{c_1}}^{A,(1,1)}
g_{\gamma_{c_1},\alpha}^{A,(2,2)}\rangle\rangle \langle\langle
g_{\beta,\gamma_{c_1}}^{A,(1,1)}
g_{\gamma_{c_1},\beta}^{A,(2,2)}\rangle\rangle
\\ &=& \label{eq:order8-3} \frac{ c_{1/2} }{8}
\left(\frac{J_0}{W}\right)^{8}\frac{R_{c_1}}{l_e} \frac{1}{k_F
  R_{\alpha,\gamma_{c,1}}} \frac{1}{k_F R_{\beta,\gamma_{c,1}}}
\frac{\Delta^4}{\left(\Delta^2-(\omega-i\eta)^2\right)^2}
\exp{\left[i\left(-\varphi_a- \varphi_b+ 2
    \varphi_{c_1}\right)\right]}
,
  \end{eqnarray}
\end{widetext}
where $(-1,-1,2,0)$ in the L.H.S. superscript refers to the signs in
the R.H.S. $\exp \left[ i \left( -\varphi_a - \varphi_b +
  2\varphi_{c_1} \right) \right]$ combination. The notation $(1,1)$ in
the subscript is the same as in the preceding
section~\ref{sec:themethods}, {\it i.e.} it stands for the
``electron-electron'' Nambu component.}

{In agreement with the diagrams on figures~\ref{fig:nambu-schematics}a
  and b, the $(1,1,-2,0)$ combination yielding $\exp \left[ i \left(
    \varphi_a + \varphi_b - 2\varphi_{c_1} \right) \right]$ is
  vanishingly small at the order $(J_0/W)^8$, if the $(1,1)$
  electron-electron component is evaluated. Conversely, the $(2,2)$
  hole-hole component of the $(-1,-1,2,0)$ $\exp \left[ i \left(
    -\varphi_a - \varphi_b + 2\varphi_{c_1} \right) \right]$
  combination is vanishingly small at the order $(J_0/W)^8$.}

{The positive sign of Eq.~(\ref{eq:order8-3}) originates from the
  product of the two $ \langle\langle
  g_{\alpha,\gamma_{c_1}}^{A,(1,1)}
  g_{\gamma_{c_1},\alpha}^{A,(2,2)}\rangle\rangle$ and $\langle\langle
  g_{\beta,\gamma_{c_1}}^{A,(1,1)}
  g_{\gamma_{c_1},\beta}^{A,(2,2)}\rangle\rangle$ transmission modes
  through the 2D metal, which both take negative values because they
  originate from taking the square of the pure imaginary complex
  number, see
  Eqs.~(\ref{eq:gA-asymptotics})-(\ref{eq:gA-asymptotics-2}).}

  {The $1/8$ coefficient in Eq.~(\ref{eq:order8-3})
    originates from the following terms:}

{(i) Each of the 2D metal transmission mode
  $\langle\langle g_{\alpha,\gamma_{c_1}}^{A,(1,1)}
  g_{\gamma_{c_1},\alpha}^{A,(2,2)}\rangle\rangle$ and $\langle\langle
  g_{\beta,\gamma_{c_1}}^{A,(1,1)}
  g_{\gamma_{c_1},\beta}^{A,(2,2)}\rangle\rangle$ yields a
  $\langle\langle \cos^2(k_F R) \rangle\rangle=1/2$ factor, see
  Eqs.~(\ref{eq:gA-asymptotics})-(\ref{eq:gA-asymptotics-2})}.

{(ii) A $1/2$ coefficient is related by convention to the
  superconducting diffusion mode $\langle\langle
  \left(g_{c_1,c_1}^{A,(1,2)}\right)^2 \rangle\rangle$ which is taken
  to be dominated by nonlocal propagation over the dirty-limit
  coherence length on the $S_{c,1}$ side of the 2D metal-$S_{c,1}$
  interface.}

{Integrating the spectral current given by Eq.~(\ref{eq:order8-3})
  over energy $\omega$ produces a positive sign because the residue at
  $\omega=-\Delta$ is positive, see
  Eqs.~(\ref{eq:I1-def})-(\ref{eq:I1-result}) in
  Appendix~\ref{app:I1}.} {In the limit of zero temperature, the above
  Eqs.~(\ref{eq:order8-1})-(\ref{eq:order8-3}) and
  Eqs.~(\ref{eq:I1-def})-(\ref{eq:useful-integral1}) in
  Appendix~\ref{app:I1} lead to
  \begin{widetext}
\begin{eqnarray}
  \label{eq:order8-4}
\int_{-\infty}^0 J_0 \langle\langle
G_{[8],S_{c,1},(\alpha,a),(1,1)}^{A,(-1,-1,2,0)}
\rangle\rangle(\omega) d\omega &=& \frac{i\pi c'_{1/2}\Delta}{32}
\left(\frac{J_0}{W}\right)^{8}\frac{\sqrt{{\cal S}_{contact}}}{l_e}
\frac{ y_0 z_0 }{(k_F R_0)^2} \exp{\left[i\left(-\varphi_a-
    \varphi_b+ 2 \varphi_{c_1}\right)\right]}.
\end{eqnarray}
  \end{widetext}
Evaluating similarly all terms in
Eqs.~(\ref{eq:I1-eq})-(\ref{eq:I4-eq}) leads to
\begin{equation}
  \label{eq:quartet-current-phase-relation}
  I_{\alpha\rightarrow a,eq}= I_{c,\,3TQ_1} \sin\varphi_{q,\,3T,\,1}
,
\end{equation}
where
\begin{equation}
  \varphi_{q,\,3T,\,1}=\varphi_a+
  \varphi_b- 2 \varphi_{c_1}
  .
\end{equation}
 The SQ$_1$ critical current $I_{c,\,3TQ_1}$ is negative, {\it i.e.}
 it is $\pi$-shifted:
\begin{eqnarray}
    \label{eq:order8-4-bis-result}
  I_{c,\,3TQ_1}= -\frac{e c'_{1/2}\pi\Delta}{4\hbar}
  \left(\frac{J_0}{W}\right)^{8}\frac{\sqrt{{\cal S}_{contact}}}{l_e}
  \frac{ y_0 z_0 }{(k_F R_0)^2} 
       .
\end{eqnarray}
}

{Finally, we define the remaining variables appearing in
  Eqs.~(\ref{eq:order8-3})-(\ref{eq:order8-4-bis-result}).}

{The coefficients $c_{1/2}$ and $c'_{1/2}$ are positive and of order
  unity.}

{The dimensionless parameters $y_0$ and $z_0$ in
  Eqs.~(\ref{eq:order8-4})-(\ref{eq:order8-4-bis-result}) depend on
  the shape of the four-terminal device, still within the short
  junction limit assumption, see subsection~\ref{sec:ingredients} for
  a discussion of the short-junction limit and
  figure~\ref{fig:device2} for the definition of $y_0$ and $z_0$.}

{The $1/k_F R_{\alpha,\gamma_{c,1}}$ and $1/k_F
  R_{\beta,\gamma_{c,1}}$ terms in Eq.~(\ref{eq:order8-3}) originate
  from ballistic propagation through the 2D metal, see
  Eqs.~(\ref{eq:gA-asymptotics})-(\ref{eq:gR-asymptotics}).  }

{In connection with subsection~\ref{sec:mode-averaging}, we assumed
  small area ${\cal S}_{contact}= \pi R_{c,1}^2$ for the circular
  contact between the 2D metal and the superconducting lead $S_{c,1}$,
  such that $R_{c,1}\alt \xi_{dirty}(0)$, where the dirty-limit
  coherence length is given by Eq.~(\ref{eq:xi-dirty-def}) and the
  geometry is shown schematically on figure~\ref{fig:device2}c. The
  assumption $R_{c,1}\alt \xi_{dirty}(0)$ yields the $R_{c,1}/l_e \sim
  \sqrt{{\cal S}_{contact}}/l_e$ scaling in Eq.~(\ref{eq:order8-3}),
  (\ref{eq:order8-4}) and Eq.~(\ref{eq:order8-4-bis-result}), see also
the discussion in the preceding subsection~\ref{sec:mode-averaging}.}

\subsubsection{Discussion}
{The following current-phase-flux relations are deduced from
  Eq.~(\ref{eq:quartet-current-phase-relation}) in the gauge given by
  Eqs.~(\ref{eq:gauge1})-(\ref{eq:gauge2}):
\begin{eqnarray}
\label{eq:quartet-current-1}
  I_{3TQ_1}(\varphi_{q,\,3T},\Phi)&=&I_{c,\,3TQ_1}
  \sin\left[\varphi_{q,\,3T}+\Phi\right]\\ I_{3TQ_2}(\varphi_{q,\,3T},\Phi)&=&I_{c,\,3TQ_2}
  \sin\left[\varphi_{q,\,3T}-\Phi\right]
  ,
\label{eq:quartet-current-2}
\end{eqnarray}
where Eq.~(\ref{eq:quartet-current-1}) and
Eq.~(\ref{eq:quartet-current-2}) correspond to the three-terminal
3TQ$_1$, 3TQ$_2$ respectively. The phase variable entering
Eq.~(\ref{eq:quartet-current-1}) is $ \varphi_{q,\,3T,\,1} \equiv
\varphi_a+\varphi_b-2\varphi_{{c_1}} \equiv \varphi_{q,\,3T}+\Phi$ and
that entering Eq.~(\ref{eq:quartet-current-2}) is $
\varphi_{q,\,3T,\,1}\equiv \varphi_a+\varphi_b-2\varphi_{{c_2}} \equiv
\varphi_{q,\,3T}-\Phi$, where $\varphi_{{c_1}}$ and $\varphi_{{c_2}}$
are given by Eqs.~(\ref{eq:gauge1})-(\ref{eq:gauge2}), and
$\varphi_{q,\,3T}$ is given by Eq.~(\ref{eq:phiq}).}

Figures~\ref{fig:nambu-schematics}a and b show two representations of
the three-terminal 3TQ$_1$:

(i) Figure~\ref{fig:nambu-schematics}a shows a representation resembling the
``diffusons'' in the theory of disordered conductors.

(ii) Figure~\ref{fig:nambu-schematics}b shows energy on the $y$-axis,
with respect to the chemical potential of the grounded $S_c$, see
also Ref.~\onlinecite{Freyn,Melin1}.

{In addition, an intuitive argument for the $\pi$-shift in
  the three-terminal 3TQ$_1$, 3TQ$_2$ current-phase relations
  Eqs.~(\ref{eq:quartet-current-phase-relation})-(\ref{eq:quartet-current-2})
  is the following \cite{Jonckheere}:}

{ The two Cooper pairs of the quartets imply taking the square of the
single-pair wave-function
\begin{eqnarray}
  \label{eq:split-pair}
  \frac{1}{\sqrt{2}}\left(c_{a,\uparrow}^+
c_{b,\downarrow}^+ - c_{a,\downarrow}^+
c_{b,\uparrow}^+\right)
\end{eqnarray}
according to
\begin{equation}
\label{eq:square}
  \frac{1}{2}\left(c_{a,\uparrow}^+
c_{b,\downarrow}^+ - c_{a,\downarrow}^+
c_{b,\uparrow}^+\right)^2
.\end{equation}
Eq.~(\ref{eq:square}) takes the form of the opposite of a pair of pair:
\begin{equation}
\label{eq:minus-sign}
\mbox{(\ref{eq:square})}=  -\left(c_{a,\uparrow}^+ c_{a,\downarrow}^+ \right)
  \left(c_{b,\uparrow}^+ c_{b,\downarrow}^+ \right)
.
\end{equation}
The minus sign appearing in the R.H.S. of Eq.~(\ref{eq:minus-sign}) is
consistent with the $\pi$-shifted critical current in
Eq.~(\ref{eq:order8-4-bis-result}), which receives interpretation of
macroscopic manifestation for the internal structure of a Cooper pair,
{\it i.e.} the orbital and spin symmetries.  }

\begin{figure}[htb]
  \includegraphics[width=.8\columnwidth]{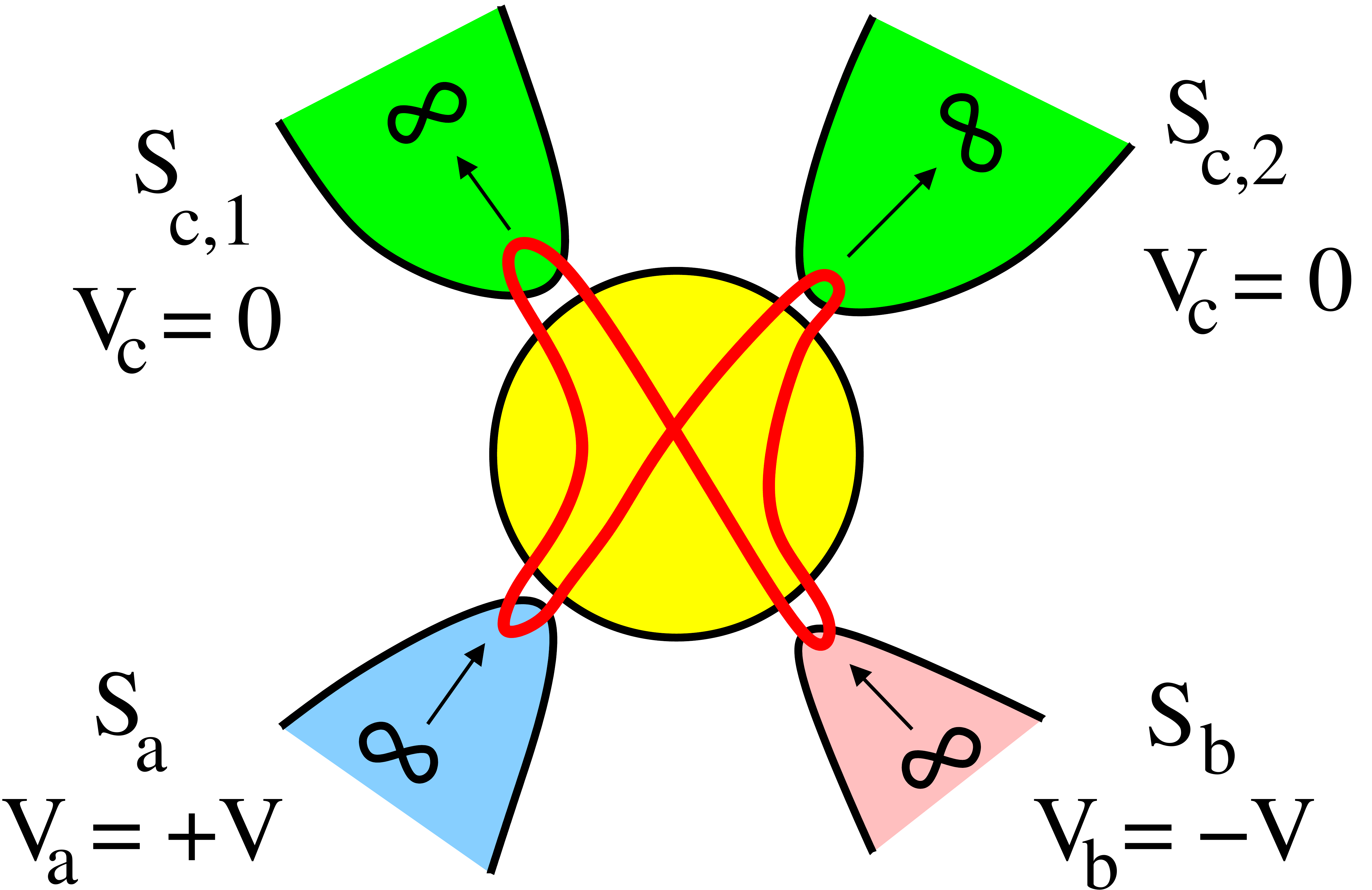}
\caption{The diagram encoding the {\it four-terminal statistical
    fluctuations of the split quartets} (4TFSQ) transmits one Cooper
  pair into $S_{{c_1}}$ and another one into $S_{{c_2}}$. The
  four-terminal 4TFSQ diagrams encode a statistical fluctuation of the
  $\varphi_{q,\,3T}$-sensitive current which does not scale with the
  number of channels. (This is because the Green's functions cannot be
  gathered in a pair-wise manner on this diagram). The four-terminal 4TFSQ
  current-phase relation is given by
  Eq.~(\ref{eq:split-quartet-current}).\label{fig:diagrams-split-quartets}}
\end{figure}

\subsection{The Four-Terminal statistical Fluctuations of the
  Split Quartet current (4TFSQ)}
\label{sec:4TFSQ-process}

Before discussing in the next subsection~\ref{sec:the4TSQ} the
four-terminal 4TSQ at the order $(J_0/W)^{12}$, {we mention now a
  simpler} ``baby-4TSQ'' at the order $(J_0/W)^8$ , see
figure~\ref{fig:diagrams-split-quartets}. The critical current of this
order-$(J_0/W)^8$ process is small, and it fluctuates around zero
value.  The reason is that the four single-particle Green's functions
crossing the 2D metal on figure~\ref{fig:diagrams-split-quartets}
cannot be gathered in a pair-wise manner if the four contacts with the
superconducting leads $S_a$, $S_b$, $S_{c,1}$ and $S_{c,2}$ make
between them distance which is much larger than the Fermi wave-length
$\lambda_F$. The current associated to the four-terminal 4TFSQ of
order $(J_0/W)^8$ on figure~\ref{fig:diagrams-split-quartets} is given
by
\begin{eqnarray}
\label{eq:split-quartet-current}
I_{4TFSQ}(\varphi_{q,\,3T})&=&I_{c,\,4TFSQ}
\sin\varphi_{q,\,4T}
,
\end{eqnarray}
where
\begin{equation}
  \label{eq:phiq-4T}
  \varphi_{q,\,4T}=\varphi_a+\varphi_{q,\,3T}-\varphi_{c,1}-\varphi_{c,2}
  .
\end{equation}
Eqs.~(\ref{eq:gauge1})-(\ref{eq:gauge2}) imply
$\varphi_{q,\,4T}=\varphi_{q,\,3T}$, where $\varphi_{q,\,3T}$ is given
by Eq.~(\ref{eq:phiq}). Overall, multichannel averaging
produces a vanishingly small critical current for the four-terminal
4TFSQ: $\langle\langle I_{c,\,4TFSQ} \rangle\rangle=0$.

\subsection{The Four-Terminal Split Quartets (4TSQ)}
\label{sec:the4TSQ}

{Now, we consider the four-terminal 4TSQ yielding nonvanishingly small
  value for critical current with multichannel interfaces. Two types
  of diagrams appear at the order $(J_0/W)^{12}$, after a first
  selection has been operated with respect to gathering the nonlocal
  Green's functions through the 2D metal in a pair-wise manner:}

{(i) {\it The diagrams containing products of three Nambu Green's
    functions within the same superconducting lead:} Their critical
  current is of order $\sqrt{{\cal S}_{contact}}/l_e$, see the
  forthcoming subsection~\ref{sec:subleading} and subsection~II\,A in
  the Supplemental Material\cite{supplemental}}.

{(ii) {\it  The remaining diagrams} provide the leading-order ${\cal
    S}_{contact}/l_e^2$ contribution to the critical current, see the
  forthcoming subsection~\ref{sec:leading} and subsection~II\,B in
  the Supplemental Material\cite{supplemental}}.

\subsubsection{The four-terminal 4TSQ current at the orders
  $(J_0/W)^{12}$ and $\sqrt{{\cal S}_{contact}}/l_e$}
\label{sec:subleading}

{We provide now the microscopic calculation for the contributions to
  $J_{a,\alpha} G^A_{\alpha,a}$ at the orders $(J_0/W)^{12}$ and
  $\sqrt{{\cal S}_{contact}}/l_e$. {The four terms given below in
    Eqs.~(\ref{eq:4TSQ-subleading1})-(\ref{eq:4TSQ-subleading4})
    correspond to the following possibilities:}

{(i) The $(1,1)$
  ``electron-electron'' or the $(2,2)$ ``hole-hole'' components of
  $J_{a,\alpha} G^A_{\alpha,a}$.}

{(ii) The $\exp\left[\pm i\left(+\varphi_a + \varphi_b - \varphi_{c,1}
    - \varphi_{c,2} \right)\right]$ factors for the $(1,1,-1,-1)$ or
  $(-1,-1,1,1)$ labels respectively.}

{We obtain the following:}
\begin{widetext}
  {
  \begin{eqnarray}
    \label{eq:4TSQ-subleading1}
J_0 \langle\langle G_{[12],(\alpha,a),(1,1),(1)}^{A,(1,1,-1,-1)}
\rangle\rangle_{total}(\omega) &=& \frac{c'_{1/2}}{8}
\left(\frac{J_0}{W}\right)^{12}\frac{\sqrt{{\cal S}_{contact}}}{l_e}
\frac{ x_0 y_0 z_0 }{(k_F R_0)^3}
\frac{\Delta^6}{\left(\Delta^2-(\omega-i\eta)^2\right)^3}
\exp{\left[i\left(\varphi_a+ \varphi_b- \varphi_{c_1}-
    \varphi_{c_2}\right)\right]}\\
    \label{eq:4TSQ-subleading2}
J_0 \langle\langle
G_{[12],(\alpha,a),(1,1),(1)}^{A,(-1,-1,1,1)}
\rangle\rangle_{total}(\omega) &=& \frac{5 c'_{1/2}}{8}
\left(\frac{J_0}{W}\right)^{12}\frac{\sqrt{{\cal S}_{contact}}}{l_e}
\frac{ x_0 y_0 z_0 }{(k_F R_0)^3}
\frac{\Delta^6}{\left(\Delta^2-(\omega-i\eta)^2\right)^3}
\exp{\left[i\left(-\varphi_a- \varphi_b+ \varphi_{c_1}+
    \varphi_{c_2}\right)\right]}\\
    \label{eq:4TSQ-subleading3}
J_0 \langle\langle
G_{[12],(\alpha,a),(2,2),(1)}^{A,(1,1,-1,-1)}
\rangle\rangle_{total}(\omega) &=& \frac{5 c'_{1/2}}{8}
\left(\frac{J_0}{W}\right)^{12}\frac{\sqrt{{\cal S}_{contact}}}{l_e}
\frac{ x_0 y_0 z_0 }{(k_F R_0)^3}
\frac{\Delta^6}{\left(\Delta^2-(\omega-i\eta)^2\right)^3}
\exp{\left[i\left(\varphi_a+ \varphi_b- \varphi_{c_1}-
    \varphi_{c_2}\right)\right]}\\ J_0 \langle\langle
G_{[12],(\alpha,a),(2,2),(1)}^{A,(-1,-1,1,1)}
\rangle\rangle_{total}(\omega) &=& \frac{c'_{1/2}}{8}
\left(\frac{J_0}{W}\right)^{12}\frac{\sqrt{{\cal S}_{contact}}}{l_e}
\frac{ x_0 y_0 z_0 }{(k_F R_0)^3}
\frac{\Delta^6}{\left(\Delta^2-(\omega-i\eta)^2\right)^3}
\exp{\left[i\left(-\varphi_a- \varphi_b+ \varphi_{c_1}+
    \varphi_{c_2}\right)\right]}
.
\label{eq:4TSQ-subleading4}
  \end{eqnarray}
  }
  \end{widetext}
{The microscopic process contributing to the $(-1,-1,1,1)$ terms given
  by Eq.~(\ref{eq:4TSQ-subleading2})
  and~Eq.~(\ref{eq:4TSQ-subleading4}) are listed in subsection~II\,A
  of the Supplemental Material\cite{supplemental}.}  {Specifically,
  Eq.~(\ref{eq:4TSQ-subleading2}) is the sum of Eqs.~(10)-(39) in the
  Supplemental Material\cite{supplemental} and
  Eq.~(\ref{eq:4TSQ-subleading4}) is the sum of Eqs.~(41)-(46), also
  in the Supplemental Material\cite{supplemental}.}

\begin{figure}[htb]
  \includegraphics[width=.8\columnwidth]{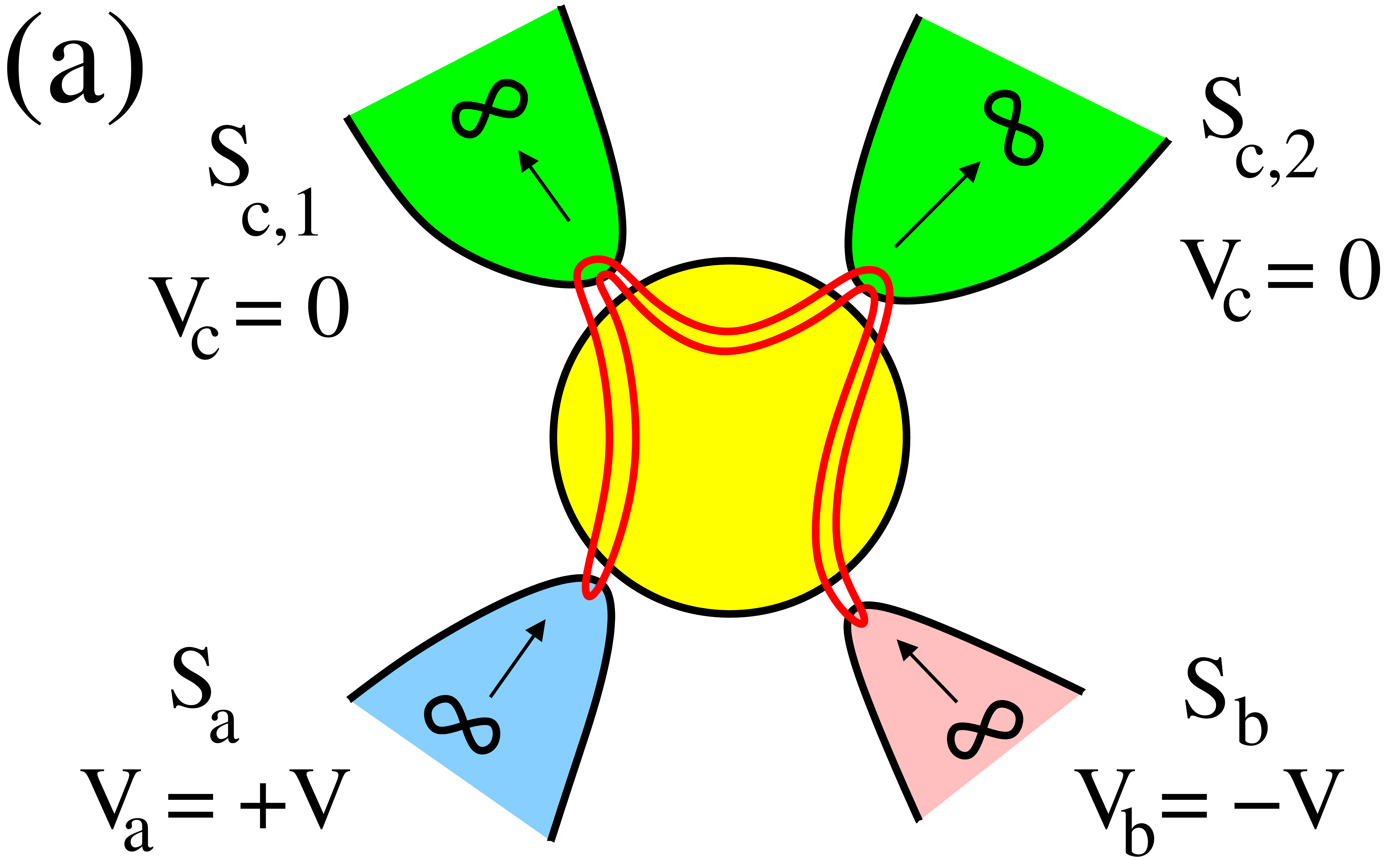}

  \bigskip
  
  \includegraphics[width=.8\columnwidth]{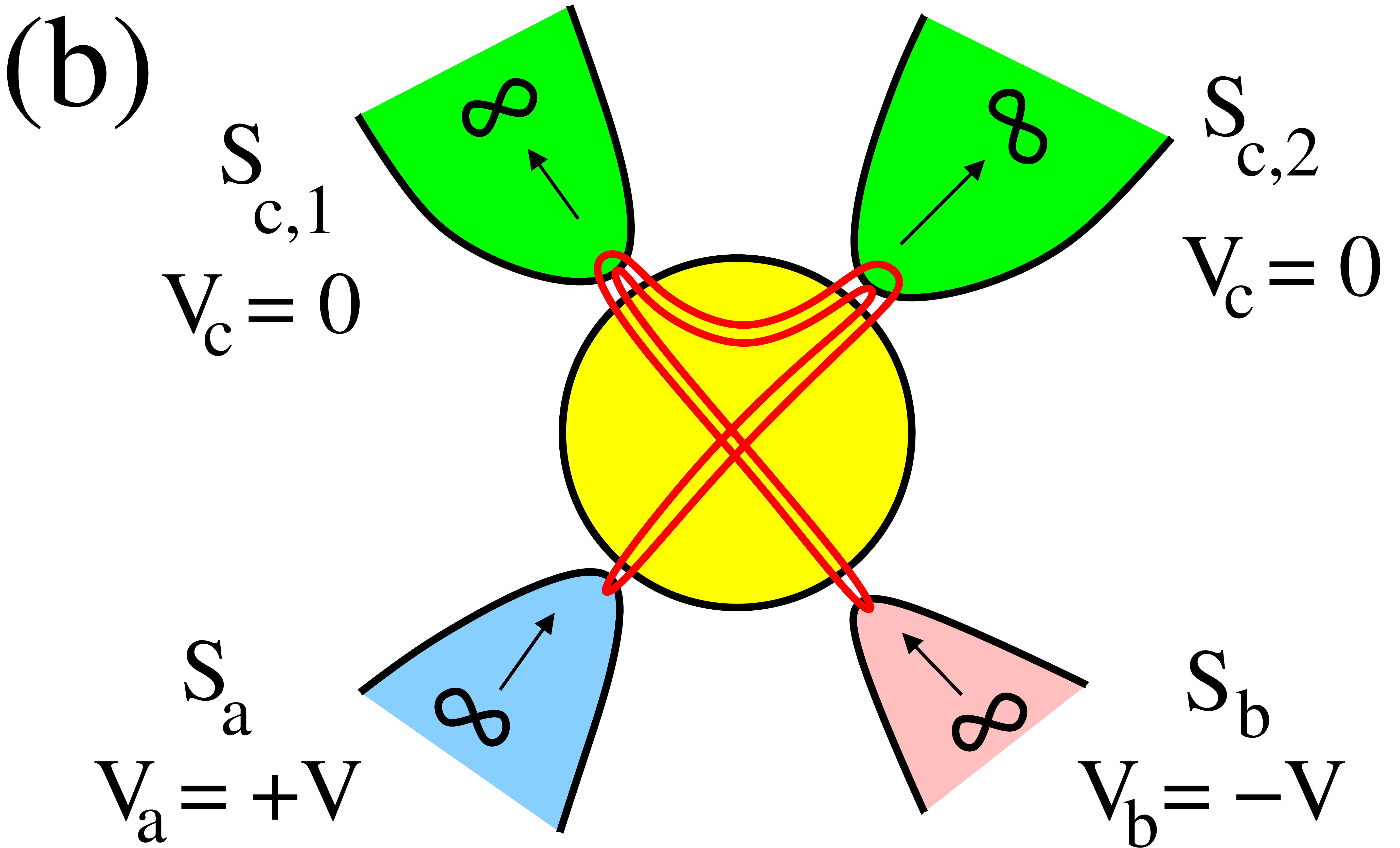}
\caption{{{\it Two of the Four-Terminal Split Quartet diagrams}
    (4TSQ): Contrary to the previous four-terminal 4TFSQ diagram (see
    figure~\ref{fig:diagrams-split-quartets}), these four-terminal
    4TSQ diagrams yield a critical current which is not a small
    statistical fluctuation.} On the contrary, the four-terminal 4TSQ
  current scales with the number of channels because the Green's
  functions are gathered in a pair-wise manner on this figure.
  \label{fig:diagrams-downmixing}}
\end{figure}

{The overall positive sign of
  Eqs.~(\ref{eq:4TSQ-subleading1})-(\ref{eq:4TSQ-subleading4})
  originates from the product of four negative contributions:}

{(i) A minus sign is associated to each of the
  three transmission modes through the 2D metal.}

{(ii) Another minus sign is due to averaging the product of three
  superconducting Green's functions, see
  Eqs.~(\ref{eq:C3})-(\ref{eq:last-eq}) in
  Appendix~\ref{app:averaging-3gs}.}

{In addition, the $\exp{\left[i\left(-\varphi_a- \varphi_b+
      \varphi_{c_1}+ \varphi_{c_2}\right)\right]}$ factor in
  Eqs.~(\ref{eq:4TSQ-subleading2}) and~(\ref{eq:4TSQ-subleading4})
  produces a ``(1,1)'' electron-electron Nambu component $5/8$
  coefficient in Eq.~(\ref{eq:4TSQ-subleading2}) which is larger than
  the ``$(2,2)$'' hole-hole component $1/8$ coefficient in
  Eq.~(\ref{eq:4TSQ-subleading4}). This is compatible with the
  observation that the ``$(2,2)$'' component associated with the
  $(-1,-1,2,0)$ combination is vanishingly small for the
  three-terminal 3TQ$_1$, see the above
  subsection~\ref{sec:butterfly-quartet-diagrams}.}

{In addition, the residue of the pole at $\omega=-\Delta$ is positive,
  see Eqs.~(\ref{eq:I2-definition})-(\ref{eq:result_I2}) in
  Appendix~\ref{app:I2} concerning the integral over the energy
  $\omega$.}

{Overall, Eqs.~(\ref{eq:I1})-(\ref{eq:I4}) and
  Eqs.~(\ref{eq:4TSQ-subleading1})-(\ref{eq:4TSQ-subleading4}) lead to
  the following current-phase relation for the 4TSQ at the orders
  $(J_0/W)^{12}$ and $\sqrt{{\cal S}_{contact}}/l_e$:
\begin{equation}
  \label{eq:4TSQ-current-phase-relation}
I_{\alpha\rightarrow a,eq}^{(1)}= I_{c,\,4TSQ}^{(1)} \sin
\varphi_{q,\,4T}
,
\end{equation}
where $\varphi_{q,\,4T}\equiv \varphi_a+\varphi_b-\varphi_{c,1}
-\varphi_{c,2}\equiv \varphi_{q,\,3T}$, as for the 4TFSQ, 
see Eqs.~(\ref{eq:split-quartet-current})-(\ref{eq:phiq-4T}).}

{The critical current $I_{c,\,4TSQ}^{(1)}$ appearing
  in Eq.~(\ref{eq:4TSQ-current-phase-relation}) is negative, {\it
    i.e.} it is $\pi$-shifted:
\begin{equation}
  \label{eq:Ic-4TSQ-result-1}
  I_{c,\,4TSQ}^{(1)}= -\frac{3 e c'_{1/2} \pi \Delta}{4 \hbar}
\left(\frac{J_0}{W}\right)^{12}\frac{\sqrt{{\cal S}_{contact}}}{l_e}
\frac{ x_0 y_0 z_0 }{(k_F R_0)^3}
.
\end{equation}
 }

\subsubsection{The four-terminal 4TSQ current
  at the orders $(J_0/W)^{12}$ and ${\cal S}_{contact}/l_e^2$}
\label{sec:leading}
{Next, we calculate the four-terminal 4TSQ critical current at the
  orders $(J_0/W)^{12}$ and ${\cal S}_{contact}/l_e^2$. Again, we
  separate between the ``electron-electron" from the ``hole-hole''
  Nambu components, and the $\exp\left[\pm i\left(+\varphi_a +
    \varphi_b - \varphi_{c,1} - \varphi_{c,2} \right)\right]$
  sensitivity on the superconducting phase variables: } {
  \begin{widetext}
    \begin{eqnarray}
  \label{eq:4TSQ-leading1}
  J_0 \langle\langle G_{[12],(\alpha,a),(1,1),(2)}^{A,(1,1,-1,-1)}
\rangle\rangle_{total}(\omega) &=& -  \frac{c'_1}{32}
\left(\frac{J_0}{W}\right)^{12}\frac{{\cal S}_{contact}}{l_e^2}
\frac{ y_0 z_0^2 + y_0^2 z_0 + x_0 z_0^2 + x_0 y_0^2 }{(k_F R_0)^3}
\frac{(\omega-i\eta)^2\Delta^4}{\left(\Delta^2-(\omega-i\eta)^2\right)^3}
e^{i\left(\varphi_a+ \varphi_b- \varphi_{c_1}-
    \varphi_{c_2}\right)}\\   \label{eq:4TSQ-leading2}
J_0 \langle\langle
G_{[12],(\alpha,a),(1,1),(2)}^{A,(-1,-1,1,1)}
\rangle\rangle_{total}(\omega) &=& -  \frac{5 c'_1}{32}
\left(\frac{J_0}{W}\right)^{12}\frac{{\cal S}_{contact}}{l_e^2}
\frac{ y_0 z_0^2 + y_0^2 z_0 + x_0 z_0^2 + x_0 y_0^2 }{(k_F R_0)^3}
\frac{(\omega-i\eta)^2\Delta^4}{\left(\Delta^2-(\omega-i\eta)^2\right)^3}
e^{i\left(-\varphi_a- \varphi_b+ \varphi_{c_1}+
    \varphi_{c_2}\right)}\\ J_0 \langle\langle
G_{[12],(\alpha,a),(2,2),(2)}^{A,(1,1,-1,-1)}
\rangle\rangle_{total}(\omega) &=& -  \frac{5 c'_1}{32}
\left(\frac{J_0}{W}\right)^{12}\frac{{\cal S}_{contact}}{l_e^2}
\frac{ y_0 z_0^2 + y_0^2 z_0 + x_0 z_0^2 + x_0 y_0^2 }{(k_F R_0)^3}
\frac{(\omega-i\eta)^2\Delta^4}{\left(\Delta^2-(\omega-i\eta)^2\right)^3}
e^{i\left(\varphi_a+ \varphi_b- \varphi_{c_1}-
    \varphi_{c_2}\right)}\\ J_0 \langle\langle
G_{[12],(\alpha,a),(2,2),(2)}^{A,(-1,-1,1,1)}
\rangle\rangle_{total}(\omega) &=& -  \frac{c'_1}{32}
\left(\frac{J_0}{W}\right)^{12}\frac{{\cal S}_{contact}}{l_e^2}
\frac{ y_0 z_0^2 + y_0^2 z_0 + x_0 z_0^2 + x_0 y_0^2 }{(k_F R_0)^3}
\frac{(\omega-i\eta)^2\Delta^4}{\left(\Delta^2-(\omega-i\eta)^2\right)^3}
e^{i\left(-\varphi_a- \varphi_b+ \varphi_{c_1}+
    \varphi_{c_2}\right)}
  \label{eq:4TSQ-leading4}
    \end{eqnarray}
  \end{widetext}
Eq.~(\ref{eq:4TSQ-leading2}) is the sum of Eqs.~(48)-(107) in subsection
II\,B of the Supplemental Material\cite{supplemental}.
Eq.~(\ref{eq:4TSQ-leading4}) is the sum of Eqs.~(109)-(120) in the
Supplemental Material\cite{supplemental}.

{The minus sign in
  Eqs.~(\ref{eq:4TSQ-leading1})-(\ref{eq:4TSQ-leading4}) is due to
  the product of three (negative) transmission modes through the 2D
  metal.}

In addition, the $\exp{\left[i\left(-\varphi_a- \varphi_b+
    \varphi_{c_1}+ \varphi_{c_2}\right)\right]}$ combination yields
the $5/32$ coefficient for the ``$(1,1)$'' component which is larger
than $1/32$ for the ``$(2,2)$'' component, see
Eqs.~(\ref{eq:4TSQ-leading2}) and~(\ref{eq:4TSQ-leading4})
respectively. This is compatible with the discussion following the
above Eqs.~(\ref{eq:4TSQ-subleading1})-(\ref{eq:4TSQ-subleading4}).}

 A (positive) residue is taken into account in the integral over
 energy, see Eqs.~(\ref{eq:I3-def})-(\ref{eq:I3-result}) in
 Appendix~\ref{app:I3}.

 It is deduced that Eqs.~(\ref{eq:I1-eq})-(\ref{eq:I4-eq})
 and Eqs.~(\ref{eq:4TSQ-leading1})-(\ref{eq:4TSQ-leading4}) lead to
 the following contribution to the equilibrium current
 $I_{\alpha\rightarrow a,eq}$ at the orders $(J_0/W)^{12}$ and ${\cal
   S}_{contact}/l_e^2$:
  \begin{equation}
    \label{eq:I-(2)}
    I_{\alpha\rightarrow a,eq}^{(2)}= I_{c,\,4TSQ}^{(2)} \sin
    \varphi_{q,\,4T}
,
  \end{equation}
  where $\varphi_{q,\,4T}$ is given by Eq.~(\ref{eq:phiq-4T}) and the
  critical current $I_{c,\,4TSQ}^{(2)}$ is positive, {\it i.e.} it is
  $0$-shifted:
\begin{equation}
  \label{eq:Ic-4TSQ-result-2}
  I_{c,\,4TSQ}^{(2)}=\frac{3 e c'_1 \Delta}{16 \hbar}
\left(\frac{J_0}{W}\right)^{12}\frac{{\cal S}_{contact}}{l_e^2}
\frac{ y_0 z_0^2 + y_0^2 z_0 + x_0 z_0^2 + x_0 y_0^2 }{(k_F R_0)^3}
  .
\end{equation}

\subsubsection{Discussion}

{ Two of the four-terminal 4TSQ diagrams appearing at the orders
  $(J_0/W)^{12}$ and ${\cal S}_{contact}/l_e^2$ are shown on
  figures~\ref{fig:diagrams-downmixing}a and~b. The diffuson and the
  energy representations on figures~\ref{fig:nambu-schematics}c, d
  respectively illustrate that the four-terminal 4TSQ of orders
  $(J_0/W)^{12}$ and ${\cal S}_{contact} /l_e^2$ involve the product
  of two superconducting diffusion modes of the
  $\langle\langle{g_{(1,1)} g_{(1,2)}}\rangle\rangle$-type. For
  instance, the superconducting diffusion modes propagating in
  $S_{c,1}$ or in $S_{c,2}$ on figures~\ref{fig:diagrams-downmixing}a
  and~b correspond to the $x_0y_0^2$ or $x_0 z_0^2$ contributions to
  Eqs.~(\ref{eq:4TSQ-leading1})-(\ref{eq:4TSQ-leading4}) respectively
  [see also Eq.~(\ref{eq:Ic-4TSQ-result-2})].}

The $0$-shifted four-terminal 4TSQ current [see
Eqs.~(\ref{eq:I-(2)})-(\ref{eq:Ic-4TSQ-result-2})] is interpreted as
the intermediate state
\begin{equation}
  \label{eq:cccc}
  c_{S_{c_1},\uparrow}^+ c_{S_{c_1},\downarrow}^+
  c_{S_{c_2},\uparrow}^+ c_{S_{c_2},\downarrow}^+
\end{equation}
made with two Cooper pairs from $S_a$ and $S_b$ biased at $\pm V$.
Anticommuting the $(S_{{c_1}},\uparrow)$ and the
$(S_{{c_2}},\uparrow)$ partners in Eq.~(\ref{eq:cccc}) leads to a
minus sign which implies $0$-shift for the four-terminal 4TSQ
current-phase relation [see
  Eqs.~(\ref{eq:I-(2)})-(\ref{eq:Ic-4TSQ-result-2})) in comparison
  with the previous $\pi$-shift of the three-terminal 3TQ$_1$ and
  3TQ$_2$ [see
    Eqs.~(\ref{eq:quartet-current-phase-relation})-(\ref{eq:order8-4-bis-result})]. Indeed,
  the three-terminal 3TQ$_1$ and 3TQ$_2$ do not contain the additional
  two-fermion exchange of the 4TSQ which is made possible by the 2D
  quantum wake, see the forthcoming section~\ref{sec:why-2D}.
  
\section{Interference between the three-terminal 3TQ$_1$, 3TQ$_2$ and
  the four-terminal 4TSQ}
\label{sec:interference-Q-4TSQ}

{We proceed by further considering that, in the gauge given by
  Eqs.~(\ref{eq:gauge1})-(\ref{eq:gauge2}), the
  $\varphi_{q,\,3T}$-sensitive critical current is the result of an
  interference between the three-terminal 3TQ$_1$, 3TQ$_2$ (see
  subsection~\ref{sec:butterfly-quartet-diagrams}), and the four-terminal
  4TSQ (see subsection~\ref{sec:the4TSQ}):
\begin{eqnarray}
  \label{eq:Ic-4TSQ}
  I_{c}(\Phi/\Phi_0)&=&\mbox{Max}_{\varphi_{q,\,3T}}
  \left[I_{3TQ_1}(\varphi_{q,\,3T},\Phi)\right.\\ &&\left.+
    I_{3TQ_2}(\varphi_{q,\,3T},\Phi)
    +I_{4TSQ}(\varphi_{q,\,3T}) \right] .\nonumber
\end{eqnarray}

The contact areas ${\cal S}_{contact}$ are considered to be large
compared to $(l_e)^2$, {\it i.e.} ${\cal S}_{contact}\gg l_e^2$. This
realistic assumption yields $|I_{c,\,4TSQ}^{(1)}|\ll
I_{c,\,4TSQ}^{(2)}$.  The four-terminal 4TSQ critical current is
approximated as $I_{4TSQ}(\varphi_{q,\,3T})\simeq
I_{c,\,4TSQ}^{(2)}$. The resulting current-phase relation
\begin{equation}
I_{4TSQ}(\varphi_{q,\,3T}) \simeq I_{c,\,4TSQ}^{(2)} \sin
\varphi_{q,\,3T}
\label{eq:Ic-approx}
\end{equation}
is independent on the value of the magnetic flux $\Phi$, see the
expression of $\varphi_{q,\,3T}$ given by Eq.~(\ref{eq:phiq}) and
$I_{c,\,4TSQ}^{(2))}$ in Eq.~(\ref{eq:Ic-4TSQ-result-2}).

Eq.~(\ref{eq:Ic-4TSQ}) is $2\pi$-periodic in $\Phi$, while the
previous $\mbox{Max}_{\varphi_{q,\,3T}} \left[I_{3TQ_1}(\varphi_{q,\,3T},\Phi) +
  I_{3TQ_2}(\varphi_{q,\,3T},\Phi)\right]$ was $\pi$-periodic.

More specifically, specializing to $\Phi/\Phi_0=0$ and to
$\Phi/\Phi_0=1/2$ leads to
\begin{equation}
  \label{eq:IqcA}
  I_{c}(0)= \left|I_{c,\,3TQ_1}+I_{c,\,3TQ_2}+I_{c,\,4TSQ}\right| ,
\end{equation}
which is different from
\begin{equation}
  \label{eq:IqcB}
  I_{c}(1/2)= \left|I_{c,\,3TQ_1}+I_{c,\,3TQ_2}-I_{c,\,4TSQ}\right| .
\end{equation}

\section{Why the four-terminal 4TSQ appear only in 2D}
\label{sec:why-2D}
The preceding section~\ref{sec:perturbation-V0+} presented the
calculation (in perturbation in $J_0/W$ and in the adiabatic limit
$V=0^+$) of the sign and the amplitude of the $\pi$-shifted
three-terminal 3TQ$_1$, 3TQ$_2$ and the $0$-shifted four-terminal
4TSQ critical currents. We proceed further with discussing why the
four-terminal 4TSQ yields a vanishingly small current if a 1D or 3D
metal is used instead of the 2D metal such as graphene gated away from
the Dirac point in the Harvard group experiment
\cite{Harvard-group-experiment}.}

{We establish a link between
  Eqs.~(\ref{eq:gA-asymptotics})-(\ref{eq:gR-asymptotics}) for the
  Green's function of a ballistic 2D metal, and the general theory of
  the ``wake'' in the solution of the even-dimensional wave-equation,
  starting in subsection~\ref{sec:wake-classical-wave-equation} with
  the classical wave equation. The 2D quantum wake in nanoscale
  electronic devices is considered in
  subsection~\ref{sec:wake-meso-nano}. Synchronizing two Josephson
  junctions with quasiparticles ``surfing'' on the 2D quantum wake is
  discussed in subsection~\ref{sec:wake-synchronization}, in
  connection with the features of the four-terminal 4TSQ diagram, see
  one of the 4TSQ diagrams in figure~\ref{fig:nambu-schematics}d. A
  summary of this section~\ref{sec:why-2D} is presented in
  subsection~\ref{sec:summary-2D-4TSQ}.}

{At this point, we also make reference to a very recent
  preprint\cite{quantum-wakes-cold-atoms} about the production of
  quantum wakes with ultracold atoms.}

\subsection{The wake effect in the classical wave equation}
\label{sec:wake-classical-wave-equation}
Volterra was the first to understand that the solutions of the
wave-equation are drastically different in even or odd space
dimension. Let us assume that an excitation is produced at given
location and time. A detector is at distance $R$ from the location of
the excitation.  In all cases, the signal reaches $R$ after the time
delay $t_0=R/v$, where $v$ is the speed of wave propagation. In odd
dimensions (such as in 1D or 3D), the detected signal consists of the
sharp pulse associated the wave-front. But in even dimension (such as
in 2D), the detected signal oscillates long after the time delay
$t_0$. This ``classical wake'' appears in even space dimensions but
not in odd dimensions, and it meets common sense regarding a boat
propagating on a calm sea.

\subsection{The 2D quantum wake in meso or nanoscale devices}
\label{sec:wake-meso-nano}
The signal at the detector mentioned above results from a convolution
of the initial excitation with the $D$-dimensional Green's
functions. It turns out that Green's functions are at the heart of the
calculation of the electronic transport properties in meso or nanoscale
devices.

The normal-state Green's function at distance $R$ is a plane-wave in
1D:
\begin{equation}
  g^A_{1D,\,(1,1)}(R,\omega)\sim \exp(ikR)
  \label{eq:g-1D}
,
\end{equation}
where $k$ is the wave-vector at the considered energy $\omega$ and
``$(1,1)$'' refers to the ``spin-up electron'' Nambu component. The
3D Green's function is also a plane wave:
\begin{equation}
  g^A_{3D,\,(1,1)}(R,\omega)\sim \frac{\exp(ikR)}{kR}
\label{eq:g-3D}
  ,
\end{equation}
which is normalized by the factor
$kR$ arising from probability conservation. In 2D, the Green's
function is a Bessel function which behaves like
\begin{equation}
  g^A_{2D,\,(1,1)}(R,\omega)\sim
  i \frac{\cos(kR-\pi/4)}{\sqrt{kR}}
\label{eq:g-2D}
\end{equation}
at large $R\gg 1/k$, see the above Eq.~(\ref{eq:gA-asymptotics}) and
Appendix~\ref{app:2D-metal-Greens-function} for the demonstration of
Eqs.~(\ref{eq:gA-asymptotics}) and~(\ref{eq:g-2D}).

\subsection{Synchronizing two Josephson junctions by the 2D quantum
  wake}
\label{sec:wake-synchronization}

The difference between the 1D or 3D $\exp(ikR)$ oscillations in
Eqs.~(\ref{eq:g-1D}) and~(\ref{eq:g-3D}), and the 2D $\cos(kR-\pi/4)$
oscillations in Eq.~(\ref{eq:g-2D}) is now discussed in connection
with synchronizing two Josephson junctions.

Specifically, we focus on the highlighted section of the four-terminal
4TSQ diagram on figure~\ref{fig:nambu-schematics}d, which involves
taking the square of the advanced Green's function according to
$\left[g^A_{(1,1)}(R)\right]^2$. Multichannel interfaces are simulated
by averaging $\left[g^A_{(1,1)}(R)\right]^2$ over $R$ around the value
$R_0$ such that $\lambda_F\ll R_0\alt l_\varphi$. Averaging over the
separation $R$ between $S_{c,1}$ and $S_{c,2}$ in an interval of width
$\Delta R \sim 2\pi/k$ around $R=R_0$ yields
\begin{eqnarray}
  \label{eq:g2-1}
  \langle\langle\left[g^A_{1D,\,(1,1)}(R)\right]^2\rangle\rangle&=&0\\
  \label{eq:g2-2}
  \langle\langle\left[g^A_{2D,\,(1,1)}(R)\right]^2\rangle\rangle&\simeq&
-\frac{1}{2W^2 k_F R_0}
  \\
  \label{eq:g2-3}
  \langle\langle\left[g^A_{3D,\,(1,1)}(R)\right]^2\rangle\rangle&=&0
  .
\end{eqnarray}
Eqs.~(\ref{eq:g2-1})-(\ref{eq:g2-3}) are deduced from
Eqs.~(\ref{eq:g-1D}), (\ref{eq:gA-asymptotics}) and~(\ref{eq:g-3D})
respectively. These Eqs.~(\ref{eq:g2-1})-(\ref{eq:g2-3}) imply short
range coupling over $\lambda_F$ if a 1D or a 3D metal is used instead
of a 3D metal, which is in agreement with the general theory of the 2D
wake mentioned above.

To summarize, it is only in 2D that the 4TSQ diagrams on
figures~\ref{fig:nambu-schematics}c and~d are nonvanishingly small,
due to the nonvanishingly small $\langle \langle \left[g^A_{(1,1)}(R)
  \right]^2 \rangle \rangle\ne 0$ connecting $S_{{c_1}}$ and
$S_{{c_2}}$ on figure~\ref{fig:nambu-schematics}d, and physically
encoding the exchange of a quasiparticle {\it via} the 2D quantum
wake.

\subsection{Summary of this section}
\label{sec:summary-2D-4TSQ}
The microscopic theory of the four-terminal 4TSQ was discussed:

(i) The four-terminal 4TSQ are specific to 2D, and they are related
to the quantum limit of the wake in the even-dimensional wave-equation.

(ii) The four-terminal 4TSQ realize {\it quantum mechanical}
synchronization between Josephson junctions by coherently exchanging a
quasiparticle between them. The quasiparticle which is exchanged
propagates on the 2D quantum wake.

(iii) The four-terminal 4TSQ couple the Andreev bound states of the
two Josephson junctions in the simple limit of equilibrium with bias
voltage $V=0$, and in the adiabatic limit with $V=0^+$ on the quartet
line, see also the remarks on the long range coupling of the
four-terminal 4TSQ in the concluding subsection~\ref{sec:final-remarks}.

Finally, we note that the four-terminal 4TSQ do not contribute to the
current in the previous Grenoble group experiment \cite{Lefloch}. In
this experiment, the intermediate region connecting the
superconducting leads consists of an evaporated ``T-shaped'' Copper
lead which is 3D, as opposed to the atomically thin 2D sheet of
graphene used in the Harvard group experiment
\cite{Harvard-group-experiment}. The 2D quantum wake is neither
expected to play a role in the Weizmann Institute group experiment
\cite{Heiblum} made with a semiconducting nanowire.

\section{Inversion between $\Phi/\Phi_0=0$ and $\Phi/\Phi_0=1/2$}
\label{sec:<>}

In this section, we show that the relative shift of $\pi$ between the
three-terminal 3TQ$_1$, 3TQ$_2$ and the four-terminal 4TSQ obtained in
the above section~\ref{sec:perturbation-V0+}, implies emergence of the
inversion $I_c(0)<I_c(1/2)$ between $\Phi/\Phi_0=0$ and
$\Phi/\Phi_0=1/2$ in the reduced flux $\Phi/\Phi_0$ dependence of the
critical current $I_c(\Phi/\Phi_0)$ given by
Eqs.~(\ref{eq:Ic-4TSQ})-(\ref{eq:Ic-approx}).

In addition, we address the reverse question of the information which
is deduced from ``{\it Observation of inversion in $I_c(\Phi/\Phi_0)$
between $\Phi/\Phi_0=0$ and $\Phi/\Phi_0=1/2$}'', regarding the sign of
the three-terminal 3TQ$_1$, 3TQ$_2$ and the four-terminal 4TSQ
current-phase relations.

The assumptions about the $0$- and $\pi$- shifted current-phase
relations are presented in subsection~\ref{sec:intro-shift-pi}. The
reasoning in itself is presented in
subsection~\ref{sec:general-physical-statements}. The consequences for
the Harvard group experiment are provided in
subsection~\ref{sec:conclusions-Harvard-group-experiment-0-pi}.

\subsection{The assumptions}

\label{sec:intro-shift-pi}

This subsection is based on the following assumptions:

(i) We have information about the $\Phi/\Phi_0$-sensitivity of the
critical current $I_{c}$, more specifically about whether
$I_c(0)$ is smaller or larger than $I_c(1/2)$.

(ii) The signs of the three-terminal 3TQ$_1$, 3TQ$_2$ and the
four-terminal 4TSQ critical currents are left a free parameters{,
  while they interfere according to the preceding
  Eq.~(\ref{eq:Ic-4TSQ}).}

\subsection{General statements}
\label{sec:general-physical-statements}

Let us now assume that inversion $I_c(0)< I_c(1/2)$ between
$\Phi/\Phi_0=0$ and $\Phi/\Phi_0=1/2$ is observed. Combining
Eqs.~(\ref{eq:IqcA}),~(\ref{eq:IqcB}) to Eq.~(121) in section~III of
the Supplemental Material \cite{supplemental} yields the following
``logical chain'':
\begin{eqnarray}
  \label{eq:logical-equivalence1}
  &&
  I_{c}(0)<I_{c}(1/2)\\
  \label{eq:logical-equivalence2}
  &\iff&
  \left|I_{c,\,3TQ_1}+I_{c,\,3TQ_2}+I_{c,\,4TSQ}\right|\\
  \nonumber
  &&<  
  \left|I_{c,\,3TQ_1}+I_{c,\,3TQ_2}-I_{c,\,4TSQ}\right|\\
  \label{eq:logical-equivalence3}
  &\iff& I_{c,\,3TQ_1}+I_{c,\,3TQ_2} \mbox{ and }
  I_{c,\,4TSQ}\\ &&\nonumber \mbox{ have opposite signs.}
\end{eqnarray}

\begin{figure*}[htb]
  \includegraphics[width=\textwidth]{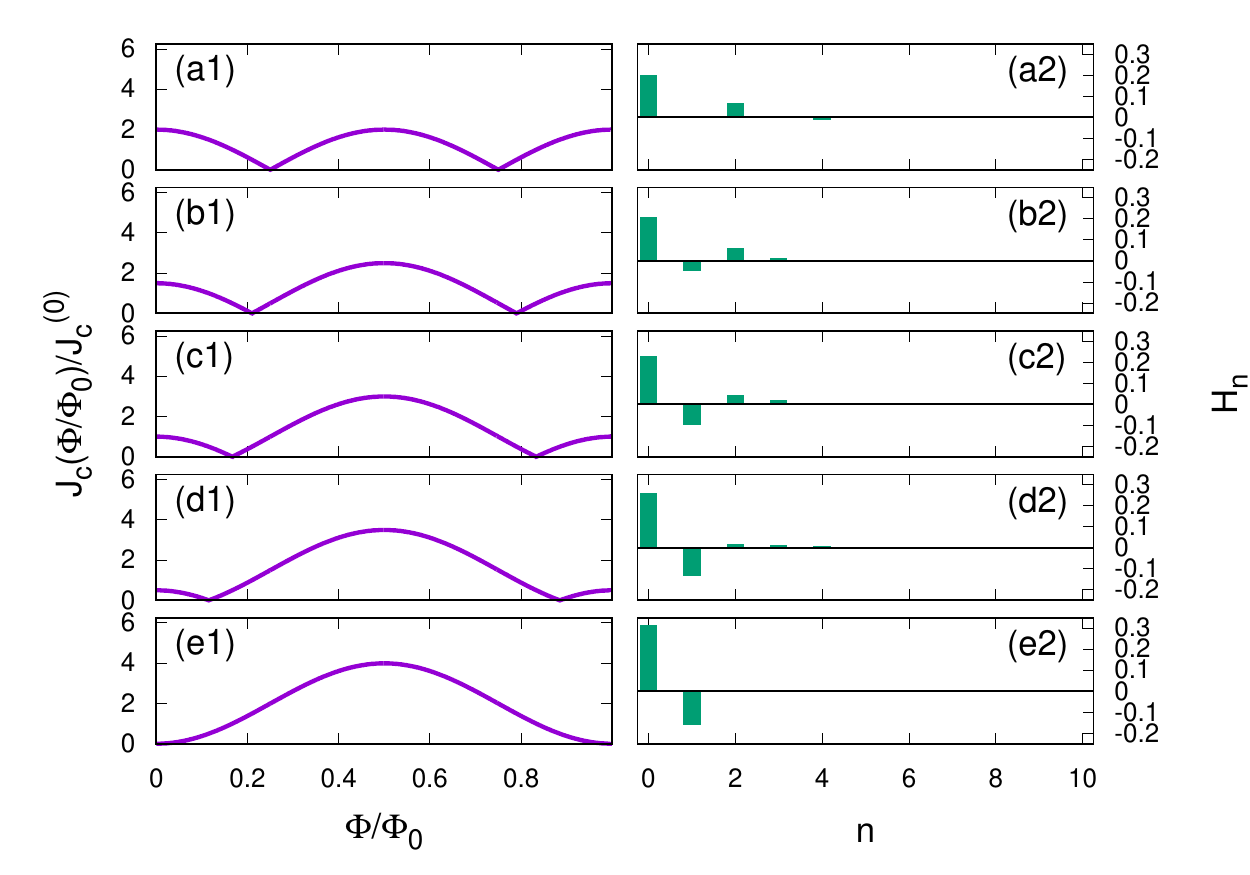}
  \caption{{\it Critical current-flux relations (panel a1-e1) and the
      corresponding Fourier coefficients (panels a2-e2):} The
    parameters $\alpha_{3TQ_1}=\alpha_{3TQ_2}=-1$ are used, and
    $\alpha_{4TSQ}=0,\,0.5,\,1,\,\,1.5,\,2$ on panels a1-a2, b1-b2,
    c1-c2, d1-d2 and e1-e2 respectively. The opposite signs of
    $\alpha_{3TQ_1}<0$, $\alpha_{3TQ_2}<0$ and $\alpha_{4TSQ}\ge 0$
    correspond to relative shift of $\pi$ between the three-terminal
    3TQ$_1$, 3TQ$_2$ and the four-terminal 4TSQ. This relative
    $\pi$-shift is the result of lowest-order perturbation theory in
    the tunnel amplitudes, see section~\ref{sec:perturbation-V0+}.
    \label{fig:figure-shift-pi}}
\end{figure*}

\begin{figure*}[htb]
  \includegraphics[width=\textwidth]{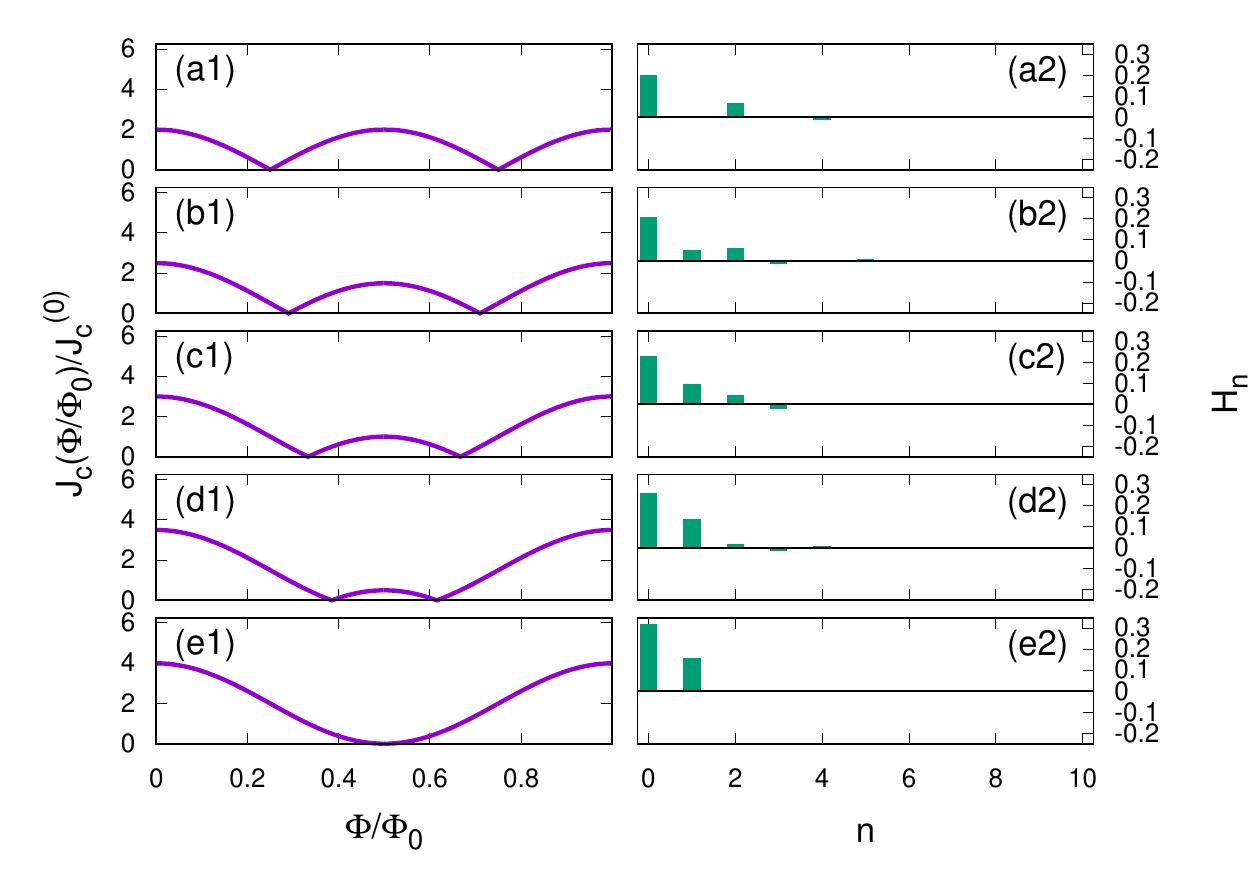}
  \caption{{\it Critical current-flux relations (panel a1-e1) and the
      corresponding Fourier coefficients (panels a2-e2):} The
    parameters $\alpha_{3TQ_1}=\alpha_{3TQ_2}=-1$ are used, and
    $\alpha_{4TSQ}=0,\,-0.5,\,-1,\,\,-1.5,\,-2$ on panels a1-a2,
    b1-b2, c1-c2, d1-d2 and e1-e2 respectively. The signs of
    $\alpha_{3TQ_1}<0$, $\alpha_{3TQ_2}<0$ and $\alpha_{4TSQ}\le 0$
    corresponds to relative $0$-shift between the three-terminal
    3TQ$_1$, 3TQ$_2$ and the four-terminal 4TSQ.
    \label{fig:figure-shift-0}}
\end{figure*}

\subsection{Conclusion on the Harvard group
  experiment\cite{Harvard-group-experiment}}
\label{sec:conclusions-Harvard-group-experiment-0-pi}

In this subsection, we present the consequences for the Harvard group
experiment \cite{Harvard-group-experiment}.

{As it is mentioned above, perturbation theory in the tunnel
  amplitudes $J_0/W$ combined to the adiabatic limit $V=0^+$ imply the
  $\pi$-shifted three-terminal 3TQ$_1$, 3TQ$_2$, and $0$-shifted
  four-terminal 4TSQ, see section~\ref{sec:perturbation-V0+}.}  {Given
  that Eq.~(\ref{eq:logical-equivalence3}) implies
  Eq.~(\ref{eq:logical-equivalence1}), we conclude that perturbation
  theory and the adiabatic limit imply ``{\it Inversion in the
    critical current $I_c(\Phi/\Phi_0)$ between $\Phi/\Phi_0=0$ and
    $\Phi/\Phi_0=1/2$}'', {\it i.e.} $I_c(0)<I_c(1/2)$.}

Conversely, ``{\it Experimental evidence for inversion}'' implies

  ``{\it Evidence that the three-terminal 3TQ$_1$,
    3TQ$_2$ are $\pi$-shifted and the four-terminal 4TSQ are $0$-shifted}'',

  or, alternatively:

  ``{\it Evidence for $0$-shifted 3TQ$_1$, 3TQ$_2$ and $\pi$-shifted
    4TSQ}.

  No information is gained about which of the three-terminal
  3TQ$_1$, 3TQ$_2$ or the four-terminal 4TSQ which is $\pi$-shifted,
  the other being $0$-shifted.}

\section{Gate voltage dependence of the
magnetic field oscillations}
\label{sec:gate-voltage0}
\subsection{Notations for the phenomenological model}
The previous calculations are summarized in the following
phenomenological form of the critical current-flux $\Phi$ relation:
\begin{eqnarray}
  \label{eq:Ic-4TSQ-phenomenological}
  J_{c}(\Phi/\Phi_0)&=& J_c^{(0)}\mbox{Max}_{\varphi_{q,\,3T}}\left\{
  \alpha_{3TQ_1} \sin\left(\varphi_{q,\,3T}+\Phi\right)\right.\\
  &+&\left.\alpha_{3TQ_2} \sin\left(\varphi_{q,\,3T}-\Phi\right)
  +\alpha_{4TSQ} \sin\varphi_{q,\,3T}\right\},\nonumber
\end{eqnarray}
which is deduced from the previous Eq.~(\ref{eq:Ic-4TSQ}).

The factorized scaling parameter $J_c^{(0)}$ is positive and it has
dimension of a critical current. The dimensionless parameters
$\alpha_{3TQ_1}$, $\alpha_{3TQ_2}$ and $\alpha_{4TSQ}$ characterize
the relative weights and signs of the three-terminal 3TQ$_1$, 3TQ$_2$
and the four-terminal 4TSQ critical currents.

The perturbative calculations presented in the above
section~\ref{sec:perturbation-V0+} lead to $\alpha_{3TQ_1}<0$,
$\alpha_{3TQ_2}<0$, and to $\alpha_{4TSQ}>0$. Following the previous
section~\ref{sec:<>}, we assume more generally that the three-terminal
$\alpha_{3TQ_1}$, $\alpha_{3TQ_2}$ and the four-terminal
$\alpha_{4TSQ}$ can have arbitrary positive or negative relative
signs.

General positive or negative signs of $\alpha_{3TQ_1}$,
$\alpha_{3TQ_2}$, and $\alpha_{4TSQ}$ could be relevant to higher
transparency of the contacts between the 2D metal and the
superconducting leads. However, it has not yet been examined whether
increasing the contact transparency {\it via} the parameter $J_0/W$
can produce change of sign in these coefficients.

\subsection{Analogy with interferometric detection of the $\pi$-shift
  \cite{Guichard}}
\label{sec:analogy}

Now, we mention a connection between

(i) The SQUID containing a $0$ and a $\pi$-junction which was realized
experimentally in Ref.~\onlinecite{Guichard}.

(ii) The relative $\pi$-shift between the three-terminal
3TQ$_1$, 3TQ$_2$ and the four-terminal 4TSQ.

More specifically:

(i) Half-period shift of the critical current magnetic oscillations is
observed in Ref.~\onlinecite{Guichard} with a SQUID containing a
$\pi$-shifted and a $0$-shifted Josephson junction, in comparison with
a SQUID containing two $0$-shifted Josephson junctions.

(ii) Half-period shift in the critical current of the four-terminal
Josephson junction shown on figures~\ref{fig:device}
and~\ref{fig:device2} is produced in our theory when changing
``relative shift of $\pi$ between the three-terminal 3TQ$_1$,
3TQ$_2$'' and the four-terminal 4TSQ into ``relative shift of $0$''.

\subsection{Gate voltage dependence of the
  critical current magnetic oscillations in the perturbative limit}
\label{sec:gate-voltage}

{Figures~\ref{fig:figure-shift-pi} and~\ref{fig:figure-shift-0} show
  on panels a1-e1 the magnetic oscillations of the critical current
  given by Eq.~(\ref{eq:Ic-4TSQ-phenomenological}), and
  their Fourier coefficients $H_n$ are shown on panels a2-e2:
  \begin{equation}
    \label{eq:Hn}
    H_n=\int \frac{d\Phi}{2\pi} 
    \cos\left(\frac{2\pi n \Phi}{\Phi_0}\right)
    J_c\left(\frac{\Phi}{\Phi_0}\right)
    ,
  \end{equation}
  where $J_c(\Phi/\Phi_0)$ is given by
  Eq.~(\ref{eq:Ic-4TSQ-phenomenological}).  The parameters
  $\alpha_{3TSQ_1}=\alpha_{3TSQ_2}=-1$ used on
  figures~\ref{fig:figure-shift-pi} and~\ref{fig:figure-shift-0} have
  the meaning of the $\pi$-shifted three-terminal 3TQ$_1$, 3TQ$_2$
  critical currents deduced from perturbation theory in $J_0/W$, see
  section~\ref{sec:perturbation-V0+}.}

{The parameter $\alpha_{4TSQ}>0$ is used on
  figure~\ref{fig:figure-shift-pi}, thus with $\pi$-shift between the
  three-terminal 3TQ$_1$, 3TQ$_2$ and the four-terminal 4TSQ:
  $\alpha_{4TSQ}=0$ (panels a1-a2), $\alpha_{4TSQ}=0.5$ (panels
  b1-b2), $\alpha_{4TSQ}=1$ (panels c1-c2), $\alpha_{4TSQ}=1.5$
  (panels d1-d2), and to $\alpha_{4TSQ}=2$ (panels e1-e2).}

{Figure~\ref{fig:figure-shift-0} shows the corresponding data with
  $\alpha_{4TSQ}<0$ {\it i.e.} with and $0$-shift between the
  three-terminal 3TQ$_1$, 3TQ$_2$ and the four-terminal 4TSQ:
  $\alpha_{4TSQ}=0$ (panels a1-a2), $\alpha_{4TSQ}=-0.5$ (panels
  b1-b2), $\alpha_{4TSQ}=-1$ (panels c1-c2), $\alpha_{4TSQ}=-1.5$
  (panels d1-d2), and to $\alpha_{4TSQ}=-2$ (panels e1-e2).  }

{Figures~\ref{fig:figure-shift-pi}~a1-e1 and
  figures~\ref{fig:figure-shift-0}~a1-e1 illustrate the logical chain
  of
  Eqs.~(\ref{eq:logical-equivalence1})-(\ref{eq:logical-equivalence3}):
  Figures~\ref{fig:figure-shift-pi}~a1-e1 with relative $\pi$-shift
  between the three-terminal 3TQ$_1$, 3TQ$_2$ and the four-terminal
  4TSQ reveal the inversion $J_c(0)<J_c(1/2)$ between
  $J_c(\Phi/\Phi_0)$ at $\Phi/\Phi_0=0$ and
  $\Phi/\Phi_0=1/2$. Conversely,
  figures~\ref{fig:figure-shift-0}~a1-e1 with relative $0$-shift
  feature the noninverted behavior $J_c(0)>J_c(1/2)$.}

{In addition, figures~\ref{fig:figure-shift-pi}
  and~\ref{fig:figure-shift-0} are deduced from each other by
  half-period shift of $\Phi/2\pi$ on the $x$-axis, which is in
  agreement with the analogous SQUID containing a $0$- and a
  $\pi$-shifted Josephson junction, see Ref.~\onlinecite{Guichard}
  and the preceding subsection~\ref{sec:analogy}.}

{ Gating the 2D metal away from the center of the band has the effect
  of increasing the density of states, which increases $J_0/W$ and
  favors the four-terminal 4TSQ over the three-terminal 3TQ$_1$,
  3TQ$_2$, because they appear in perturbation at the different orders
  $(J_0/W)^8$ and $(J_0/W)^{12}$ respectively, see
  section~\ref{sec:perturbation-V0+}. }

  {It is deduced from figures~\ref{fig:figure-shift-pi}~a2-e2 and
    figures~\ref{fig:figure-shift-0}~a2-e2 that tuning gate voltage
    away from the Dirac point increases $|H_1|$ and reduces $H_2$
    [where $H_1$ and $H_2$ are defined as the $n=1,\,2$ in
      Eq.~(\ref{eq:Hn})]}, {which favors the $\Phi_0$ harmonics over
    the $2\Phi_0$ one.  Figures~\ref{fig:figure-shift-pi}
    and~\ref{fig:figure-shift-0} reveal in addition the expected
    negative $H_1<0$ for relative $\pi$-shift between the
    three-terminal 3TQ$_1$, 3TQ$_2$ and the four-terminal 4TSQ, and
    $H_1>0$ for a relative $0$-shift, which is in agreement with the
    perturbative calculations of section~\ref{sec:perturbation-V0+}.}

  We conclude this section with underlying that our theory is in a
  qualitative agreement with the Harvard group experimental data
  \cite{Harvard-group-experiment} regarding the gate voltage
  dependence of the critical current magnetic oscillations on the
  quartet line.}  Figures~\ref{fig:figure-shift-pi}
and~\ref{fig:figure-shift-0} are related to figure~3 in the recent
experimental preprint of the Harvard
group\cite{Harvard-group-experiment}.

\section{Generalization to arbitrary interface transparencies and 
finite bias voltage}
\label{sec:beyond-perturbation-theory}

The three-terminal 3TQ$_1$, 3TQ$_2$ transmit even number of Cooper
pairs into $S_{{c_1}}$ or $S_{{c_2}}$ while the four-terminal 4TSQ
transmit odd number of Cooper pairs. This characterization based on
the parity of the number of Cooper pairs transmitted into $S_{c,1}$ or
$S_{c,2}$ is now generalized in the following subsection to arbitrary
interface transparencies and finite bias voltage.

Given the arguments of subsection~\ref{sec:connection-ball}, we
replace the ``realistic model I of clean interfaces and
superconductors in the dirty limit'' by the ``physically motivated
approximation of the model~III'', {\it i.e.} clean interfaces and
superconductors in the ballistic limit, and averaging over $\{k_F
R_{k,l}\}$.

{We start in subsection~\ref{sec:generalized-AB-adiabatic} with
  demonstrating the generalized Ambegaokar-Baratoff formula in the
  $V=0^+$ adiabatic limit at arbitrary interface transparencies.  The
  next subsection~\ref{sec:AB-V} generalizes this argument to
  finite voltage $V$ on the quartet line, instead of the previous
  $V=0^+$ adiabatic limit. Discussion of the Harvard group experiment
  \cite{Harvard-group-experiment} is presented in
  subsections~\ref{sec:Generalized-AB} and \ref{sec:AB-V-exp}.}

\subsection{Generalized Ambegaokar-Baratoff formula
  in the $V=0^+$ adiabatic limit}
\label{sec:generalized-AB-adiabatic}

{We start in this subsection with the $V=0^+$ adiabatic limit.
  Subsection~\ref{sec:classification-and-AB-formula-V0plus}
  demonstrates the generalized Ambegaokar-Baratoff for the quartet
  current-flux relation, see the forthcoming
  Eq.~(\ref{eq:Iqc-non-perturbatif}). Subsection~\ref{sec:Generalized-AB}
  presents experimental consequences.}

\subsubsection{Demonstration of the generalized Ambegaokar-Baratoff formula
at $V=0^+$}
\label{sec:classification-and-AB-formula-V0plus}

{Now, we calculate the quartet current in the $V=0^+$ adiabatic limit,
  for arbitrary interface transparencies, and within the model~III
  presented in the above subsection~\ref{sec:connection-ball}.}
   
{The first term $J_{a,\alpha}\hat{G}_{\alpha,a}^A$
  appearing in Eq.~(\ref{eq:I1-eq}) is written as
\begin{eqnarray}
  J_{a,\alpha}\hat{G}_{\alpha,a}^A&=&
  \sum_n \sum_{m_a,m_b,m_{c_1},m_{c_2}}
  X_n^{(m_a,m_b,m_{c_1},m_{c_2}),A}\\
  \nonumber
  &&\times J_a^{2 m_a} J_b^{2 m_b}
  J_{{c_1}}^{2 m_{{c_1}}} J_{{c_2}}^{2 m_{{c_2}}}
  \exp\left(i n \varphi_{q,\,3T} \right)
  .
\end{eqnarray}
Conversely, $J_{a,\alpha}\hat{G}_{\alpha,a}^R$ involving the retarded
Green's function takes the form
\begin{eqnarray}
  J_{a,\alpha}\hat{G}_{\alpha,a}^R&=&
  \sum_n \sum_{m_a,m_b,m_{c_1},m_{c_2}}
  X_n^{(m_a,m_b,m_{c_1},m_{c_2}),R}\\&&\times J_a^{2 m_a} J_b^{2 m_b}
  J_{{c_1}}^{2 m_{{c_1}}} J_{{c_2}}^{2 m_{{c_2}}}
  \exp\left(i n \varphi_{q,\,3T} \right)
  .
  \nonumber
\end{eqnarray}
The bare Green's functions [{\it i.e.}
  Eqs.~(\ref{eq:gA-asymptotics})-(\ref{eq:gR-asymptotics}) and
  Eq.~(\ref{eq:gA-supra-general-ballistique})] are used to produce a
relation between the ``advanced'' and the ``retarded'' Green's
functions by taking the complex conjugate and changing the sign of the
superconducting phases.}{ This symmetry is then generalized to the
  fully dressed advanced and retarded Green's functions by making use
  of the Dyson Eq.~(\ref{eq:otimes}). The resulting
  $\hat{G}^A(\omega,R_0,\psi_F,\varphi_N) =
  \left[\hat{G}^R(\omega,R_0,\psi_F,-\varphi_N)\right]^*$ leads to
\begin{equation}
  X_n^{(m_a,m_b,m_{c_1},m_{c_2}),R}
  =\left[X_n^{(m_a,m_b,m_{c_1},m_{c_2}),A}\right]^*
  .
\end{equation}
 Thus,
\begin{eqnarray}
  \label{eq:A_minus_R}
&&  J_{a,\alpha}\hat{G}_{\alpha,a}^A-J_{a,\alpha}\hat{G}_{\alpha,a}^R
  = 2i
  \sum_n \sum_{m_a,m_b,m_{c_1},m_{c_2}}\\
  &&\times
  \mbox{Im}\left[X_n^{(m_a,m_b,m_{c_1},m_{c_2}),R}\right] J_a^{2 m_a} J_b^{2 m_b}
  J_{{c_1}}^{2 m_{{c_1}}} J_{{c_2}}^{2 m_{{c_2}}}
  \exp\left(i n \varphi_{q,\,3T} \right)
  \nonumber
  ,
\end{eqnarray}
where the variable $n$ stands for $n\equiv n_a=n_b$, see the notations
in Eq.~(\ref{eq:nanbnc}). Eqs.~(\ref{eq:I1-eq})-(\ref{eq:I4-eq}) imply
the following decomposition of the critical current in the $V=0^+$
adiabatic limit on the quartet line:
\begin{eqnarray}
  \label{eq:Iqc-non-perturbatif}
  I'_{c}(\Phi/\Phi_0)&=&\mbox{Max}_{\varphi_c}
  \sum_{n,p} X(2n,p) \times \\ &&\sin
  \left[ \left(2 n - p\right) \left(\varphi_c-\frac{\Phi}{2}\right) +
    p \left( \varphi_c+\frac{\Phi}{2}\right)\right]
  ,\nonumber
\end{eqnarray}
where the quartet phase is expressed in the gauge given by
Eqs.~(\ref{eq:gauge1}) and~(\ref{eq:gauge2}). Eq.~(\ref{eq:A_minus_R})
shows that the coefficients $X(2n,p)$ appearing in the
Ambegaokar-Baratoff formula Eq.~(\ref{eq:Iqc-non-perturbatif}) are
real-valued. A number $n$ of Cooper pairs is taken from the
superconducting lead $S_a$ biased at $V_a=+V$, and $n$ others pairs
are taken from $S_b$ biased at $V_b=-V$. The integer $p$ in
Eq.~(\ref{eq:Iqc-non-perturbatif}) denotes partition between the $p$
pairs transmitted into $S_{{c_2}}$ contact and the remaining $2n-p$
pairs transmitted into $S_{{c_1}}$.}

\subsubsection{Experimental consequences}
\label{sec:Generalized-AB}
{In this subsection, we proceed further with the same
  assumptions as in the preceding
  subsection~\ref{sec:classification-and-AB-formula-V0plus}, and
  establish a link between:}

(i) Emergence of different values for the critical current at fluxes
$\Phi/\Phi_0=0$ and $\Phi/\Phi_0=1/2$ [{\it i.e.}  $I_c(0)\ne
  I_c(1/2)$].

(ii) Evidence for interference between quantum
processes transmitting even or odd numbers of Cooper pairs into
$S_{{c_1}}$ or $S_{{c_2}}$.}

Specifically, we make the change of variables $\varphi_c\rightarrow
\varphi_c +\Phi/2$ in Eq.~(\ref{eq:Iqc-non-perturbatif}), which is
equivalent to changing the gauge from
Eqs.~(\ref{eq:gauge1})-(\ref{eq:gauge2}) to
$\varphi_{{c_1}}=\varphi_c$ and $\varphi_{{c_2}}=\varphi_c+\Phi$:
\begin{equation}
  \label{eq:Iqc-non-perturbatif2}
  I'_{c}(\Phi/\Phi_0)=\mbox{Max}_{\varphi_c}
  \sum_{n,p} X(2n,p) \sin
  \left[ \left(2 n - p\right) \varphi_c+
    p \left( \varphi_c+\Phi\right)\right]
  .
\end{equation}
This Eq.~(\ref{eq:Iqc-non-perturbatif2}) simplifies as
\begin{equation}
  I'_{c}(\Phi/\Phi_0)=\mbox{Max}_{\varphi_c} \sum_{n,p} X(2n,p) \sin \left[ 2 n
    \varphi_c+ p \Phi\right].
\end{equation}
It deduced that
\begin{eqnarray}
  I'_{c}(0)&=&\mbox{Max}_{\varphi_c} \sum_{n,p} X(2n,p) \sin \left[ 2
    n \varphi_c \right]\\ I'_{c}(1/2)&=&\mbox{Max}_{\varphi_c}
  \sum_{n,p} X(2n,p) (-)^p \sin \left[ 2 n \varphi_c \right]
  .
\end{eqnarray}
Separating the terms with $p$ even or odd according to
\begin{eqnarray}
  \label{eq:Y-even}
  Y_{even}(\varphi_c)=\sum_{p \, even} \sum_n X(2n,p) \sin \left[ 2 n
    \varphi_c \right]\\
  Y_{odd}(\varphi_c)=\sum_{p \, odd} \sum_n X(2n,p) \sin \left[ 2 n
    \varphi_c \right]
  \label{eq:Y-odd}
\end{eqnarray}
leads to
\begin{eqnarray}
  \label{eq:Iseconde-1}
  I'_{c}(0)&=& \mbox{Max}_{\varphi_c} \left[ Y_{even}(\varphi_c) +
    Y_{odd}(\varphi_c) \right]\\
  \label{eq:Iseconde-2}
  I'_{c}(1/2)&=&
  \mbox{Max}_{\varphi_c} \left[ Y_{even}(\varphi_c) -
    Y_{odd}(\varphi_c) \right] .
\end{eqnarray}

The following logical link is deduced within the assumptions
mentioned above:

``{\it Experimental observation for different values of the critical
  current between reduced fluxes $\Phi/\Phi_0=0$ and $\Phi/\Phi_0=1/2$
  at arbitrary transparency}'' [{\it i.e.}  $I'_{c}(0)\ne I'_{c}(1/2)$
  in Eqs.~(\ref{eq:Iseconde-1}) and~(\ref{eq:Iseconde-2})]

  is equivalent to

  ``{\it
  Evidence for interference between processes transmitting even or
  odd number of Cooper pairs into $S_{{c_1}}$ and $S_{{c_2}}$}''.

\subsection{Generalization to finite bias voltage on the quartet line}
\label{sec:AB-V}

Now, we generalize to finite bias voltage $V$ and
arbitrary interface transparencies. Disorder is treated within the
model~III introduced in subsection~\ref{sec:connection-ball}.

Specifically, we show in subsection~\ref{sec:AB-V-1} that the
Ambegaokar-Baratoff formula Eq.~(\ref{eq:Iqc-non-perturbatif}) holds
at finite~$V$ within our treatment.  Consequences for the proposed
interpretation of the Harvard group experiment are discussed in
subsection~\ref{sec:AB-V-exp}.

\subsubsection{Demonstration of the Ambegaokar-Baratoff formula
  at finite bias voltage}

\label{sec:AB-V-1}

{Now, at finite bias voltage~$V$ on the quartet line, we show that the
  currents transmitted at the $S_{c,1}$ or $S_{c,2}$ contacts take
  the form of the generalized Ambegaokar-Baratoff formula
  Eq.~(\ref{eq:Iqc-non-perturbatif}), where the coefficients $X(2n,p)$
  appearing in the $V=0^+$ Eq.~(\ref{eq:Iqc-non-perturbatif}) are
  replaced by their values $X(2n,p,eV/\Delta)$ at finite bias
  voltage~$V$.}

{The Keldysh Green's function given by
  Eq.~(\ref{eq:expression-de-Gpm}) is written as
  $\hat{G}^{+,-}=\hat{G}^{+,-}_A+\hat{G}^{+,-}_B$, where the
  ``quasiequilibrium'' and the ``nonequilibrium'' $\hat{G}^{+,-}_A$
  and $\hat{G}^{+,-}_B$ are given by
\begin{eqnarray}
  \label{eq:Gpm-A}
  \hat{G}^{+,-}_A&=& \hat{n}_F \hat{G}^A - \hat{G}^R \hat{n}_F\\
  \hat{G}^{+,-}_B&=& \hat{G}^R
  \left[\hat{J} \hat{n}_F - \hat{n}_F \hat{J}\right]
  \hat{G}^A
  \label{eq:Gpm-B}
\end{eqnarray}
respectively.  } {The matrices appearing in
  Eqs.~(\ref{eq:Gpm-A})-(\ref{eq:Gpm-B}) are now defined both in Nambu
  and in the infinite set of harmonics of the Josephson
  frequencies. In the following, we use the notations $\check{
    \varphi}=(\varphi_a,\varphi_b,\varphi_{c,1},\varphi_{c,2})$ for
  the superconducting phases and
  ${\check{n}}=(n_a,n_b,n_{c,1},n_{c,2})$ for labeling the multiples
  of the voltage frequency $eV/\hbar$. The $\check{n}$ vector belongs
  to the set $S_{\check{n}}$ of quadruplets which fulfill the
  constraints
\begin{eqnarray}
  \label{eq:nanbnc1nc2}
&&n_a+n_b+n_{c,1}+n_{c,2} = 0\\
&&n_a = n_b,
\label{eq:nanb}
\end{eqnarray}
see the discussion following Eq.~(\ref{eq:nanbnc}).}

{We start with the quasiequilibrium contribution $I_A$
  deduced from Eqs.~(\ref{eq:I1})-(\ref{eq:I4}):
  \begin{eqnarray}
    \nonumber
    &&-I_A \equiv - I_{A,\gamma_{c_1} \rightarrow c_1}=\frac{e}{\hbar}
    \sum_p \int d\omega \\ \label{eq:I1-A}
    &&\left\{
    \left[\hat{J}_{c_{1,p},\gamma_{c_{1,p}}}
      \hat{G}^{+,-}_{A,\gamma_{c_{1,p}},c_{1,p}}\right]_{(1,1)/(0,0)}(\omega,\check{\varphi},\{\psi_{F,k,l}\},V)\right. \\&&-
    \left[\hat{J}_{c_{1,p},\gamma_{c_{1,p}}}
      \hat{G}^{+,-}_{A,\gamma_{c_{1,p}},c_{1,p}}\right]_{(2,2)/(0,0)}(\omega,\check{\varphi},\{\psi_{F,k,l}\},V)\\&&
    -\left[\hat{J}_{\gamma_{c_{1,p}},c_{1,p}}
      \hat{G}^{+,-}_{A,c_{1,p},\gamma_{c_{1,p}}}\right]_{(1,1)/(0,0)}(\omega,\check{\varphi},\{\psi_{F,k,l}\},V)\\&&
    + \left.  \left[\hat{J}_{\gamma_{c_{1,p}},c_{1,p}}
      \hat{G}^{+,-}_{A,c_{1,p},\gamma_{c_{1,p}}}\right]_{(2,2)/(0,0)}(\omega,\check{\varphi},\{\psi_{F,k,l}\},V)\right\}
    ,
\label{eq:I4-A}
  \end{eqnarray}
  where $\hat{G}^{+,-}_A$ is given by Eq.~(\ref{eq:Gpm-A}).}  {The
  notation $\{\psi_{F,k,l}\}$ stands for the phases oscillating at the
  scale of the Fermi wave-length in a multichannel configuration, as
  they appear in the 2D metal and superconductor Green's functions,
  see Eqs.~(\ref{eq:gA-asymptotics})-(\ref{eq:gR-asymptotics}) and
  Eq.~(\ref{eq:gA-supra-general-ballistique}) respectively. Namely,
  $\psi_{F,k,l}=k_F R_{k,l}-\pi/4$ for the 2D metal, see
  Eqs.~(\ref{eq:gA-asymptotics})-(\ref{eq:gR-asymptotics}), and
  $\psi_{F,k,l}=k_F R_{k,l}$ for the ballistic 3D superconductors, see
  Eq.~(\ref{eq:gA-supra-general-ballistique}).}

{The Dyson Eq.~(\ref{eq:otimes}) implies that the fully dressed
  advanced and retarded Green's functions take the following form:
\begin{eqnarray}
  \label{eq:GAn}
  \hat{G}^A\left(\omega,\check{\varphi},\{\psi_{F,k,l}\},V\right) &=&
  \sum_{\check{n}\in S_{\check{n}}}
  \hat{G}^A_{\check{n}}\left(\omega,\{\psi_{F,k,l}\},V\right) e^{i
    {\check{n}}.\check{\varphi}}
  \\ \hat{G}^R\left(\omega,\check{\varphi},\{k_F
      R_{k,l}\}\right) &=& \sum_{\check{n}\in S_{\check{n}}}
      \hat{G}^R_{\check{n}}\left(\omega,\{\psi_{F,k,l}\}\right)
      e^{i {\check{n}}.\check{\varphi}} .
      \label{eq:GRn}
\end{eqnarray}
} {In order to relate $\hat{G}^A_{\check{n}}$ to
    $\hat{G}^R_{\check{n}}$ in Eqs.~(\ref{eq:GAn}) and~(\ref{eq:GRn}),
    we note that the bare Green's functions given by
    Eqs.~(\ref{eq:gA-asymptotics})-(\ref{eq:gR-asymptotics}) and
    Eq.~(\ref{eq:gA-supra-general-ballistique}) are such that
\begin{equation}
  \label{eq:T-Nambu}
  \hat{\cal T}_{Nambu} \hat{g}^{A,R}
  \left(\omega,\check{\varphi},\{\psi_{F,k,l}\},V\right)
  =
\hat{g}^{A,R}\left(\omega,-\check{\varphi},\{-\psi_{F,k,l}\},-V\right)
,
\end{equation}
where the transformation $\hat{\cal T}_{Nambu}$
exchanges the ``1'' and ``2'' Nambu components for ``spin-up
electron'' and ``spin-down hole'' respectively. The Dyson equation
given by Eq.~(\ref{eq:otimes}) yields
\begin{eqnarray}
  \label{eq:T-Nambu2-1}
  &&\hat{\cal T}_{Nambu} \hat{J}_{a,\gamma_{c_1}} \hat{G}^{A,R}_{\gamma_{c_1},c_1}
  \left(\omega,\check{\varphi},\{\psi_{F,k,l}\},V\right)\\
  &=&\hat{J}_{a,\gamma_{c_1}}
  \hat{G}^{A,R}_{\gamma_{c_1},c_1}\left(\omega,-\check{\varphi},
\{-\psi_{F,k,l}\},-V\right)
  .
  \nonumber
\end{eqnarray}
Combining Eq.~(\ref{eq:Gpm-A}) to
Eq.~(\ref{eq:T-Nambu2-1}) leads to
\begin{eqnarray}
  \nonumber && \sum_p \int d\omega n_F(\omega) \left\{
  \langle\langle\hat{J}_{c_{1,p},\gamma_{c_{1,p}}}
  \hat{G}^{A}_{\gamma_{c_{1,p}},c_{1,p}}\left(\omega,\{\psi_{F,k,l}\},V,\check{\varphi}\right)
  \rangle\rangle_{(1,1)}\right.\\ && - \nonumber
  \left.\langle\langle\hat{J}_{c_{1,p},\gamma_{c_{1,p}}}
  \hat{G}^{A}_{\gamma_{c_{1,p}},c_{1,p}}\left(\omega,\{\psi_{F,k,l}\},-V,
  \check{\varphi}\}\right)\rangle\rangle_{(2,2)}\right\}
  \\ &&= 2i \sum_p \int d\omega n_F(\omega)
  \sum_{\check{n}\in S_{\check{n}}}\\\nonumber&& \langle\langle
  \hat{J}_{c_{1,p},\gamma_{c_{1,p}}}\mbox{Im}
  \left[G^A_{\gamma_{c_{1,p}},c_{1,p},{\check{n}}}\left(\omega,\{\psi_{F,k,l}\},V
    \right)\right] \rangle\rangle_{(1,1)}
  \sin\left({\check{n}}.\check{\varphi}\right) .
  \label{eq:second-term}
\end{eqnarray}
Within the considered model~III, averaging over disorder is mimicked
by integrating over the phases $\{\psi_{F,k,l}\}$ in the
$\left[0,2\pi\right]$ interval. The terms which are odd in $k_F
R_{k,l}$ do not contribute to this integral, and thus
\begin{eqnarray}
&&\langle\langle \hat{J}_{c_{1,p},\gamma_{c_{1,p}}}
  \hat{G}^{A}_{\gamma_{c_{1,p}},c_{1,p}}(\omega,\check{\varphi},\{\psi_{F,k,l}\},V)\rangle\rangle_{(1,1)}\\ &=& \langle\langle
  \hat{J}_{c_{1,p},\gamma_{c_{1,p}}}
  \hat{G}^{A}_{\gamma_{c_{1,p}},c_{1,p}}(\omega,-\check{\varphi},\{\psi_{F,k,l}\},-V)\rangle\rangle_{(2,2)} \nonumber
\end{eqnarray}
is independent on whether
$\hat{G}^{A}_{\gamma_{c_{1,p}},c_{1,p}}(\omega,-\check{\varphi},\{\psi_{F,k,l}\},V)$
or
$\hat{G}^{A}_{\gamma_{c_{1,p}},c_{1,p}}(\omega,-\check{\varphi},\{-\psi_{F,k,l}\},V)$
is averaged over $\{\psi_{F,k,l}\}$.} In addition, the calculation is
specific to the ``quartet current'' $I_q$ which is even if the
voltage~$V$ changes sign.

{The subtracted ``retarded'' terms are deduced from the
  ``advanced'' ones by taking the complex conjugate and changing
  $\check{\varphi}$ into $-\check{\varphi}$, see
  subsection~\ref{sec:classification-and-AB-formula-V0plus}. We deduce
  the following expression of $\langle\langle I_A \rangle\rangle$:
  \begin{eqnarray}
    \label{eq:AB-A}
    \langle\langle I_{q,A} \rangle\rangle&=&\frac{2e}{\hbar} \sum_p \int
    d\omega n_F(\omega) \sum_{\check{n}\in S_{\check{n}}}
    \\ &&\nonumber\mbox{Im} \langle\langle
    J_{c_{1,p},\gamma_{c_{1,p}}}
    G^A_{{\check{n}},\gamma_{c_{1,p}},c_{1,p}}
    \left(\omega,\{\psi_{F,k,l}\},V\right)\rangle\rangle
    \sin\left({\check{n}}.\check{\varphi}\right)\\
    &+& \left( V \rightarrow -V \right)
    \nonumber
    ,
\end{eqnarray}
which takes the form of the Ambegaokar-Baratoff formula
Eq.~(\ref{eq:Iqc-non-perturbatif}) for $I_{q,A}(V)=
[I_A(V)+I_A(-V)]/2$.}

{Now, Eqs.~(\ref{eq:I1})-(\ref{eq:I4}) and Eq.~(\ref{eq:Gpm-B}) yield
  the following ``nonequilibrium'' contribution $I_B$ to the current:
  \begin{eqnarray}
    \nonumber && -I_B\equiv- I_{B,\gamma_{c_1} \rightarrow
      c_1}=\frac{e}{\hbar} \sum_p \int d\omega\\ \label{eq:I1-B}&&
    \left\{ \left[\hat{J}_{c_{1,p},\gamma_{c_{1,p}}}
      \hat{G}^{+,-}_{B,\gamma_{c_{1,p}},c_{1,p}}\right]_{(1,1)/(0,0)}(\omega,\check{\varphi},\{\psi_{F,k,l}\},V)\right. \\&&-
    \left[\hat{J}_{c_{1,p},\gamma_{c_{1,p}}}
      \hat{G}^{+,-}_{B,\gamma_{c_{1,p}},c_{1,p}}\right]_{(2,2)/(0,0)}(\omega,\check{\varphi},\{\psi_{F,k,l}\},V)\\&&
    -\left[\hat{J}_{\gamma_{c_{1,p}},c_{1,p}}
      \hat{G}^{+,-}_{B,c_{1,p},\gamma_{c_{1,p}}}\right]_{(1,1)/(0,0)}(\omega,\check{\varphi},\{\psi_{F,k,l}\},V)\\&&
    + \left.  \left[\hat{J}_{\gamma_{c_{1,p}},c_{1,p}}
      \hat{G}^{+,-}_{B,c_{1,p},\gamma_{c_{1,p}}}\right]_{(2,2)/(0,0)}(\omega,\check{\varphi},\{\psi_{F,k,l}\},V)\right\}
    ,
\label{eq:I4-B}
  \end{eqnarray}
  where $\hat{G}^{+,-}_B$ is given by Eq.~(\ref{eq:Gpm-B}).}

{We make use of the transformation ${\cal T}_{Nambu}$ given by
  Eq.~(\ref{eq:T-Nambu}) to obtain
  \begin{eqnarray}
    \nonumber &&\left[\hat{J}_{c_{1,p},\gamma_{c_{1,p}}}
      \hat{G}^{+,-}_{B,\gamma_{c_{1,p}},c_{1,p}}\right]_{(1,1)/(0,0)}(\omega,\check{\varphi},\{\psi_{F,k,l}\},V)\\\nonumber
    &-& \left[\hat{J}_{c_{1,p},\gamma_{c_{1,p}}}
      \hat{G}^{+,-}_{B,\gamma_{c_{1,p}},c_{1,p}}\right]_{(2,2)/(0,0)}(\omega,\check{\varphi},\{\psi_{F,k,l}\},-V)\\\nonumber
    &=& 2i \sum_{\check{n}\in S_{\check{n}}}
    \left[\hat{J}_{c_{1,p},\gamma_{c_{1,p}}}
      \hat{G}^{+,-}_{B,\gamma_{c_{1,p}},c_{1,p},\check{n}}\right]_{(1,1)/(0,0)}(\omega,\check{\varphi},\{\psi_{F,k,l}\},V)\\ &&\times
    \sin\left(\check{n}.\check{\varphi}\right) .  \label{eq:I-Gpm-B}
\end{eqnarray}}

{Now, we note that $\hat{J}^\dagger=\hat{J}$ combined to
\begin{equation}
\left[\hat{g}^{A,R}
  \left(\omega,\check{\varphi},\{\psi_{F,k,l}\},V\right)
  \right]^\dagger
  =
  \hat{g}^{R,A}
  \left(\omega,\check{\varphi},\{\psi_{F,k,l}\},V\right)
\end{equation}
and to the Dyson Eq.~(\ref{eq:otimes}) leads to
$\left(\hat{G}^{A,R}\right)^\dagger=\hat{G}^{R,A}$. Conversely,
combining to Eq.~(\ref{eq:Gpm-B}) yields
$\left(\hat{G}^{+,-}_B\right)^\dagger =\hat{G}^{+,-}_B$.  We deduce
that $\langle\langle I_{q,B} \rangle\rangle$ takes the following form of the Ambegaokar-Baratoff formula
\begin{eqnarray}
    \label{eq:AB-B}
    \langle\langle I_{q,B}\rangle\rangle&=&\frac{2e}{\hbar} \sum_p \int
    d\omega n_F(\omega) \sum_{\check{n}\in S_{\check{n}}}
    \\ &&\nonumber\mbox{Im} \langle\langle
    J_{c_{1,p},\gamma_{c_{1,p}}}
    G^{+,-}_{B,{\check{n}},\gamma_{c_{1,p}},c_{1,p}}
    \left(\omega,\{\psi_{F,k,l}\},V\right)\rangle\rangle
    \sin\left({\check{n}}.\check{\varphi}\right)\\
    &+& \left( V \rightarrow -V \right)
    \nonumber
    ,
\end{eqnarray}
where $I_{q,B}(V)=[I_B(V)+I_B(-V)]/2$.}

{It is concluded that both Eq.~(\ref{eq:AB-A}) for the
  ``quasi-equilibrium quartet current'' $\langle\langle I_{q,A}
  \rangle\rangle$ entering or exiting $S_{c,1}$ and
  Eq.~(\ref{eq:AB-B}) for the ``nonequilibrium quartet current''
  $\langle\langle I_{q,B}\rangle\rangle$ take the form of the
  ``generalized Ambegaokar-Baratoff formula''
  Eq.~(\ref{eq:Iqc-non-perturbatif}) where the coefficients $X(2n,p)$
  acquire a dependence on the voltage~$V$.}

\subsubsection{Conclusion on the Harvard group experiment
  \cite{Harvard-group-experiment}}
\label{sec:AB-V-exp}

It deduced that the assumption of arbitrary interface transparencies
and finite bias voltage~$V$, combined to mimicking disorder in the
superconducting leads by averaging over $\{k_F R_{k,l}\}$, leads to
the following statement:

``{\it Experimental evidence for $I'_{c}(0)\ne I'_{c}(1/2)$}'' implies
``{\it Evidence for transmission of odd number of Cooper pairs into
  $S_{{c_1}}$ or $S_{{c_2}}$}''.

{This statement implies ``{\it Evidence for microscopic processes containing
  odd number of electron-hole or hole-electron conversions in lead
  $S_{c,1}$.}''}

{Going one step further, we note that multiple quartet superconducting
  diffusion modes of the $\langle\langle g_{(1,2)}g_{(1,2)}
  \rangle\rangle$-type in Eqs.~(\ref{eq:diff-mode1})
  and~(\ref{eq:F-3TQ}) necessarily imply even numbers of electron-hole
  or hole-electron conversions. Thus, the requirement of odd number of
  electron-hole or hole-electron conversions in lead $S_{c,1}$ implies
  that at least one $\langle\langle g_{(1,1)} g_{(1,2)}\rangle\rangle$
  mode of the 4TSQ-type is involved in the corresponding diagram, see
  Eqs.~(\ref{eq:diff-mode1}) and~(\ref{eq:F-4TSQ}).}

{This argument relies on gathering the nonlocal Green's functions in a
  pair-wise manner. It would break down for localized contacts such
  that $r_0\alt l_e$, because the unpaired ``local'' electron-hole
  conversions would have to be taken into account on the same footing
  as the pairs of nonlocal Green's functions.}

{The paper is concluded with the following remark regarding the
  Harvard group experiment \cite{Harvard-group-experiment}:}

``{\it Experimental evidence for different values of the critical
  currents between $\Phi/\Phi_0=0$ and $\Phi/\Phi_0=1/2$ '',
  i.e. $I'_{c}(0)\ne I'_{c}(1/2)$}

implies

``{\it Evidence for the four-terminal 4TSQ}''.

This statement holds for arbitrary device parameters and it was
demonstrated within the physically motivated approximation of the
model~III discussed in the above subsection~\ref{sec:connection-ball}.

\section{Conclusions}
\label{sec:conclusions}

Now, we provide a summary of the paper in
subsection~\ref{sec:summary-of-the-paper}, specific conclusions on the
Harvard group experiment in subsection~\ref{sec:specific-conclusion}
and final remarks and outlook in subsection~\ref{sec:final-remarks}.

\subsection{Summary of the paper}
\label{sec:summary-of-the-paper}
In this paper, we provided a possible mechanism for the inversion in
the critical current $I_c(\Phi/\Phi_0)$ on the quartet line in a
four-terminal Josephson junction (see figures~\ref{fig:device}
and~\ref{fig:device2}), in connection with the recent Harvard group
experiment \cite{Harvard-group-experiment}.

{The Harvard group experiment \cite{Harvard-group-experiment} uses
  graphene gated away from the Dirac point, which was modeled as a
  simple 2D metal with circular Fermi surface. We took the two
  dimensions of the graphene sheet into account while ignoring the
  effects related to the Dirac cones.}

{Specifically, we calculated microscopically the Josephson relations
  from lowest-order perturbation theory in the tunnel amplitudes,
  assuming in addition the adiabatic limit. We found that the three-
  and four-terminal quartet channels interfere with each other in the
  critical current on the quartet line. The ``standard''
  three-terminal 3TQ$_1$ transmit two pairs into $S_{c,1}$ and the
  three-terminal 3TQ$_2$ transmit two pairs into $S_{c,2}$. The
  nonstandard four-terminal 4TSQ transmit two Coopers in the same
  quantum process but in a split manner, {\it i.e.} one pair into
  $S_{c,1}$ and the other one into $S_{c,2}$.}

{We found that the four-terminal 4TSQ do not contribute to the
  dc-current if a 1D or a 3D metal and multichannel contacts are used
  instead of the considered 2D metal. The importance of 2D is related
  to the general properties of the solutions of the wave equation,
  which imply a wake in even dimension (such as 2D), but not in odd
  dimension (such as 1D or 3D). We demonstrated that the ``2D quantum
  wake'' can synchronize two Josephson junctions by the exchange of a
  quasiparticle at the $S_{{c_1}}$ and $S_{{c_2}}$ contacts, yielding
  a nonvanishingly small four-terminal 4TSQ critical current.}

{We demonstrated that, with a 2D metal, lowest-order perturbation
  theory and the adiabatic limit produce $\pi$- and $0$-shifted
  current-phase relations for the three-terminal 3TQ$_1$, 3TQ$_2$ and
  for the four-terminal 4TSQ respectively. This implies inversion
  $I_{c}(0)<I_c(1/2)$ between the critical currents $I_c(0)$ and
  $I_c(1/2)$ at fluxes $\Phi/\Phi_0=0$ and $\Phi/\Phi_0=1/2$
  respectively. In turn, experimental evidence for inversion implies
  $(\pi,0)$ or $(0,\pi)$ shifts for the three-terminal 3TQ$_1$,
  3TQ$_2$ and the four-terminal 4TSQ respectively. This type of
  experiment cannot determine which of the three-terminal 3TQ$_1$,
  3TQ$_2$ or the four-terminal 4TSQ is $\pi$-shifted, the other being
  $0$-shifted.}

We proposed an analogy with experiments on a SQUID containing
$\pi$- and $0$-shifted Josephson junctions \cite{Guichard}. In this
analogy with Ref.~\onlinecite{Guichard}, the $\pi$- and $0$-shifted
three-terminal 3TQ$_1$, 3TQ$_2$ and the four-terminal 4TSQ play the
role of the $\pi$- or $0$-shifted two-terminal Josephson junctions
inserted in the SQUID loop respectively.

  In addition, the perturbative calculation predicts that the relative
  weight of the three-terminal 3TQ$_1$, 3TQ$_2$ and the four-terminal
  4TSQ changes with gate voltage on the 2D metal, in a way which is
  compatible with the experimental data of the Harvard group
  \cite{Harvard-group-experiment}.

{We also generalized our theory to arbitrary interface transparencies
  and finite bias voltage $V$ on the quartet line. This generalization
  was based on the even or odd parity of the number of Cooper pairs
  transmitted into $S_{c,1}$ or $S_{c,2}$ by the three-terminal
  3TQ$_1$, 3TQ$_2$ and the four-terminal 4TSQ.}

{ We treated the ingredients of ``arbitrary interface transparencies''
  and ``finite bias voltage'' within a physically-motivated
  approximation for disorder. The current entering or exiting
  $S_{c,1}$ takes the form of the generalized Ambegaokar-Baratoff
  relation from which we could infer that the 4TSQ imply different
  values for critical current $I_c(\Phi/\Phi_0)$ at $\Phi/\Phi_0=0$
  and $\Phi/\Phi_0=1/2$.} We argued within this framework of
physically-motivated approximation that ``{\it $I_{c}(0)\ne
  I_{c}(1/2)$}'' implies ``{\it Evidence for the nonstandard
  four-terminal 4TSQ}''.

\subsection{Specific conclusion on the Harvard
  group experiment \cite{Harvard-group-experiment}}

\label{sec:specific-conclusion}

{To summarize, our theory teaches the following on the Harvard group
  experimental data \cite{Harvard-group-experiment}:}

(i) Perturbation theory in the tunnel amplitudes combined to the
$V=0^+$ adiabatic limit produces the ``{\it Inversion in
  $I_c(\Phi/\Phi_0)$ between $\Phi/\Phi_0=0$ and $\Phi/\Phi_0=1/2$}''
which is observed in the Harvard group experiment
\cite{Harvard-group-experiment}.

{(ii) The calculated gate voltage dependence of the critical current
  oscillations as a function of magnetic field is compatible with the
  Harvard group experimental results \cite{Harvard-group-experiment}.}

{(iii) We argued that ``{\it Experimental evidence for $I_c(0)\ne
    I_c(1/2)$}'' implies ``{\it Evidence for the four-terminal
    4TSQ}''. Thus, our model implies that the Harvard group experiment
  \cite{Harvard-group-experiment} is evidence for the four-terminal
  4TSQ.}

\subsection{Final remarks and outlook}
\label{sec:final-remarks}
{ The four-terminal 4TSQ shown on figures~\ref{fig:nambu-schematics}c,
  d and on figure~\ref{fig:diagrams-downmixing} are robust against
  taking the long junction limit $R_0/x_0\alt l_\varphi$ along the
  $x$-axis direction, where $l_\varphi$ is the mesoscopic phase
  coherence length of the 2D metal. It is assumed in addition that the
  device remains short along the $y$-direction, {\it i.e.}
  $R_0/y_0\alt \xi_{dirty}(0)$.} The device geometry is shown on
figure~\ref{fig:device2}.

{ More precisely, the nonlocal $\langle\langle {g_{(1,1)}g_{(1,1)}}
  \rangle\rangle$ mode connects $S_{c,1}$ and $S_{c,2}$ through the 2D
  metal by the quantum wake, see the highlighted section of the
  four-terminal 4TSQ diagram on
  figure~\ref{fig:nambu-schematics}d. Eq.~(\ref{eq:gA-asymptotics})
  provides the expression of each single-particle Green's function
  $g_{(1,1)}$ entering the $\langle\langle{ g_{(1,1)}g_{(1,1)}}
  \rangle\rangle$ mode through the 2D metal. Both nonlocal $g_{(1,1)}$
  are in the ``electron-electron'' channel and they are both evaluated
  at the same wave-vector $k_e=k_F+\omega/v_F$. Thus, $\langle\langle{
    g_{(1,1)}g_{(1,1)}} \rangle\rangle$ is not washed out by multichannel
  averaging if the energy $\omega$ is larger than the Thouless energy
  $\hbar v_F/R$ associated to the separation $R$ between the
  contacts.} This is why the four-terminal 4TSQ critical current given
by the diagram on figure~\ref{fig:diagrams-downmixing}d remains large as
long as the device dimension $R_0/x_0$ along the $x$-axis is in the
mesoscopic domain, {\it i.e.} $R_0/x_0\alt l_\varphi$.

{In addition, the $V=0$ limit of phase-biased superconductors (instead
  of the $V=0^+$ adiabatic limit of a voltage-biased device) also
  involves long-distance coupling between the Andreev bound states
  associated to each pair of Josephson junctions sharing a 2D metal as
  a common weak link, according to to $R_0/x_0\alt l_\varphi$
  mentioned above in the geometry on figure~\ref{fig:device2}.  }

{These arguments show that the four-terminal 4TSQ constitute a
  nonstandard ``mesoscopic'' channel of quantum coherent
  synchronization, which operates in between the quartets at the
  smallest scale and the early 1980s synchronization of macroscopic
  Josephson circuits \cite{NJ1,NJ2}. An interesting complementary
  point of view is to approach this mesoscopic regime from the
  classical limit, {\it i.e.} to incorporate quantum fluctuation in
  the classical circuit models.}

{We also note that comparing Eq.~(\ref{eq:order8-4-bis-result}) for
  the three-terminal 3TQ$_1$ critical current $I_{c,\,3TQ_1}$ to
  Eq.~(\ref{eq:Ic-4TSQ-result-2}) for the four-terminal 4TSQ critical
  current $I_{c,\,4TSQ}^{(2)}$ leads to the following order of
  magnitude for their ratio:
  \begin{equation}
    K=\left|\frac{I_{c,\,4TSQ}^{(2)}}{I_{c,\,3TQ_1}}\right|
    \approx
    \left(\frac{J_0}{W}\right)^{4}\frac{\sqrt{{\cal S}_{contact}}}{l_e}
    .
  \end{equation}
  It was assumed implicitly in this paper~I that perturbation theory
  is converging, which implies $K< 1$. Paper~III of the series will
  address resummations of the perturbative expansions if the diffusion
  modes proliferate for $K>1$.}

{Finally, we point out that, in the presence of Coulomb interactions,
  the charging energy is larger for the three-terminal 3TQ$_1$ and
  3TQ$_2$ (involving four fermions in $S_{c,1}$ or four fermions in
  $S_{c,2}$) than for the four-terminal 4TSQ (involving one pair in
  $S_{c,1}$ and another one in $S_{c,2}$). Thus, static Coulomb
  interactions favor the nonstandard four-terminal 4TSQ over the
  three-terminal 3TQ$_1$, 3TQ$_2$. It would be interesting to
  address dynamical Coulomb blockade \cite{Levy-Yeyati} for the device
  in figures~\ref{fig:device} and~\ref{fig:device2}.}

\section*{Acknowledgements}
The author acknowledges the stimulating collaboration with the Harvard
group (K.~Huang, Y.~Ronen and P.~Kim) to
subsections~\ref{sec:butterfly-quartet-diagrams}
and~\ref{sec:4TFSQ-process}. The author wishes to thank R.~Danneau and
B.~Dou\c{c}ot for their collaboration on an early attempt to find
signatures of the 2D quantum wake in the signal of multiple Andreev
reflection through bilayer graphene. R.~Danneau also provided useful
comments on an early version of the manuscript. The author
acknowledges fruitful discussions with D.~Feinberg. The author thanks
the Infrastructure de Calcul Intensif et de Donn\'ees (GRICAD) for use
of the resources of the M\'esocentre de Calcul Intensif de
l’Universit\'e Grenoble-Alpes (CIMENT).

\appendix

\section{Green's function of a 2D metal}
\label{app:2D-metal-Greens-function}
We start this Appendix with the Fourier transform of the Green's
function between two tight-binding sites separated by distance $R$:
\begin{eqnarray}
  \label{eq:gA11}
  &&g^{A,(1,1)/(2,2)}(R,\omega)= \\
  &&\int_{-\pi}^\pi d\theta
  \int_0^{+\infty} \frac{k dk}{(2\pi)^2} \exp\left(ikR\cos\theta\right)
  g^{A,(1,1)/(2,2)}({\bf k},\omega)
  ,
  \nonumber
\end{eqnarray}
where
\begin{eqnarray}
  g^{A,(1,1)}({\bf k},\omega)
  &=&\frac{1}{\omega-\xi_{\bf k}-i\eta}\\
  g^{A,(2,2)}({\bf k},\omega)
  &=&\frac{1}{\omega+\xi_{\bf k}-i\eta}
  \label{eq:gA-complement2}
  ,
\end{eqnarray}
with $\xi_{\bf k}$ the kinetic energy of the 2D plane-wave state ${\bf
  k}$ with respect to the chemical potential $\mu=\hbar^2 k_F^2/2m$,
and where $k_F$ and $m$ are the Fermi wave-vector and the band-mass
respectively. The superscripts ``$(1,1)$'' or ``$(2,2)$'' in
Eqs.~(\ref{eq:gA11})-(\ref{eq:gA-complement2}) refer to propagation in
the electron-electron or hole-hole channels respectively.

Eq.~(\ref{eq:gA11}) is written as
\begin{equation}
  \label{eq:gA11-2}
  g^{A}(R,\omega)=\int_0^{+\infty} \frac{k dk}{(2\pi)}
  \frac{J_0(kR)}{\omega-\epsilon \xi_{\bf k}-i\eta}
    ,
\end{equation}
where the Bessel function
\begin{equation}
  J_0(x)=\frac{1}{2\pi} \int_{-\pi}^{\pi}
  \exp\left(ix\cos\theta\right) d\theta
\end{equation}
was introduced in Eq.~(\ref{eq:gA11-2}). The parameter $\epsilon$ takes
the values $\epsilon=\pm 1$ for the $(1,1)$ and $(2,2)$ components respectively.
We consider a pole at
\begin{equation}
  k\equiv k_0\simeq k_F+\frac{\epsilon (\omega-i\eta)}{v_F}
\end{equation}
and the residue is evaluated according to
\begin{equation}
  \omega-\epsilon \xi_{k_0+\delta k}-i\eta
    \simeq
    -\frac{\hbar^2}{2m} \epsilon k_0 \delta k
    \simeq - \epsilon v_F k_0 \delta k
    .
\end{equation}
Considering in addition that $\mbox{sign} \left(\mbox{Im} k_0\right)=-
\epsilon$ and using contour integration in the complex $k$ plane leads
to
\begin{eqnarray}
  \label{eq:Green2D-1}
  g^{A,(1,1)}(R)&=&g^{A,(2,2)}(R)=\frac{i}{W} J_0(k_F
  R)\\ g^{R,(1,1)}(R)&=&g^{R,(2,2)}(R)=-\frac{i}{W} J_0(k_F R)
  \label{eq:Green2D-2}
      ,
\end{eqnarray}
where the limit $\omega R /v_F\alt 1$ is considered. Noting that
$\omega$ has typical order of magnitude set by the superconducting
gap, the condition $\omega R /v_F\alt 1$ yields the short junction
limit $R \alt \xi_{ball}(0)$ for each pair of superconductor-2D
metal-superconductor Josephson junction, where the ballistic
superconducting coherence length is given by Eq.~(~\ref{eq:xi-ball-0})
and Eq.~(\ref{eq:xi-ball}).

Eqs.~(\ref{eq:Green2D-1})-(\ref{eq:Green2D-2}) lead to
Eqs.~(\ref{eq:gA-asymptotics})-(\ref{eq:gR-asymptotics}) if the
realistic condition $2\pi/k_F\ll R \alt \xi_{ball}(0)$ is
fulfilled on the separation $R$.

\section{Superconducting diffusion modes in the dirty limit}
\label{app:dirty-limit}
The goal of this Appendix is to treat disorder in the superconducting
leads of the considered four-terminal device. The calculations are
based on Ref.~\onlinecite{Smith-Ambegaokar}.

Averaging the one-particle Green's functions over disorder in the Born
approximation is summarized in subsection~\ref{sec:1part-GF}.
Subsection~\ref{sec:2part-GF} presents the calculation of the
disorder-averaged superconducting modes in the dirty limit.

The superconducting modes are formed with products of Nambu Green's
functions scattering together on the same configuration of disorder,
see subsection~\ref{sec:mode-averaging}.

Now, the superconducting diffusion modes are calculated in the ladder
approximation.

\subsection{One-particle Green's function averaged over disorder}
\label{sec:1part-GF}

We start with the expression of the average over disorder of the
one-particle Green's functions \cite{Abrikosov}.

Using the notations of Ref.~\onlinecite{Smith-Ambegaokar}, the average
superconducting Green's function takes the following form in the Born
approximation:
\begin{equation}
  \hat{g}^{A,R}(\xi_{\bf k},\omega)=\frac{ \hat{\cal N}_{\bf k}(\omega)}
      {{\cal D}_{\bf k}(\omega)}
  ,
\end{equation}
with
\begin{eqnarray}
  \hat{\cal N}_{\bf k}(\omega)&=&\overline{\omega} \mp i\eta
    +\xi_{\bf k} \hat{\tau}_3 +
    \overline{\Delta} \hat{\tau}_1\\
    {\cal D}_{\bf k}(\omega)&=&
  \left(\overline{\omega} \mp i\eta\right)^2 -
    \overline{\Delta}^2 - \xi_{\bf k}^2
  ,
\end{eqnarray}
where $\omega$ and ${\bf k}$ refer to the energy and wave-vector
respectively.  The notations $\overline{\omega}$, $\overline{\Delta}$
and $s(\omega)$ stand for
\begin{eqnarray}
  \label{eq:overline-omega}
  \overline{\omega}&=&\omega\left(1+\frac{1}{s(\omega) \tau}\right)\\
  \label{eq:overline-Delta}
  \overline{\Delta}&=&\Delta\left(1+\frac{1}{s(\omega) \tau}\right)\\
  s^2(\omega)&=&\Delta^2 - \omega^2
  .
\end{eqnarray}

\subsection{Product of two Green's function averaged over disorder}
\label{sec:2part-GF}

Now, we evaluate the average over disorder of the product of two Nambu
Green's functions. An intermediate result on two integrals is
presented in subsection~\ref{sec:useful-integral}.  The resummations of
the ladders is presented in subsection~\ref{sec:matrix-series}.

\subsubsection{Intermediate result on evaluation of useful integrals}
\label{sec:useful-integral}
Following Ref.~\onlinecite{Smith-Ambegaokar}, we evaluate the
following integrals at energy $|\omega|<\Delta$ within the gap:
\begin{eqnarray}
  \label{eq:calI-A-A}
  &&  {\cal I}_\alpha^{A,A}({\bf q},\omega) \\
\nonumber
  &=& v^2
  \int \frac{d {\bf k}}{(2\pi)^3}\hat{\tau}_3
  \hat{g}^A({\xi}_{\bf k},\omega) \hat{\tau}_\alpha
  \hat{g}^A(\xi_{{\bf k}+{\bf q}},\omega)\hat{\tau}_3\\
    \label{eq:calI-A-R}
  && {\cal I}_\alpha^{A,R}({\bf q},\omega) \\ \nonumber &=& v^2 \int
    \frac{d {\bf k}}{(2\pi)^3}\hat{\tau}_3 \hat{g}^A({\xi}_{\bf
      k},\omega) \hat{\tau}_\alpha \hat{g}^R(\xi_{{\bf k}+{\bf
        q}},\omega)\hat{\tau}_3
    ,
\end{eqnarray}
where $\hat{\tau}_\alpha$ are the $2\times 2$ Pauli matrices.

The first step of the calculation is to combine the relation
\begin{equation}
\label{eq:integral-d3k}
\int \frac{d{\bf k}}{(2\pi)^3} = \frac{1}{8\pi^2}
\int_{-1}^1 du \int_{-\infty}^{+\infty} k^2 dk \exp(ikRu) 
\end{equation}
to contour integration for the integral over $k$.

The poles of $\hat{g}^A({\xi}_{\bf k},\omega)$ are given by the
solutions of ${\cal D}_{\bf k}^A(\omega)=0$ at $k=k_0$, namely
\begin{equation}
\xi_{{\bf k}_0}^2=
\overline{\omega}^2 - \overline{\Delta}^2
-2i\eta \overline{\omega}=0
.
\end{equation}
A branch-cut along the $\omega<0$ axis leads to
\begin{equation}
  \label{eq:xi-k-0}
  \xi_{{\bf k}_0}^{(\epsilon)}(\omega)=i\epsilon
  \sqrt{\overline{\Delta}^2 -
    \overline{\omega}^2} ,
\end{equation}
with
\begin{equation}
  \label{eq:epsilon}
  \epsilon=\mbox{sgn}\left(-\eta\overline{\omega}\right)
  .
\end{equation}

The solutions of Eq.~(\ref{eq:xi-k-0}) for $k_0>0$ or $k_0<0$ are
given by
\begin{equation}
  \label{eq:k0-eps-eps_prime}
  k_0^{(\epsilon,\epsilon')}(\omega)=\epsilon' \left( k_F + \frac{i\epsilon
    \overline{s}(\omega)}{\hbar v_F}\right) ,
\end{equation}
with $\epsilon'=\pm$.

No reason is advocated for why a constraint holds between
$\mbox{Re}\left[k_0(\omega)\right]$ and $\mbox{Im}
\left[k_0(\omega)\right]$. Instead, the four solutions are relevant to
a double interface, {\it i.e.} those with exponential growth or decay
along the positive or negative directions of propagation. For
instance, the four waves are involved in the solution of the
Bogoliubov-de Gennes equations within the Blonder-Tinkham-Klapwijk
approach \cite{BTK} for a normal metal-superconductor-normal metal
double junction \cite{Floser}, with separation between the contacts
comparable to the superconducting coherence length. Thus, the four
wave-vectors $k_0^{(\epsilon,\epsilon')}$ in
Eq.~(\ref{eq:k0-eps-eps_prime}) are taken into account in the
following, where $\epsilon$ and $\epsilon'$ are free to take the
values $\epsilon=\pm$ and $\epsilon'=\pm$.

To obtain the residue, we evaluate ${\cal D}_{{\bf k}}(\omega)$ for
$k=k_0^{(\epsilon,\epsilon')}(\omega)+\delta k$ in the limit $\delta k\rightarrow 0$:
\begin{equation}
  {\cal D}_{k_0+\delta k}(\omega)\simeq -2\frac{\hbar^2}{m}
  k_0^{(\epsilon,\epsilon')}(\omega) \xi_{k_0}^{(\epsilon)}(\omega)
  \delta k .
\end{equation}
Next, we expand ${\cal D}_{{\bf k}_0+{\bf q}}(\omega)$ according to
\begin{equation}
  {\cal D}_{{\bf k}_0+{\bf q}}(\omega) \simeq - \frac{\hbar^2}{m}
  k_0^{(\epsilon,\epsilon')}(\omega) q u \left\{2 \xi_{{\bf
      k}_0}^{(\epsilon)}(\omega) +\frac{\hbar^2}{m}
  k_0^{(\epsilon,\epsilon')}(\omega) q u \right\}
\end{equation}
and we expand $\hat{\cal N}_{{\bf k}_0+{\bf q}}(\omega)$ according to
\begin{equation}
  \label{eq:expansion-N}
  \hat{\cal N}_{{\bf k}_0+{\bf q}}(\omega) \simeq
  \hat{\cal N}_{{\bf k}_0}(\omega)+ q u \left( \frac{\hbar^2 k_0^{(\epsilon,\epsilon')}(\omega)}{m}
  \hat{\tau}_3\right)
  .
\end{equation}
Evaluating the contribution proportional to $q u$ in
Eq.~(\ref{eq:expansion-N}) leads to a term which is subleading in the
limit of a dirty superconductor {\it i.e.} if the elastic scattering
time is much smaller than $\hbar/\Delta$.

Contour integration leads to the dominant contribution of the pole at
$k_0^{(\epsilon,\epsilon')}(\omega)$ in $\hat{g}^A(\xi_{\bf
  k},\omega)$ in Eqs.~(\ref{eq:calI-A-A})-(\ref{eq:calI-A-R}):
\begin{widetext}
  \begin{equation}
    \label{eq:general-expression}
{\cal I}_\alpha^{A,A}({\bf q},\omega,\hat{\tau}_\alpha) =
{\cal I}_\alpha^{A,R}({\bf q},\omega,\hat{\tau}_\alpha) 
= \frac{i \epsilon \epsilon'v^2}{4\pi} \int_{-1}^1 du \sum_{\epsilon,\epsilon'=\pm}
\frac{ \hat{\tau}_3\left(\overline{\omega} + \xi_{k_0}^{(\epsilon)}
\hat{\tau}_3 + \overline{\Delta}\hat{\tau}_1\right)
\hat{\tau}_\alpha
\left(\overline{\omega} + \tilde{\xi}_{k_0}^{(\epsilon)}
\hat{\tau}_3 + \overline{\Delta}\hat{\tau}_1\right)\hat{\tau}_3}
  {\left(\frac{-2 \hbar^2 k_0^{(\epsilon,\epsilon')}(\omega)}{m} \right)^2
    \left(1+ \frac{\hbar^2 k_0^{(\epsilon,\epsilon')}(\omega) q u}
         {2 m \xi_{k_0}^{(\epsilon)}}\right) q u}
  ,
  \end{equation}
\end{widetext}
to which is added the ``$u\rightarrow -u$'' contribution from the pole
$k_1$ of $\hat{g}^{A,R}(\xi_{{\bf k}+{\bf q}},\omega)$ in
Eqs.~(\ref{eq:calI-A-A})-(\ref{eq:calI-A-R}), defined as ${\cal
  D}_{{\bf k}_1+{\bf q}}^A(\omega)=0$.

Selecting $\tilde{\xi}_{k_0}^{(\epsilon)}=\left(\xi_{k_0}^{(\epsilon)}
\right)^*$ in Eq.~(\ref{eq:general-expression}) involves the product
mentioned above of``forward-moving exponentially decay'' and
``backward-moving exponentially growth'', which leads to
\begin{equation}
{\cal I}_\alpha^{A,A}({\bf q},\omega,\hat{\tau}_\alpha) = {\cal
  I}_\alpha^{A,R}({\bf q},\omega,\hat{\tau}_\alpha)
.
\label{eq:equality}
\end{equation}

This identity is compatible with the real-valued $\hat{g}^A_{{\bf
    x}_1,{\bf x}_2}(\omega)=\hat{g}^R_{{\bf x}_1,{\bf x}_2}(\omega)$
if $|\omega|<\Delta$ and $\eta=0^+$, see
Eq.~(\ref{eq:gA-supra-general-ballistique}). Namely, the real-space
Dyson equations used to describe scattering on disorder take real
values, which is compatible with the preceding Eq.~(\ref{eq:equality})
obtained from integration over the wave-vector ${\bf k}$.

The next step is to evaluate the numerator of
Eq.~(\ref{eq:general-expression}) for all Pauli matrices
$\hat{\tau}_\alpha$. We separate between the following contributions:
\begin{equation}
  \left(\overline{\omega} + i \overline{s} \hat{\tau}_3
  +\overline{\Delta} \hat{\tau}_1\right) \hat{\tau}_\alpha
  \left(\overline{\omega} - i \overline{s} \hat{\tau}_3
  +\overline{\Delta} \hat{\tau}_1\right)
  =\hat{X}^{(1)}+\hat{X}^{(2)}
  ,
\end{equation}
with
\begin{eqnarray}
  \hat{X}^{(1)}(\hat{\tau}_\alpha)&=& \left(\overline{\omega}
  +\overline{\Delta} \hat{\tau}_1\right) \hat{\tau}_\alpha
  \left(\overline{\omega} +\overline{\Delta} \hat{\tau}_1\right) +
  \overline{s}^2 \hat{\tau}_3 \hat{\tau}_\alpha\hat{\tau}_3
  \\ \hat{X}^{(2)}(\hat{\tau}_\alpha)&=& i \overline{s} \left[
    \hat{\tau}_3 \hat{\tau}_\alpha \left( \overline{\omega}
    +\overline{\Delta} \hat{\tau}_1\right) - \left( \overline{\omega}
    +\overline{\Delta} \hat{\tau}_1\right) \hat{\tau}_3
    \hat{\tau}_\alpha\right]
  .
\end{eqnarray}
A straightforward calculation leads to
\begin{eqnarray}
  \hat{X}^{(1)}(\hat{I})&=& 2 \overline{\Delta}^2 \hat{I} + 2 \overline{\omega}
  \overline{\Delta} \hat{\tau}_1\\
  \hat{X}^{(1)}(\hat{\tau}_1)&=& 2 \overline{\omega} \overline{\Delta} \hat{I} +
  2\overline{\omega}^2 \hat{\tau}_1 \\
  \hat{X}^{(1)}(\hat{\tau}_3)&=&0\\
  \hat{X}^{(1)}(\hat{\tau}_3\hat{\tau}_1)&=&-2\left(\overline{\Delta}^2
  - \overline{\omega}^2 \right)\hat{\tau}_3\hat{\tau}_1
  .
\end{eqnarray}
Expanding Eq.~(\ref{eq:general-expression}) to order $u^2$ and
integrating over the variable $u$ leads to the following contributions
to Eq.~(\ref{eq:general-expression}):
\begin{eqnarray}
  \nonumber
{\cal I}^{A,A,(1)}({\bf q},\omega,\hat{I}) &=&
{\cal I}^{A,R,(1)}({\bf q},\omega,\hat{I}) \\
\label{eq1}
&=& {\cal A}\left[\overline{\Delta}^2 \hat{I} - \overline{\omega}
  \overline{\Delta} \hat{\tau}_1 \right]\\
  \nonumber
{\cal I}^{A,A,(1)}({\bf q},\omega,\hat{\tau}_1) &=&
{\cal I}^{A,R,(1)}({\bf q},\omega,\hat{\tau}_1) \\
\label{eq2}
&=& {\cal A}\left[\overline{\omega} \overline{\Delta} \hat{I} -
  \overline{\omega}^2 \hat{\tau}_1 \right]\\
  \nonumber
{\cal I}^{A,A,(1)}({\bf q},\omega,\hat{\tau}_3) &=&
{\cal I}^{A,R,(1)}({\bf q},\omega,\hat{\tau}_3) \\
&=& 0\\
  \nonumber
{\cal I}^{A,A,(1)}({\bf q},\omega,\hat{\tau}_3\hat{\tau}_1) &=&
{\cal I}^{A,R,(1)}({\bf q},\omega,\hat{\tau}_3\hat{\tau}_1) \\
&=&-{\cal A}\left(\overline{\Delta}^2-\overline{\omega}^2
\right) \hat{\tau}_3\hat{\tau}_1
,
\end{eqnarray}
with
\begin{equation}
  \label{eq:calA}
        {\cal A}= \frac{k_F^3}{16\pi \left[\overline{s}(\omega)\right]^3
          \epsilon_F}
        \left[1-\frac{v_F^2 q^2}
          {12 \left[\overline{s}(\omega)\right]^2}\right]
 .
\end{equation}
The notation $\overline{s}(\omega)$ in Eq.~(\ref{eq:calA}) stands for
$\left[\overline{s}(\omega))\right]^2=
\overline{\Delta}^2-\overline{\omega}^2$, where $\overline{\omega}$ and
$\overline{\Delta}$ are given by Eqs.~(\ref{eq:overline-omega})
and~(\ref{eq:overline-Delta}) respectively.

\subsubsection{Summing the $2\times 2$ matrix geometric
  series in the ladder approximation}
\label{sec:matrix-series}
Iterating Eqs.~(\ref{eq1}) and~(\ref{eq2}) to form the first ``rungs''
of the superconducting diffusion modes in the ladder approximation
yields
\begin{eqnarray}
  \nonumber &&{\cal I}^{A,A,(1)}({\bf q},\omega,{\cal
    I}^{A,A,(1)}({\bf q},\omega,\hat{I})) \\ &=&
\label{eq1bis}
 {\cal A}\left[\overline{\Delta}^2 - \overline{\omega}^2\right]
      {\cal I}^{A,A,(1)}({\bf q},\omega,\hat{I})\\
  \nonumber
&&      {\cal I}^{A,A,(1)}({\bf q},\omega,{\cal I}^{A,A,(1)}({\bf q},\omega,\hat{\tau}_1))\\
      &=&
\label{eq2bis}
      {\cal A}\left[\overline{\Delta}^2 - \overline{\omega}^2\right]
      {\cal I}^{A,A,(1)}({\bf q},\omega,\hat{\tau}_1)
      ,
\end{eqnarray}
where ${\cal A}$ is given by Eq.~(\ref{eq:calA}).

We find the following at the next order:
\begin{eqnarray}
  \nonumber
      &&{\cal I}^{A,A,(1)}({\bf q},\omega,{\cal I}^{A,A,(1)}({\bf q},\omega,{\cal I}^{A,A,(1)}({\bf q},\omega,\hat{I}))) \\
      &=&
\label{eq1ter}
\left[ {\cal A}\left[\overline{\Delta}^2 - \overline{\omega}^2\right]\right]^2
      {\cal I}^{A,A,(1)}({\bf q},\omega,\hat{I})\\
  \nonumber
&&      {\cal I}^{A,A,(1)}({\bf q},\omega,{\cal I}^{A,A,(1)}({\bf q},\omega,\hat{\tau}_1)))\\
      &=&
\label{eq2ter}
\left[{\cal A}\left[\overline{\Delta}^2 - \overline{\omega}^2\right]
  \right]^2
      {\cal I}^{A,A,(1)}({\bf q},\omega,\hat{\tau}_1)
      .
\end{eqnarray}

Summing all terms to infinite order leads to the disorder-averaged
superconducting diffusion modes:
\begin{eqnarray}
  \label{eq:disorder1}
  \overline{\hat{g}\hat{I}\hat{g}}(q,\omega)&=&
  \frac{1}{16\pi W}
  \frac{1}{2\sqrt{\Delta^2-\omega²}+{\cal D} q^2}\\
  \nonumber
&&\times  \frac{\Delta^2 \hat{I}+\omega\Delta \hat{\tau}_1}
      {\Delta^2-\omega^2}\\
      \label{eq:disorder2}
  \overline{\hat{g}\hat{\tau}_1\hat{g}}(q,\omega)&=&
  \frac{1}{16\pi W}
  \frac{1}{2\sqrt{\Delta^2-\omega²}+{\cal D} q^2}\\
\nonumber
  &&\times
  \frac{\omega \Delta \hat{I}-\omega^2 \hat{\tau}_1}
       {\Delta^2-\omega^2}\\
       \overline{\hat{g}\hat{\tau}_3\hat{g}}(q,\omega)&=&0\\
       \label{eq:disorder3}
       \left|\overline{\hat{g}\hat{\tau}_3\hat{\tau}_1\hat{g}}(q,\omega)\right|
       &\ll& \left|\overline{\hat{g}\hat{I}\hat{g}}(q,\omega)\right|\\
       \left|\overline{\hat{g}\hat{\tau}_3\hat{\tau}_1\hat{g}}(q,\omega)\right|
       &\ll& \left|\overline{\hat{g}\hat{\tau}_1\hat{g}}(q,\omega)\right|
       ,
       \label{eq:disorder4}
\end{eqnarray}
where the diffusion constant is ${\cal D}=v_F^2 \tau/3$.

Expanding the Nambu components of Eq.~(\ref{eq:disorder1}) leads to
\begin{eqnarray}
  \label{eq:Nambu-expansion1}
\left[\overline{\hat{g}\hat{I}\hat{g}}(q,\omega)\right]_{(1,1)}&=&
\overline{\hat{g}_{(1,1)}\hat{g}_{(1,1)}}(q,\omega) \\&&+ \nonumber
\overline{\hat{g}_{(1,2)}\hat{g}_{(2,1)}}(q,\omega)\\
\left[\overline{\hat{g}\hat{I}\hat{g}}(q,\omega)\right]_{(1,2)}&=&
\overline{\hat{g}_{(1,1)}\hat{g}_{(1,2)}}(q,\omega) \\&&+\nonumber
\overline{\hat{g}_{(1,2)}\hat{g}_{(2,2)}}(q,\omega)\\
\left[\overline{\hat{g}\hat{I}\hat{g}}(q,\omega)\right]_{(1,1)}&=&
\overline{\hat{g}_{(2,1)}\hat{g}_{(1,1)}}(q,\omega) \\&&+\nonumber
\overline{\hat{g}_{(2,2)}\hat{g}_{(2,1)}}(q,\omega)\\
\label{eq:Nambu-expansion2}
\left[\overline{\hat{g}\hat{I}\hat{g}}(q,\omega)\right]_{(1,2)}&=&
\overline{\hat{g}_{(2,1)}\hat{g}_{(1,2)}}(q,\omega) \\&&+\nonumber
\overline{\hat{g}_{(2,2)}\hat{g}_{(2,2)}}(q,\omega)
.
\end{eqnarray}
Similarly, expanding the Nambu components of Eq.~(\ref{eq:disorder1})
leads to
\begin{eqnarray}
  \label{eq:Nambu-expansion1bis}
\left[\overline{\hat{g}\hat{\tau}_1\hat{g}}\right]_{(1,1)}(q,\omega)&=&
\overline{\hat{g}_{(1,1)}\hat{g}_{(2,1)}}(q,\omega) \\&&+\nonumber
\overline{\hat{g}_{(1,2)}\hat{g}_{(1,1)}}(q,\omega)\\
\left[\overline{\hat{g}\hat{\tau}_1\hat{g}}\right]_{(1,2)}(q,\omega)&=&
\overline{\hat{g}_{(1,1)}\hat{g}_{(2,2)}}(q,\omega) \\&&+\nonumber
\overline{\hat{g}_{(1,2)}\hat{g}_{(1,2)}}(q,\omega)\\
\left[\overline{\hat{g}\hat{\tau}_1\hat{g}}\right]_{(1,1)}(q,\omega)&=&
\overline{\hat{g}_{(2,1)}\hat{g}_{(2,1)}}(q,\omega) \\&&+\nonumber
\overline{\hat{g}_{(2,2)}\hat{g}_{(1,1)}}(q,\omega)\\
\label{eq:Nambu-expansion2bis}
\left[\overline{\hat{g}\hat{\tau}_1\hat{g}}\right]_{(1,2)}(q,\omega)&=&
\overline{\hat{g}_{(2,1)}\hat{g}_{(2,2)}}(q,\omega) \\&&+\nonumber
\overline{\hat{g}_{(2,2)}\hat{g}_{(1,2)}}(q,\omega)
.
\end{eqnarray}
Combining Eqs.~(\ref{eq:disorder1})-(\ref{eq:disorder4}) to
Eqs.~(\ref{eq:Nambu-expansion1})-(\ref{eq:Nambu-expansion2}) and to
Eqs.~(\ref{eq:Nambu-expansion1bis})-(\ref{eq:Nambu-expansion2bis}),
and next replacing Eqs.~(\ref{eq:disorder3})-(\ref{eq:disorder4}) by
$\overline{\hat{g}\hat{\tau}_3\hat{\tau}_1\hat{g}}(q,\omega)=0$ yields
\begin{eqnarray}
  \nonumber
  &&\overline{{g}_{(1,1)}{g}_{(1,1)}}(q,\omega)=\overline{{g}_{(2,2)}{g}_{(2,2)}}(q,\omega)\\ \label{eq:mode-EC}
  &=&\frac{1}{16\pi W} \frac{1}{2\sqrt{|\Delta|^2-\omega²}+{\cal D}
    q^2} \frac{|\Delta|^2} {|\Delta|^2-\omega^2}\\\nonumber
  &&\overline{{g}_{(1,2)}{g}_{(2,1)}}(q,\omega)=\overline{{g}_{(2,1)}{g}_{(1,2)}}(q,\omega)\\\label{eq:mode-CAR}
  &=&\frac{1}{16\pi W} \frac{1}{2\sqrt{|\Delta|^2-\omega²}+{\cal D}
    q^2} \frac{|\Delta|^2} {|\Delta|^2-\omega^2} ,
\end{eqnarray}
where Eqs.~(\ref{eq:mode-EC}) and~(\ref{eq:mode-CAR}) are relevant to
elastic cotunneling (EC) \cite{Hekking,Melin-Feinberg} and crossed
Andreev reflection (CAR) \cite{Hekking,Feinberg,Melin-Feinberg} in a
three-terminal normal metal-superconductor-normal metal device, with
contacts separated by distance comparable to the superconducting
coherence length.

Eqs.~(\ref{eq:disorder1})-(\ref{eq:disorder4}),
Eqs.~(\ref{eq:Nambu-expansion1})-(\ref{eq:Nambu-expansion2}) and
Eqs.~(\ref{eq:Nambu-expansion1bis})-(\ref{eq:Nambu-expansion2bis})
yield
\begin{eqnarray}
  \nonumber
  &&\overline{{g}_{(1,2)}{g}_{(1,2)}}(q,\omega)\\
    \label{eq:g12g12-1}
  &=&\frac{1}{16\pi W}
  \frac{1}{2\sqrt{|\Delta|^2-\omega²}+{\cal D} q^2} \frac{|\Delta|^2
    \exp\left(2i\varphi_N\right)}
       {|\Delta|^2-\omega^2}\\ \nonumber &&\overline{{g}_{(2,1)}{g}_{(2,1)}}(q,\omega)\\
         \label{eq:g12g12-2}
       &=&\frac{1}{16\pi W} \frac{1}{2\sqrt{|\Delta|^2-\omega²}+{\cal
             D} q^2} \frac{|\Delta|^2\exp\left(-2i\varphi_N\right)}
               {|\Delta|^2-\omega^2} ,
\end{eqnarray}
where $\varphi_N$ is the macroscopic phase variable of the considered
superconductor $S_N$.  Eq.~(\ref{eq:g12g12-1}) is relevant to double
crossed Andreev reflection \cite{Freyn,Melin1} and to the three-terminal
3TQ$_1$, 3TQ$_2$ in the presence of biasing at opposite voltages,
with distance between the interfaces approximately set by the
superconducting coherence length.  

Eqs.~(\ref{eq:disorder1})-(\ref{eq:disorder4}),
Eqs.~(\ref{eq:Nambu-expansion1})-(\ref{eq:Nambu-expansion2}) and
Eqs.~(\ref{eq:Nambu-expansion1bis})-(\ref{eq:Nambu-expansion2bis})
imply
\begin{eqnarray}
  \nonumber
  &&\overline{{g}_{(1,1)}{g}_{(2,2)}}(q,\omega)
  =
  \overline{{g}_{(2,2)}{g}_{(1,1)}}(q,\omega)\\
  &=&\frac{1}{16\pi W}
  \frac{1}{2\sqrt{|\Delta|^2-\omega²}+{\cal D} q^2} \frac
       {2\omega^2-|\Delta|^2}{|\Delta|^2-\omega^2}
            \label{eq:g11g22}
            ,
\end{eqnarray}
which is relevant to double elastic cotunneling \cite{Freyn} (dEC) in
a three-terminal Josephson junction biased at identical voltages.

Finally, the following superconducting diffusion modes are relevant to
the four-terminal 4TSQ which is the subject of this paper~I, and it
is also relevant to the normal metal-superconductor-superconductor
double junction of Ref.~\onlinecite{NSS}:
\begin{eqnarray}
    \nonumber
  &&\overline{{g}_{(1,1)}{g}_{(1,2)}}(q,\omega)=
  \overline{{g}_{(2,2)}{g}_{(1,2)}}(q,\omega)\\
\nonumber
  &=&\overline{{g}_{(1,2)}{g}_{(1,1)}}(q,\omega)=
    \overline{{g}_{(1,2)}{g}_{(2,2)}}(q,\omega)\\ 
    &=&\frac{1}{16\pi W} \frac{1}{2\sqrt{|\Delta|^2-\omega²}+{\cal D}
      q^2} \frac{\omega|\Delta| \exp\left(i\varphi_N\right)}
             {|\Delta|^2-\omega^2}
  \label{eq:g11g12-1}
\end{eqnarray}
and
\begin{eqnarray}
  \nonumber
  &&\overline{{g}_{(2,2)}{g}_{(2,1)}}(q,\omega)=
  \overline{{g}_{(1,1)}{g}_{(2,1)}}(q,\omega)\\
  \nonumber
  &=&\overline{{g}_{(2,1)}{g}_{(2,2)}}(q,\omega)=
  \overline{{g}_{(2,1)}{g}_{(1,1)}}(q,\omega)\\
  &=&\frac{1}{16\pi W} \frac{1}{2\sqrt{|\Delta|^2-\omega²}+{\cal D}
    q^2} \frac{\omega|\Delta| \exp\left(-i\varphi_N\right)}
           {|\Delta|^2-\omega^2}
           .
  \label{eq:g11g12-2}
\end{eqnarray}

Eqs.~(\ref{eq:F-3TQ}) and~(\ref{eq:F-4TSQ}) in
subsection~\ref{sec:mode-averaging} are deduced from
Eqs.~(\ref{eq:g12g12-1}) and~(\ref{eq:g11g12-1}) respectively. 

\section{The ballistic limit}
\label{app:clean-limit}
This Appendix addresses the limit of ballistic superconducting leads,
which is relevant to the model~III introduced in
subsection~\ref{sec:connection-ball}.

The expression of the ballistic superconducting diffusion modes is
provided in
subsection~\ref{sec:ballistic-modes}. Subsection~\ref{sec:explanation}
provides an explanation to different signs in the dirty and ballistic
limits of $\langle\langle g_{(1,1)} g_{(1,2)} \rangle \rangle$.

\subsection{Expression of the superconducting modes of a ballistic
  superconductor}
\label{sec:ballistic-modes}

Now, we provide the expression of the ballistic superconducting modes
propagating in the superconducting lead $S_N$ with phase $\varphi_N$,
on the basis of averaging over $k_F R_1$ the product of Nambu
superconducting Green's functions given in
Eq.~(\ref{eq:gA-supra-general-ballistique}):
\begin{widetext}
\begin{eqnarray}
  \label{eq:clean1}
  \overline{{g}^A_{(1,1)}(R_{\alpha,\beta},\omega)
    {g}^A_{(1,1)}(R_{\beta,\alpha},\omega)} =
  \overline{{g}^A_{(2,2)}(R_{\alpha,\beta},\omega)
    {g}^A_{(2,2)}(R_{\beta,\alpha},\omega)} &=& \frac{1}{2 W^2 (k_F
    R_1)^2}\frac{\Delta^2}{\Delta^2-\omega^2}
  \exp{\left(-\frac{2R_1}{\xi_{ball}(\omega)}\right)}\\ \label{eq:clean2}
  \overline{{g}^A_{(1,2)}(R_{\alpha,\beta},\omega)
    {g}^A_{(2,1)}(R_{\beta,\alpha},\omega)} =
  \overline{{g}^A_{(2,1)}(R_{\alpha,\beta},\omega)
    {g}^A_{(2,1)}(R_{\beta,\alpha},\omega)} &=& \frac{1}{2 W^2 (k_F
    R_1)^2} \frac{\Delta^2}{\Delta^2-\omega^2}
  \exp{\left(-\frac{2R_1}{\xi_{ball}(\omega)}\right)}\\
\label{eq:clean3}
\overline{{g}^A_{(1,2)}(R_{\alpha,\beta},\omega)
  {g}^A_{(1,2)}(R_{\beta,\alpha},\omega)} &=& \frac{1}{2 W^2 (k_F
  R_1)^2}\frac{\Delta^2}{\Delta^2-\omega^2}
\exp{\left(-\frac{2R_1}{\xi_{ball}(\omega)}\right)} \exp{\left(2 i
  \varphi_N\right)}\\\label{eq:clean4}
\overline{{g}^A_{(2,1)}(R_{\alpha,\beta},\omega)
  {g}^A_{(2,1)}(R_{\beta,\alpha},\omega)} &=& \frac{1}{2 W^2 (k_F
  R_1)^2}\frac{\Delta^2}{\Delta^2-\omega^2}
\exp{\left(-\frac{2R_1}{\xi_{ball}(\omega)}\right)} \exp{\left(-2 i
  \varphi_N\right)}\\ \label{eq:clean5}
\overline{{g}^A_{(1,1)}(R_{\alpha,\beta},\omega)
  {g}^A_{(2,2)}(R_{\beta,\alpha},\omega)} =
\overline{{g}^A_{(2,2)}(R_{\alpha,\beta},\omega)
  {g}^A_{(1,1)}(R_{\beta,\alpha},\omega)} &=& \frac{1}{2 W^2 (k_F
  R_1)^2}\frac{2\omega^2-\Delta^2}{\Delta^2-\omega^2}
\exp{\left(-\frac{2R_1}{\xi_{ball}(\omega)}\right)}\\
\nonumber
\overline{{g}^A_{(1,1)}(R_{\alpha,\beta},\omega)
  {g}^A_{(1,2)}(R_{\beta,\alpha},\omega)} =
\overline{{g}^A_{(2,2)}(R_{\alpha,\beta},\omega)
  {g}^A_{(1,2)}(R_{\beta,\alpha},\omega)} &=&
\overline{{g}^A_{(1,2)}(R_{\alpha,\beta},\omega)
  {g}^A_{(1,1)}(R_{\beta,\alpha},\omega)}=\overline{{g}^A_{(1,2)}(R_{\alpha,\beta},\omega)
  {g}^A_{(1,2)}(R_{\beta,\alpha},\omega)} \\\label{eq:clean6}
&=& \frac{1}{2 W^2 (k_F
  R_1)^2}\frac{-\omega\Delta}{\Delta^2-\omega^2}
\exp{\left(-\frac{2R_1}{\xi_{ball}(\omega)}\right)} \exp{\left(i
  \varphi_N\right)}\\ \nonumber
\overline{{g}^A_{(1,1)}(R_{\alpha,\beta},\omega)
  {g}^A_{(2,1)}(R_{\beta,\alpha},\omega)} =
\overline{{g}^A_{(2,2)}(R_{\alpha,\beta},\omega)
  {g}^A_{(2,1)}(R_{\beta,\alpha},\omega)} &=&
\overline{{g}^A_{(2,1)}(R_{\alpha,\beta},\omega)
  {g}^A_{(1,1)}(R_{\beta,\alpha},\omega)}=\overline{{g}^A_{(2,1)}(R_{\alpha,\beta},\omega)
  {g}^A_{(2,1)}(R_{\beta,\alpha},\omega)} \\&=& \frac{1}{2 W^2 (k_F
  R_1)^2}\frac{-\omega\Delta}{\Delta^2-\omega^2}
\exp{\left(-\frac{2R_1}{\xi_{ball}(\omega)}\right)} \exp{\left(-i
  \varphi_N\right)}\label{eq:clean7}
\end{eqnarray}
,
\end{widetext}
where $R_1\equiv R_{\alpha,\beta}=R_{\beta,\alpha}$ denotes the
separation between the tight-binding sites $\alpha$ and $\beta$. The
overline in Eqs.~(\ref{eq:clean1})-(\ref{eq:clean7}) denotes averaging
over the $k_F R_1$ oscillations in the superconducting Green's
function, and the $1/2$ coefficient appearing in front of
Eqs.~(\ref{eq:clean1})-(\ref{eq:clean7}) originates from
$\overline{\cos^2(k_F R_1)}=\overline{\sin^2(k_F R_1)}=1/2$, see the
ballistic-limit superconducting Green's function given by
Eq.~(\ref{eq:gA-supra-general-ballistique}).

Eqs.~(\ref{eq:clean1}) and~(\ref{eq:clean2}) for elastic cotunneling
(EC) \cite{Hekking,Melin-Feinberg} and crossed Andreev reflection
(CAR) \cite{Feinberg,Hekking,Melin-Feinberg} in the ballistic limit
are associated to Eqs.~(\ref{eq:mode-EC}) and~(\ref{eq:mode-CAR}) in
the dirty limit respectively. Eq.~(\ref{eq:clean3})-(\ref{eq:clean4})
for the quartets \cite{Freyn,Melin1} corresponds to
Eqs.~(\ref{eq:g12g12-1})-(\ref{eq:g12g12-2}).  Eq.~(\ref{eq:clean5})
for double elastic cotunneling (dEC)\cite{Freyn} is associated to
Eqs.~(\ref{eq:g11g22}).  Eqs.~(\ref{eq:clean6})-(\ref{eq:clean7})
which are relevant to the four-terminal 4TSQ and to a normal
metal-superconductor-superconductor double junction \cite{NSS} have
Eqs.~(\ref{eq:g11g12-1})-(\ref{eq:g11g12-2}) as their dirty-limit
analog.

\subsection{Discussion of the opposite signs
  of Eqs.~(\ref{eq:g11g12-1})-(\ref{eq:g11g12-2})
  Eqs.~(\ref{eq:clean6})-(\ref{eq:clean7}) }
\label{sec:explanation}

{ The $\langle\langle{g_{(1,1)}g_{(1,2)}}\rangle\rangle$ mode relevant
  to the four-terminal 4TSQ is found to have opposite signs in the
  dirty and ballistic limits, see
  Eqs.~(\ref{eq:g11g12-1})-(\ref{eq:g11g12-2}) in the dirty limit and
  Eqs.~(\ref{eq:clean6})-(\ref{eq:clean7}) in the ballistic limit
  respectively. We provide now an explanation to the different signs
  appearing in the dirty and ballistic limits.}

{ Specifically, we mimic the disorder scattering
  potential $v$ [see Eqs.~(\ref{eq:calI-A-A})-(\ref{eq:calI-A-R})] by
  a tunnel barrier. Namely, we replace a $S_aIS_0IS_b$ double junction
  (where $I$ is an insulator) by a $S_aIS_1IS_2IS_b$ triple junction
  where $S_1$ and $S_2$ are two ballistic superconductors separated by
  an insulating tunnel barrier.}

\begin{figure}[htb]
  \includegraphics[width=\columnwidth]{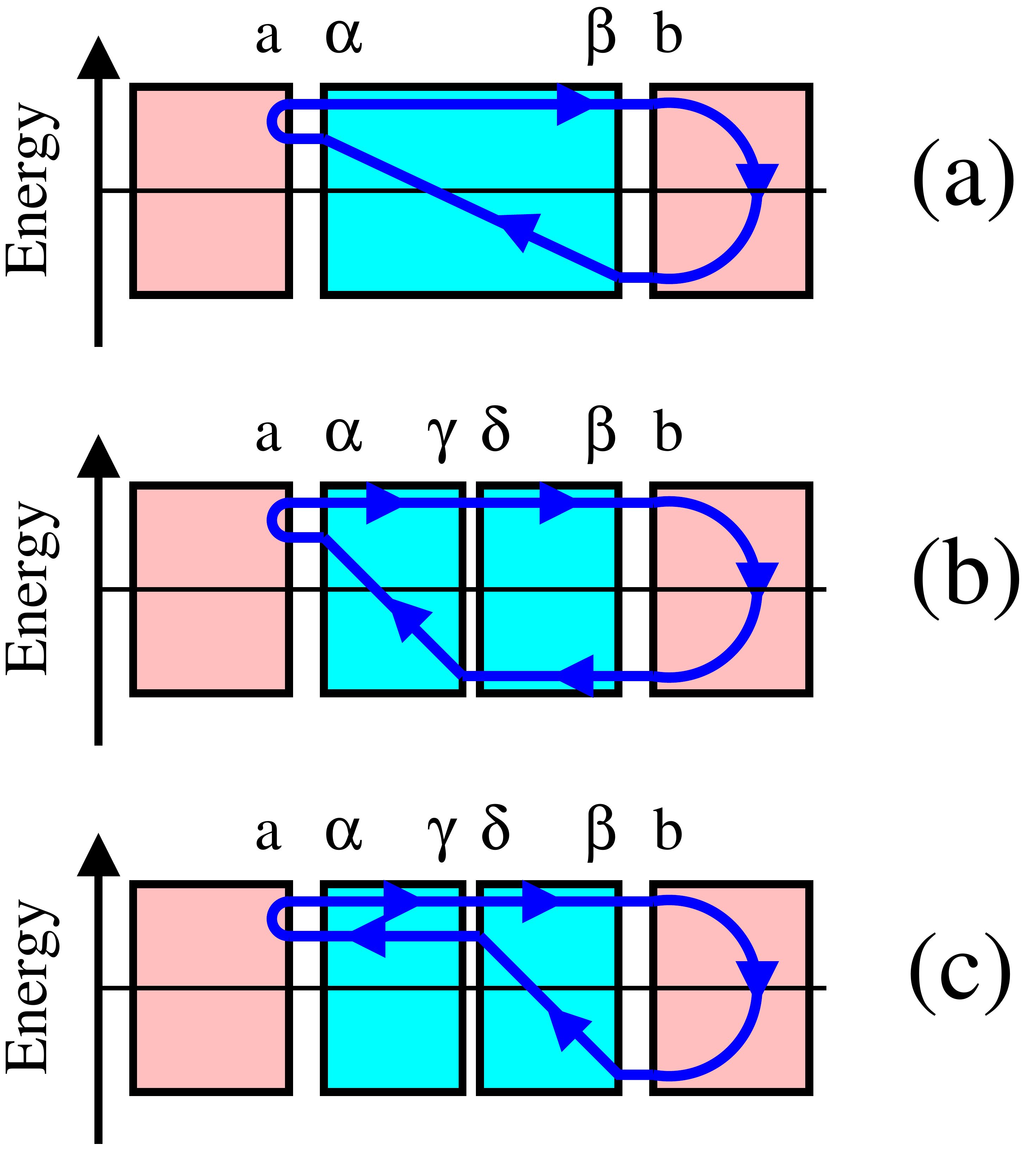}
  \caption{{Panel a shows the $\langle\langle g_{(1,1)}g_{(1,2)}
      \rangle\rangle$ mode at lowest order in the tunnel amplitudes
      in a $S_aIS_0IS_b$ double junction [${\cal D}^{A,(0)}$ term in
        Eq.~(\ref{eq:D4TSQ0})]. Panels b and c show schematically to
      the ${\cal D}^{A,(1)}$ and ${\cal D}^{A,(2)}$ terms in a
      $S_aIS_1IS_2IS_b$ triple junction, see
      Eqs.~(\ref{eq:DA1-1})-(\ref{eq:DA1-2}) and
      Eqs.~(\ref{eq:DA2-1})-(\ref{eq:DA2-2}) respectively.}
    \label{fig:4TSQ-diagrams}}
\end{figure}

{We start with the simple $\langle\langle
  g_{(1,1)}g_{(1,2)}\rangle\rangle$ four-terminal 4TSQ diagram in a
  $S_aIS_0IS_b$ double junction, see
  figure~\ref{fig:4TSQ-diagrams}a:
  \begin{eqnarray}
    \nonumber
 &&         {\cal D}^{A,(0)}=\\
         \label{eq:D4TSQ0}
          &&\langle\langle
          J_{a,\alpha}^{(1,1)}g_{\alpha,\beta}^{A,(1,1)}
          J_{\beta,b}^{(1,1)} g_{b,b}^{A,(1,2)} J_{b,\beta}^{(2,2)}
          g_{\beta,\alpha}^{A,(2,1)} J_{\alpha,a}^{(1,1)}
          g_{a,a}^{A,(1,1)} \rangle\rangle \\ &&= - J_0^4
          \langle\langle g_{a,a}^{A,(1,1)} \rangle\rangle
          \langle\langle g_{b,b}^{A,(1,2)} \rangle\rangle
          \langle\langle g_{\alpha,\beta}^{A,(1,1)}
          g_{\beta,\alpha}^{A,(2,1)} \rangle\rangle .
          \label{eq:D4TSQ0-bis}
  \end{eqnarray}
  Thus, ${\cal D}^{A,(0)}$ in
  Eq.~(\ref{eq:D4TSQ0})-(\ref{eq:D4TSQ0-bis}) is given by
  \begin{eqnarray}
    {\cal D}^{A,(0)}=- J_0^4
    \langle\langle g_{a,a}^{A,(1,1)} \rangle\rangle
    \langle\langle g_{b,b}^{A,(1,2)} \rangle\rangle
                   {\cal A}_{ \alpha,\gamma}
                   ,
  \end{eqnarray}
  where Eq.~(\ref{eq:clean7}) was written as
  \begin{eqnarray}
    \label{eq:modeA1}
    &&\langle\langle g_{\alpha,\gamma}^{A,(1,1)}
    g_{\gamma,\alpha}^{A,(2,1)} \rangle\rangle \\&=&\nonumber
    -{\cal A}_{\alpha,\gamma}
    \frac{(\omega-i\eta)|\Delta|}{|\Delta|^2-(\omega-i\eta)^2}
    \exp\left(-i\varphi_c\right)
    ,
  \end{eqnarray}
  and the ``geometrical'' prefactor ${\cal A}_{\alpha,\gamma}$ is
  given by
  \begin{equation}
    {\cal A}_{\alpha,\gamma}=\frac{1}{2W^2} \frac{1}{(k_F
      R_{\alpha,\gamma})^2}
    \exp\left(-\frac{2R_{\alpha,\gamma}}{\xi_{ball}(\omega-i\eta)}\right)
    .
  \end{equation}
}

{Now, we consider a $S_aIS_1IS_2IS_b$ triple junction
  and start with the diagrams ${\cal D}^{A,(1)}$ and ${\cal
    D}^{A,(2)}$ appearing at the lowest order $(J_0/W)^6$ in the tunnel
  amplitudes, see figures~\ref{fig:4TSQ-diagrams}b and c
  respectively:
  \begin{widetext}
    \begin{eqnarray}
      \label{eq:DA1-1}
    {\cal D}^{A,(1)}&=& \langle\langle
    J_{a,\alpha}^{(1,1)}g_{\alpha,\gamma}^{A,(1,1)}
    J_{\gamma,\delta}^{(1,1)} g_{\delta,\beta}^{A,(1,1)}
    J_{\beta,b}^{(1,1)} g_{b,b}^{A,(1,2)} J_{b,\beta}^{(2,2)}
    g_{\beta,\delta}^{A,(2,2)} J_{\delta,\gamma}^{(2,2)}
    g_{\gamma,\alpha}^{A,(2,1)} J_{\alpha,a}^{(1,1)} g_{a,a}^{A,(1,1)}
    \rangle\rangle \\ &=& J_0^{12} \langle\langle g_{a,a}^{A,(1,1)}
    \rangle\rangle \langle\langle g_{b,b}^{A,(1,2)} \rangle\rangle
    \langle\langle g_{\alpha,\gamma}^{A,(1,1)} g_{\gamma,\alpha}^{A,(2,1)}
    \rangle\rangle \langle\langle g_{\delta,\beta}^{A,(1,1)}
    g_{\beta,\delta}^{A,(2,2)} \rangle\rangle .
    \label{eq:DA1-2}
  \end{eqnarray}
and
\begin{eqnarray}
  \label{eq:DA2-1}
    {\cal D}^{A,(2)}&=& \langle\langle
    J_{a,\alpha}^{(1,1)}g_{\alpha,\gamma}^{A,(1,1)}
    J_{\gamma,\delta}^{(1,1)} g_{\delta,\beta}^{A,(1,1)}
    J_{\beta,b}^{(1,1)} g_{b,b}^{A,(1,2)} J_{b,\beta}^{(2,2)}
    g_{\beta,\delta}^{A,(2,1)} J_{\delta,\gamma}^{(1,1)}
    g_{\gamma,\alpha}^{A,(1,1)} J_{\alpha,a}^{(1,1)} g_{a,a}^{A,(1,1)}
    \rangle\rangle \\ &=& J_0^{12} \langle\langle g_{a,a}^{A,(1,1)}
    \rangle\rangle \langle\langle g_{b,b}^{A,(1,2)} \rangle\rangle
    \langle\langle g_{\alpha,\gamma}^{A,(1,1)} g_{\gamma,\alpha}^{A,(1,1)}
    \rangle\rangle \langle\langle g_{\delta,\beta}^{A,(1,1)}
    g_{\beta,\delta}^{A,(2,1)} \rangle\rangle .
\label{eq:DA2-2}
  \end{eqnarray}
  \end{widetext}
Eq.~(\ref{eq:clean5}) leads to
\begin{eqnarray}
  \label{eq:modeA2}
  &&\langle\langle g_{\delta,\beta}^{A,(1,1)} g_{\beta,\delta}^{A,(2,2)}
  \rangle\rangle\\ &=&{\cal A}_{\beta,\delta}
  \frac{2(\omega-i\eta)^2-|\Delta|^2}{|\Delta|^2-(\omega-i\eta)^2} .
\nonumber
\end{eqnarray}}

{We deduce the following:
  \begin{eqnarray}
    \label{eq:D0-result}
  {\cal D}^{A,(0)} &=& - {\cal B}
  \frac{(\omega-i\eta)^2|\Delta|^2}{\left(|\Delta|^2-(\omega-i\eta)^2
  \right)^2}\\
  {\cal D}^{A,(1)} &=& {\cal C}
  \frac{(\omega-i\eta)^2|\Delta|^2\left(2(\omega-i\eta)^2
-|\Delta|^2\right)}{\left(|\Delta|^2-(\omega-i\eta)^2\right)^3}\\
{\cal D}^{A,(2)} &=& {\cal C}
\frac{(\omega-i\eta)^2|\Delta|^4}{\left(|\Delta|^2-(\omega-i\eta)^2\right)^3}
  ,
\end{eqnarray}
where ${\cal B}$ and ${\cal C}$ have the same sign.
Thus,
\begin{eqnarray}
  \label{eq:D1-plus-D2-result}
{\cal D}^{A,(1)} +
{\cal D}^{A,(2)}
=
\frac{2 {\cal C} (\omega-i\eta)^4|\Delta|^2}
     {\left(|\Delta|^2-(\omega-i\eta)^2
       \right)^3}
\end{eqnarray}
has sign which is opposite to that of ${\cal D}_0^A$.}

{It is concluded to change of sign in $\langle\langle
  g_{(1,1)}g_{(1,2)}\rangle\rangle$ connecting $S_a$ and $S_b$,
  between the following two situations:}

{(i) Eq.~(\ref{eq:D0-result}) for ${\cal D}^{A,(0)}$ in a $S_LIS_0IS_R$
  double ballistic tunnel junction.}

{(ii) Eq.~(\ref{eq:D1-plus-D2-result}) for ${\cal D}^{A,(1)} + {\cal
    D}^{A,(2)}$ in a $S_aIS_1IS_2IS_b$ triple ballistic tunnel
  junction.}

{This resolves the apparent paradox that emerged between the preceding
  calculations of the $\langle\langle g_{(1,1)} g_{(1,2)} \rangle\rangle$
  mode:}

{(iii) Eqs.~(\ref{eq:clean6})-(\ref{eq:clean7}) for $\langle \langle
  g_{(1,1)} g_{(1,2)} \rangle \rangle$ in the ballistic limit.}

{(iv) Eqs.~(\ref{eq:g11g12-1})-(\ref{eq:g11g12-2}) for
  impurity scattering the $\langle \langle g_{(1,1)} g_{(1,2)} \rangle
  \rangle$ superconducting diffusion mode at the order $v^2$.}

\section{Details on the calculation of the current-phase relations}
  \label{app:details-calculation-of-currents} 

  \subsection{Demonstration of Eq.~(\ref{eq:order8-4-bis-result})
    for $I_{c,\,3TQ_1}$}
  \label{app:I1}
  We evaluate the following integral:
  \begin{equation}
    \label{eq:I1-def}
    I_1=\int_{-\infty}^0 \frac{\Delta^4}{\left(\Delta^2-(\omega-i\eta)^2
      \right)^2} d\omega
    .
  \end{equation}
  We expand according to $\omega-i\eta=-\Delta+\epsilon$. Assuming
  $|\epsilon|\ll\Delta$ leads to
  \begin{equation}
    \label{eq:I1-result}
    \frac{\Delta^4}{\left(\Delta^2-(\omega-i\eta)^2
      \right)^2}
    \simeq
    \frac{\Delta^2}{4\epsilon^2}+ \frac{\Delta}{8\epsilon}+ ...
    .
\end{equation}
If $\eta\ll\Delta$,
contour integration yields
\begin{equation}
  I_1=\frac{i\pi \Delta}{4}
  .
  \label{eq:useful-integral1}
\end{equation}
Eq.~(\ref{eq:useful-integral1}) is used in
subsection~\ref{sec:butterfly-quartet-diagrams} to deduce the quartet
current-phase relations
Eqs.~(\ref{eq:quartet-current-phase-relation})-(\ref{eq:order8-4-bis-result})
from Eq.~(\ref{eq:order8-4}).

\subsection{Demonstration of Eq.~(\ref{eq:Ic-4TSQ-result-1})
  for $I_{c,\,4TSQ}^{(1)}$}
\label{app:I2}
Now, we evaluate
\begin{equation}
  \label{eq:I2-definition}
  I_2=\int_{-\infty}^0 \frac{\Delta^6}{\left(\Delta^2-(\omega-i\eta)^2
    \right)^3} d\omega
  .
\end{equation}
Expanding around $\omega-i\eta=-\Delta+\epsilon$ with $|\epsilon|\ll\Delta$
leads to
\begin{equation}
\frac{\Delta^6}{\left(\Delta^2-(\omega-i\eta)^2
  \right)^3}\simeq
\frac{\Delta^3}{8\epsilon^3}
+ \frac{3\Delta^2}{16\epsilon^2}
+ \frac{3\Delta}{16\epsilon}
+ ...
\end{equation}
The residue of the simple pole is $3\Delta/16$. Contour integration
yields
\begin{equation}
  \label{eq:result_I2}
  I_2=\frac{3i\pi\Delta}{8}
\end{equation}
if $\eta\ll\Delta$. Eq.~(\ref{eq:result_I2}) is used in
subsection~\ref{sec:subleading} to deduce the four-terminal 4TSQ
current-phase relations
Eqs.~(\ref{eq:4TSQ-current-phase-relation})-(\ref{eq:Ic-4TSQ-result-1})
from Eq.~(\ref{eq:4TSQ-subleading4}).

\subsection{Demonstration of Eq.~(\ref{eq:Ic-4TSQ-result-2})
  for $I_{c,\,4TSQ}^{(2)}$}
\label{app:I3}
The integral
\begin{equation}
  \label{eq:I3-def}
  I_3=\int_{-\infty}^0 \frac{(\omega-i\eta)^2
    \Delta^4}{\left(\Delta^2-(\omega-i\eta)^2 \right)^3}  d\omega
\end{equation}
is equal to:
\begin{equation}
  \label{eq:I3-result}
  I_3=I_2
\end{equation}
if $\eta\ll\Delta$, where $I_2$ is given by Eq.~(\ref{eq:result_I2}).
The leading-order four-terminal 4TSQ current-phase relations
Eqs.~(\ref{eq:I-(2)})-(\ref{eq:Ic-4TSQ-result-2}) are deduced from
Eq.~(\ref{eq:4TSQ-leading4}) by making use of
Eqs.~(\ref{eq:I3-def})-(\ref{eq:I3-result}), see
subsection~\ref{sec:leading}.

\subsection{Averaging the product of three superconducting
  Green's functions}
\label{app:averaging-3gs}

The average over disorder of the product of three Green's functions
\begin{equation}
  \label{eq:C3}
  {\cal C}_3=\langle\langle{g_{(1,1)} g_{(1,2)} g_{(2,2)}}\rangle\rangle
\end{equation}
appears in the expression of the four-terminal 4TSQ at the orders
$(J_0/W)^{12}$ and $\sqrt{{\cal S}_{contact}}/l_e$, see Eqs.~(9)-(46)
  in subsection~II\,A of the Supplemental Material\cite{supplemental}.

Each of the three Green's function appearing in Eq.~(\ref{eq:C3}) can
propagate locally or nonlocally within the superconducting
lead $S_N$ having superconducting phase variable $\varphi_N$.

The Wick theorem leads to
\begin{eqnarray}
          \nonumber
      {\cal C}_3&=&\langle\langle{g_{(1,1)} g_{(1,2)}}\rangle\rangle
        \langle\langle g_{(2,2)}\rangle\rangle +
        \langle\langle{g_{(1,1)} g_{(2,2)}}\rangle\rangle
        \langle\langle g_{(1,2)} \rangle\rangle\\&+&
        \langle\langle g_{(1,1)}\rangle\rangle \langle\langle{g_{(1,2)}
          g_{(2,2)}}\rangle\rangle ,
  \label{eq:C3-bis}
\end{eqnarray}
where Eqs.~(\ref{eq:overline-omega}) and~(\ref{eq:overline-Delta})
imply the following for the local Green's functions of the
disordered superconductor $S_N$:
\begin{eqnarray}
  \langle\langle g_{(1,1)}\rangle\rangle&=&\langle\langle
  g_{(2,2)}\rangle\rangle
  =\frac{1}{W} \frac{-(\omega-i\eta)}
  {\sqrt{|\Delta|^2-(\omega-i\eta)^2}}\\
  \langle\langle g_{(1,2)}\rangle\rangle&=&
  =\frac{1}{W} \frac{|\Delta|}
  {\sqrt{|\Delta|^2-(\omega-i\eta)^2}}\exp\left(i\varphi_N\right)
  .
\end{eqnarray}
Combining with the energy dependence of Eq.~(\ref{eq:g11g22}) and
Eq.~(\ref{eq:g11g12-1}) implies the minus sign in
\begin{equation}
  \label{eq:last-eq}
  {\cal C}_3\sim
  -\frac{\Delta^3}{W^2 \left(|\Delta|^2-(\omega-i\eta)^2\right)^{3/2}}
  \exp\left(i\varphi_N\right)
  ,
\end{equation}
which is taken into account in the four-terminal 4TSQ critical current
at the orders $(J_0/W)^{12}$ and $\sqrt{{\cal S}_{contact}}/l_e$, see
Eq.~(\ref{eq:Ic-4TSQ-result-1}) in the paper and subsection~II\,A in
the Supplemental Material\cite{supplemental}.

\end{document}